\documentclass[a4paper]{article}
\usepackage{epsfig}
\usepackage{graphicx}  % standard LaTeX graphics tool
\usepackage{color}     % for color figures

\usepackage{color}
\usepackage{lscape}
\usepackage{amssymb}
\date{\today}
 \normalsize

\newcommand{\bmat}{\left(\begin{array}}
\newcommand{\emat}{\end{array}\right)}
\newcommand{\be}{\begin{equation}}
\newcommand{\ee}{\end{equation}}
\newcommand{\bea}{\begin{eqnarray}}
\newcommand{\eea}{\end{eqnarray}}

\def\gtwid{\mathrel{\raise.3ex\hbox{$>$\kern-.75em\lower1ex\hbox{$\sim$}}}}
\def\ltwid{\mathrel{\raise.3ex\hbox{$<$\kern-.75em\lower1ex\hbox{$\sim$}}}}
\def\gev{{\rm \, Ge\kern-0.125em V}}
\def\tev{{\rm \, Te\kern-0.125em V}}

\def    \be            {\begin{equation}}
\def    \ee            {\end{equation}}
\def    \bea           {\begin{eqnarray}}
\def    \eea           {\end{eqnarray}}

\def\g{\gamma}
\def\d{\delta}
\def\n{\nu}

\def\om{\omega}

\def\sig{\sigma}

\def\th{\theta}

\def\nn{\nonumber}
\def\d{\delta}
\def\D{\Delta}
\def\s{\sigma}
\def\r{\rho}
\def\t{\theta}

\def\me{\langle m\rangle_e}
\def\mee{\langle m\rangle_{ee}}
\begin{document}
\renewcommand{\thefootnote}{\fnsymbol{footnote}}
%\rightline{IPPP/03/52} \rightline{DCPT/03/104}
\vspace{.3cm}

\title{\Large\bf The One-zero Textures of Majorana  Neutrino Mass Matrix and Current Experimental Tests }

\author
{ \it \bf  E. I. Lashin$^{1,2,3}$\thanks{elashin@ictp.it} and N.
Chamoun$^{4,5}$\thanks{nchamoun@th.physik.uni-bonn.de} ,
\\
\small$^1$ Ain Shams University, Faculty of Science, Cairo 11566,
Egypt.\\
\small$^2$ Centre for Theoretical Physics, Zewail City of Science and Technology,\\
\small Sheikh Zayed, 6 October City, 12588, Giza, Egypt.\\
\small$^3$ The Abdus Salam ICTP, P.O. Box 586, 34100 Trieste, Italy. \\
\small$^4$  Physics Department, HIAST, P.O.Box 31983, Damascus,
Syria.\\
 \small$^5$  Physikalisches Institut der Universit$\ddot{a}$t Bonn, Nu${\ss}$alle 12, D-53115 Bonn, Germany. }

\maketitle

\begin{center}
\small{\bf Abstract}\\[3mm]
\end{center}
We carry out a  complete and systematic study of all the possible one-zero textures for the neutrino mass matrix in light of the recent neutrino oscillation data which hint to a relatively large non-vanishing value for the
 smallest mixing angle ($\theta_z$). We find that all the six possible one-zero textures are able to accommodate the experimental data, with five cases allowing also for non-invertible mass matrices. We present symmetry realizations for all the one-zero singular and non-singular patterns in type-I and type-II seesaw mechanism schemes.

\vspace{1.1cm}{\bf Keywords}: Neutrino masses,
\vspace{1.1cm}{\bf PACS numbers}: 14.60.Pq; 11.30.Hv; 14.60.St
\begin{minipage}[h]{14.0cm}
\end{minipage}
\vskip 0.3cm \hrule \vskip 0.5cm
%%%%%%%%%%%%%%%%%%%%%%%%%%%%%%%%%%
\section{Introduction}
The experimental observation of solar and atmospheric neutrino oscillations, and thus of neutrino masses, in the Super-Kamiokande \cite{SK} experiment is an experimental indication that the Standard Model (SM) of particle physics is incomplete. In the flavor basis where the charged lepton mass matrix is diagonal, the (effective) neutrino mass matrix $M_\nu$ has nine free
parameters: three masses ($m_1$, $m_2$ and $m_3$), three mixing angles ($\theta_x$,
$\theta_y$ and $\theta_z$) and three phases (two Majorana-type $\rho$, $\sigma$ and one
Dirac-type $\delta$). Experimental data \cite{SNO,KM,K2K,CHOOZ} put constraints on the values of the masses and the mixing angles; whereas no definite experimental measurements of the phases exist up till now.
The recent global analysis of oscillation data\cite{fog1} that includes the latest T2K\cite{t2k} and MINOS\cite{mino} results reveals a relatively large value for the mixing angle $\theta_z$, in contrast with the old global analysis\cite{fog2} that was consistent with vanishing $\theta_z$. The Daya Bay experiment \cite{daya} confirm the non-vanishing $\t_z$  and is in agreement with the findings of \cite{fog1}. This in turn opens the door for studying the CP violation in neutrino oscillations with profound implications for our understanding of matter-antimatter asymmetry in the universe. The non-vanishing $\theta_z$ provide a guaranteed, albeit small, contribution to the third neutrino mass $m_3$ for both single and double beta decay searches that must be accounted for in detailed analysis. From a theoretical point of view, the relatively large mixing angle $\theta_z$ may call for new ideas in model building.

Several schemes have been proposed in the literature to reduce the number of free parameters, some of which consist in equalling some elements, or some combinations of elements, of $M_\nu$ to zero. It was found that
 three independent zeros-texture can not accommodate the data, whereas nine patterns of two independent zeros-texture, out of fifteen possible, can do this \cite{FGM,Xing}). The recent analysis of two zero-texures\cite{xing2011}, based on  the latest T2K and Minos oscillation data,  constrain further the number of viable patterns to be equal to seven. In \cite{Xing03b}, a specific model realizing any of the
possible six patterns of one zero-texture was provided. However, it
led always to one of the neutrino masses equaling to zero.
Based on preceding works, such as \cite{Preceding}, the analysis of the phenomenological implications of single-zero textures was carried out in \cite{rod1}, and it showed the viability of the six possible one zero-textures combined with several interesting conclusions. It is of great importance to reexamine the one-zero textures, in light of the new oscillation data, in order to decide whether the six possible cases are still viable or not,
 and the purpose of this paper is to present such a study.

The seesaw mechanism of either type-I or -II, in addition to its role in understanding the scale of the neutrino masses, can help, when combined with
discrete Abelian flavor symmetry, in realizing the zero texture patterns in the neutrino mass matrix. More specifically, the light neutrino mass matrix  $M_\nu$ in the framework of the type-I seesaw mechanism is given by: \bea
M_\n &=& M_D M_R^{-1} M_D^T.
\label{see-saw}
\eea
where $M_D$ is the Dirac neutrino mass matrix, and $M_R$ is the
Majorana mass matrix of the right handed singlet neutrinos. It has been noted \cite{Kageyama,Grimus} that the zeros of $M_D$ and $M_R$ can find their way to enforce zeros in $M_\nu$, in such a way
that one can construct theoretical models enforcing specific patterns with zeros located at certain positions of $M_R$ and $M_D$ in
order to get the desired texture for $M_\nu$. However, the zeros of $M_D$ and $M_R$ may not only show up as zeros in $M_\nu$, and one
interesting possibility is that they appear as zero minors in $M_\nu$.

Phenomenological analysis of vanishing minors was first studied in \cite{Lavoura} assuming the invertibility of $M_\nu$, in which case
they amount to zeros in the inverse neutrino mass matrix $M_\nu^{-1}$. This condition was relaxed in
\cite{LashinChamoun1} and \cite{LashinChamoun2} where a phenomenological analysis of single or double vanishing minors was carried out.
Moreover, one can generalize the zero-textures in other ways. The simultaneous existence of a vanishing one minor and a texture zero element
was studied in \cite{dev1}, whereas two vanishing subtraces texture was studied in \cite{SayedTrace}.

In this work, carrying out an updated phenomenological analysis of the one-zero textures in view of the recent oscillation data, we
have paralleled the procedure
done in \cite{LashinChamoun2}, for both singular and invertible neutrino mass matrices, without assuming any specific model. Our present analysis is different in one respect that we let
 $\d m^2$, characterizing the solar neutrino mass-squared difference,  vary randomly within its experimentally allowed range, instead of fixing it to its central value as was done in\cite{LashinChamoun2}. Thus we need, in addition to the `two real' conditions corresponding to the one-zero element texture, six parameters in order to determine the neutrino mass matrix. We thus vary the six mixing and phase angles ($\theta_x, \theta_y, \theta_z, \delta, \rho$ and $\sigma$) within their experimentally acceptable regions, and in this way we can determine in the parameter space of ($\theta_x, \theta_y, \theta_z, \delta, \rho, \sigma, \d m^2$) the regions consistent with the other experimental results. We found that all the six patterns could accommodate the data. Moreover, five textures allow also for singular models, where one of the masses equals zero, making room for the data.

We use the type-I seesaw formula to present some theoretical realizations of all the one-zero textures, singular or not, where, in both cases, we can call for non-diagonal form Dirac neutrino mass matrix, in order to reproduce just one zero element in $M_\nu$. We find also a simple and direct way to realize the one-zero textures using the type-II seesaw scheme. In addition, we note a correspondence-mapping between the one vanishing minor analysis and the one-zero texture analysis amounting to inverting the masses and reflecting the phases. This serves qualitatively to relate the two phenomenological analyses. However, this map is not valid for the singular case, and whereas the singular
vanishing minor model presented in \cite{LashinChamoun2} was not very predictive, the one-zero singular models are quite
rich in phenomenology, and can not be related to the analysis of \cite{LashinChamoun2}.

The plan of the paper is as follows: in section $2$, we review
the standard notation for the neutrino mass matrix
and its relation to the experimental constraints. In section $3$,
we present the one-zero texture in $M_\nu$ and compute the expressions of
the two neutrino mass ratios. In section $4$, we classify all the patterns and present
the results and the phenomenological analysis of each case. Section~$5$ is devoted to the study of
singular one-zero textures. We present symmetry
realizations of all patterns, based on type-I or -II seesaw schemes, in section $6$ and end up with conclusions in section $7$.

\section{Standard notation}

In the flavor basis, where the charged lepton mass matrix is diagonal, one can diagonalize the symmetric
neutrino mass matrix $M_\nu$ by a unitary transformation,
\be
V^{\dagger} M_\nu\; V^{*} \; = \; \left (\matrix{ m_1 & 0 & 0 \cr 0 & m_2 & 0
\cr 0 & 0 & m_3 \cr} \right ), \;
\label{diagM}
\ee
with $m_i$ (for $i=1,2,3$) real and positive. We introduce the mixing
angles $(\theta_x, \theta_y, \theta_z)$ and the phases ($\delta,\rho,\sigma$)
such that \cite{Xing}:
\bea
V &=& UP \nn\\
P &=& \mbox{diag}(e^{i\rho},e^{i\sigma},1) \nn\\
U \; &=& \;
\left ( \matrix{ c_x c_z & s_x c_z & s_z \cr - c_x s_y
s_z - s_x c_y e^{-i\delta} & - s_x s_y s_z + c_x c_y e^{-i\delta}
& s_y c_z \cr - c_x c_y s_z + s_x s_y e^{-i\delta} & - s_x c_y s_z
- c_x s_y e^{-i\delta} & c_y c_z \cr } \right ) \; ,
\label{defv}
\eea
(with $s_x \equiv \sin\theta_x \ldots$)
to have
\begin{equation}
M_\nu \; =\; U \left ( \matrix{ \lambda_1 & 0 & 0 \cr 0 &
\lambda_2 & 0 \cr 0 & 0 & \lambda_3 \cr} \right ) U^T. \;
\label{massdef}
\end{equation}
with
\begin{equation}
\lambda_1 \; =\; m_1 e^{2i\rho} \; , ~~~ \lambda_2 \; =\; m_2
e^{2i\sigma} \; , ~~~ \lambda_3 = m_3. \;
\end{equation}

In this parametrization, the mass matrix elements
are given by:
\bea
M_{\n\,11}&=& m_1 c_x^2 c_z^2 e^{2\,i\,\r} + m_2 s_x^2 c_z^2 e^{2\,i\,\s}
+ m_3\,s_z^2,\nn\\
%%%%%%%%%%%%%%%%%%%%%%%%%%%%%%%%%%%%%
M_{\n\,12}&=& m_1\left( - c_z s_z c_x^2 s_y e^{2\,i\,\r}
- c_z c_x s_x c_y e^{i\,(2\,\r-\d)}\right)
+ m_2\left( - c_z s_z s_x^2 s_y e^{2\,i\,\s}
+ c_z c_x s_x c_y e^{i\,(2\,\s-\d)}\right) + m_3 c_z s_z s_y,\nn\\
%%%%%%%%%%%%%%%%%%%%%%%%%%%%%%%%%%
M_{\n\,13}&=& m_1\left( - c_z s_z c_x^2 s_y e^{2\,i\,\r}
+ c_z c_x s_x s_y e^{i\,(2\,\r-\d)}\right)
+ m_2\left( - c_z s_z s_x^2 c_y e^{2\,i\,\s}
- c_z c_x s_x s_y e^{i\,(2\,\s-\d)}\right) + m_3 c_z s_z c_y,\nn\\
%%%%%%%%%%%%%%%%%%%%%%%%%%%%%%%%%
M_{\n\,22}&=& m_1 \left( c_x s_z s_y  e^{i\,\r}
+ c_y s_x e^{i\,(\r-\d)}\right)^2 + m_2 \left( s_x s_z s_y  e^{i\,\s}
- c_y c_x e^{i\,(\s-\d)}\right)^2 + m_3 c_z^2 s_y^2, \nn\\
%%%%%%%%%%%%%%%%%%%%%%%%%%%%%%%%%%%%%%%%%%%%
M_{\n\,33}&=& m_1 \left( c_x s_z c_y  e^{i\,\r}
- s_y s_x e^{i\,(\r-\d)}\right)^2 + m_2 \left( s_x s_z c_y  e^{i\,\s}
+ s_y c_x e^{i\,(\s-\d)}\right)^2 + m_3 c_z^2 c_y^2, \nn\\
%%%%%%%%%%%%%%%%%%%%%%%%%%
M_{\n\, 23} &=& m_1\left( c_x^2 c_y s_y s_z^2  e^{2\,i\,\r}
 + s_z c_x s_x (c_y^2-s_y^2) e^{i\,(2\,\r-\d)} - c_y s_y s_x^2 e^{2\,i\,(\r-\d)}\right)
\nn\\
&& +  m_2\left( s_x^2 c_y s_y s_z^2  e^{2\,i\,\s}
 + s_z c_x s_x (s_y^2-c_y^2) e^{i\,(2\,\s-\d)} - c_y s_y c_x^2 e^{2\,i\,(\s-\d)}\right)
+ m_3 s_y c_y c_z^2.
\label{melements}
\eea
We note here that under the
transformation given by
\bea
T_1: &&\t_y\rightarrow {\pi \over 2} - \t_y\;\; \mbox{and}\;\; \d\rightarrow \d \pm \pi,
\label{sy1}
\eea
the mass matrix elements are transformed amidst themselves swapping the indices $2$ and $3$ and keeping the index $1$ intact:
\bea
M_{\n 11} \leftrightarrow M_{\n 11}, && M_{\n 12} \leftrightarrow M_{\n 13}\nn\\
M_{\n 22} \leftrightarrow M_{\n 33}, && M_{\n 23} \leftrightarrow M_{\n23}.
\label{tr1}
\eea
Under the mapping given by:
\bea
T_2: \r \rightarrow \pi - \r, & \s \rightarrow \pi - \s, & \d \rightarrow 2\,\pi-\d,
\label{sy2}
\eea
the mass matrix is transformed into its complex conjugate i.e
\bea
M{_\n}_{ij}\left(T_2(\d,\r,\s)\right) = M^*_{\n ij} \left((\d,\r,\s)\right)
\eea

The above two symmetries $T_{1,2}$ are
very useful in classifying the models and in connecting the phenomenological analysis of patterns related
by these symmetries.

 It is important to relate our convention for parameterizing  the mixing matrix in Eq.~(\ref{defv}) to the corresponding one
 used in the recent data analysis of \cite{fog1}. The mixing angles $(\theta_x, \theta_y, \theta_z)$ match  exactly their corresponding ones but with different nomenclatures; strictly speaking,
 \begin{equation}
\theta_x \; \equiv \; \theta_{12} \; , ~~~~~ \theta_y \;
\equiv \; \theta_{23} \; , ~~~~~ \theta_z \; \equiv \;
\theta_{13}.
\end{equation}
The matrix $U$ in Eq.~(\ref{defv}) can be decomposed as
\be
U = R_y\; R_z(\delta)\; R_x,
\label{decU}
\ee
where $R_x$ and $R_y$ represent rotations around the third and first axes:
\bea
R_x =
\left ( \matrix{ c_x  & s_x  & 0 \cr
- s_x & c_x & 0 \cr
0 & 0 &  1\cr } \right ), & &
R_y =
\left ( \matrix{ 1  & 0  & 0 \cr
0 & c_y & s_y \cr
0 & - s_y &  c_y\cr } \right ).
\nn\\
\label{ryx}
\eea
 As to $R_z(\delta)$, it depends on the convention. For illustration purposes and in order to compare our parametrization convention
 with the one adopted in \cite{fog1}, we use a superscript tilde-mark to denote the latter one.  Thus,  $R_z(\delta)$  takes
 the following two forms:
\bea
R_z(\delta) =
\left ( \matrix{ c_z & 0  & s_z \cr
0 & e^{-i\,\d} & 0 \cr
-s_z & 0 &  c_z\cr } \right ),
&&
\widetilde{R_z}(\delta) =
\left ( \matrix{ c_z & 0  & s_z \cr
0 & 1 & 0 \cr
- s_z\,e^{i\,\d} & 0 &  c_z\,e^{i\,\d}\cr } \right ).
\label{rzd}
\eea
and the diagonal phase matrix $P$ in Eq.~(\ref{defv}), which contains the Majorana phases,
is given by
\bea
P = \mbox{diag}(e^{i\rho},e^{i\sigma},1), && \widetilde{P} = \mbox{diag}(1,e^{i{\phi_2\over 2}},e^{i{\phi_3\over2}}).
\label{majp}
\eea
The relation between phases in the two parameterizations can be found by means of using rephasing  invariant quantities such as the Jarlskog 'J' quantity \cite{jarlskog} measuring the CP violation in neutrino oscillations and other two quantities ($S_1 \mbox{ and } S_2$) \cite{petcov1} related to the Majorana nature of the massive neutrinos. The explicit expressions for these rephasing invariant quantities are,
\bea
J= \mbox{Im}\left(V_{\mu 2}^* \, V_{e 3}^*\, V_{\mu 3}\, V_{e 2}\right), & S_1 = \mbox{Im}\left(V_{e1}\, V_{e 3}^*\right),&
S_2 = \mbox{Im}\left(V_{e2}\, V_{e 3}^*\right).
\eea
We find then that the Dirac phase is the same for both parameterizations, whereas the Majorana phases are related as,
\bea
\rho = -{\phi_3\over 2}, & & \sigma = {\phi_2 -\phi_3\over 2}.
\eea
All probability transition amplitudes relevant to neutrino oscillations in free space are calculated and found to be equal for both  parameterizations and yielding the same expressions in terms of mixing angles and Dirac phase($\d$). This confirms the identification of mixing angles and Dirac phase for both parameterizations.

Two independent neutrino mass-squared differences, characterizing respectively solar and
atmospheric neutrino mass-squared differences, are defined as\cite{fog1},
\begin{equation}
\delta m^2 \; \equiv \; m_2^2-m_1^2 \; , \; \left|\Delta m^2\right|  \;\equiv \; \left|m_3^2-{1\over 2}\left(m_1^2+m_2^2\right)\right|\;\; ,
\label{sqm}
\end{equation}
and the hierarchy of solar and
atmospheric neutrino mass-squared differences is characterized by
the parameter:
\begin{equation}
R_\nu \; \equiv \;  \frac{\delta m^2} {\left|\Delta
m^2\right|}.
\label{rnu}
\end{equation}
Two parameters which put bounds on the neutrino mass scales, by the reactor nuclear experiments on beta-decay kinematics and
neutrinoless double-beta decay, are the
effective electron-neutrino mass:
\begin{equation}
\langle
m\rangle_e \; = \; \sqrt{\sum_{i=1}^{3} \displaystyle \left (
|V_{e i}|^2 m^2_i \right )} \;\; ,
%       (2.23)
\end{equation}
and the effective Majorana mass term
$\langle m \rangle_{ee} $:
\begin{equation} \label{mee}
\langle m \rangle_{ee} \; = \; \left | m_1
V^2_{e1} + m_2 V^2_{e2} + m_3 V^2_{e3} \right | \; = \; \left | M_{\n 11} \right |.
%       (2.30)
\end{equation}
Another parameter with an upper bound coming from cosmological observations is the `sum'
parameter $\Sigma$:
\be
\Sigma = \sum_{i=1}^{3} m_i.
\ee

Moreover, the above mentioned Jarlskog rephasing invariant quantity
is given by:
\begin{equation}\label{jg}
J = s_x\,c_x\,s_y\, c_y\, s_z\,c_z^2 \sin{\delta}
\end{equation}

There are no experimental bounds on the phase angles, and we take the principal
value range for $\d, 2\r$ and $2 \s$ to be $[0,2\pi]$.
As to the other oscillation parameters, the experimental constraints give the following
 values with 1, 2, and 3-$\sigma$ errors \cite{fog1}:
\begin{table}[h]
 \begin{center}
{\small
 \begin{tabular}{c c c c c}
\hline
\hline
\mbox{Parameter} &$\mbox{Best fit}$ & $1\s$ \mbox{range} & $2\s$ \mbox{range} & $3\s$ \mbox{range}\\
\hline
 $\d m^2 (10^{-5}\mbox{eV}^2)$ &$7.58$ & $\left[7.32, 7.80\right]$ &$\left[7.16, 7.99\right]$ &
 $\left[6.99, 8.18\right]$ \\
 \hline
 $\left|\D m^2\right|(10^{-3}\mbox{eV}^2)$& $2.35$ &$\left[2.26, 2.47\right]$ &$\left[2.17, 2.57\right]$ &
 $\left[2.06, 2.67\right]$ \\
 \hline
$\th_x$ & $33.58^o$ &$\left[32.96^o, 35.00^o\right]$ &$\left[31.95^o, 36.09^o\right]$ & $\left[30.98^o, 37.11^o\right]$ \\
\hline
$\th_y$ &$40.40^o$ &$\left[38.65^o, 45.00^o\right]$ &$\left[36.87^o, 50.77^o\right]$ & $\left[35.67^o,53.13^o\right]$\\
\hline
$\th_z$ & $8.33^o$ &$\left[7.71^o, 10.30^o\right]$ &$\left[6.29^o, 11.68^o\right]$ &$\left[4.05^o, 12.92^o\right]$ \\
\hline
$R_{\nu}$ & $0.0323$ & $\left[0.0296 , 0.0345\right]$&$\left[0.0279 , 0.0368\right]$ & $\left[0.0262 , 0.0397\right]$\\
\hline
 \end{tabular}
 }
 \end{center}
  \caption{\small \label{tab1} The latest global-fit results of three neutrino mixing angles $(\th_x, \th_y, \th_z)$ and two neutrino mass-squared differences $\d m^2$ and $\D m^2$ as defined in Eq.~(\ref{sqm}). Here, it is assumed that $\cos{\d}= \pm 1$ and that new reactor fluxes have been used\cite{fog1}.}
 \end{table}

As to the non oscillation parameters $\me$, $\Sigma$ and $\mee$, we adopt the less conservative 2-$\sigma$ range as reported in \cite{fog3}
for the first two, while for $\mee$ we use values found in \cite{Cuoricino}:
\bea
\langle m\rangle_e &<& 1.8\; \mbox{eV}, \nonumber \\
\Sigma &<& 1.19 \;\mbox{eV}, \nonumber \\
\langle m\rangle_{ee} & < & 0.34-0.78\; \mbox{eV}.
\label{nosdata}
\eea

\section{Neutrino mass matrices with one zero texture}

Since the neutrino matrix is symmetric then we have six independent possibilities for one
vanishing entry, each of which is called one-zero texture.

The vanishing condition at the location $(a,b)$ is written
as:
\be
\label{van1}
 M_{\nu\,ab}  = 0
\ee
which amounts to
\bea
\label{van2}
\sum_{j=1}^{3}
U_{aj}U_{bj}\; \lambda_j &=& 0.
\eea
This leads to:
\bea
\frac{m_1}{m_3} &=&
\frac{Re(A_3) Im(A_2 e^{2i\sigma})-Re(A_2
e^{2i\sigma}) Im(A_3)}
{Re(A_2 e^{2i\sigma}) Im(A_1 e^{2i\rho})-Re(A_1 e^{2i\rho}) Im(A_2 e^{2i\sigma})} \nn \\
\frac{m_2}{m_3} &=& \frac{Re(A_1 e^{2i\rho}) Im(A_3)-Re(A_3) Im(A_1 e^{2i\rho})}
{Re(A_2 e^{2i\sigma}) Im(A_1 e^{2i\rho})-Re(A_1 e^{2i\rho}) Im(A_2 e^{2i\sigma})}
\label{massratio}
\eea
where
\bea
\label{Ah}
A_j&=&
U_{aj}\;U_{bj}, \;\;\;\;\;\;\;\; (\mbox{no sum over } j).
\eea

As mentioned in the introduction, one can completely reconstruct the neutrino mass matrix in terms of the six mixing and phase angles, assuming a given one-zero texture and fixing a value for $\d m^2$ within its acceptable range. We span ($\theta_x, \theta_y, \theta_z, \d m^2$) over their experimentally allowed regions, whereas the phases ($\delta, \rho,$ and $\sigma$) are varied in their full ranges. We thus can determine in the parameter space
 the acceptable regions compatible with the other experimental constraints as given in Table~(\ref{tab1})
 and Eq.~(\ref{nosdata}). One can then illustrate graphically all the possible
correlations, in the three levels of $\s$-error, between any two physical neutrino parameters. We chose to
plot for each pattern and for each type of hierarchy thirty correlations at the 3-$\s$ error level involving
the parameters $(m_1,m_2, m_3,\t_x,\t_y,\t_z,\r,\s,\d,J,m_{ee})$ and the lowest neutrino mass ({\bf LNM}).
Moreover, for each parameter, one can determine the extremum values it can take according to the considered
precision level, and we listed in tables these predictions for all the patterns and for the three $\s$-error
levels.

We found that the resulting mass patterns could be classified into
three categories:
\begin{itemize}
\item Normal hierarchy: characterized by $m_1 < m_2 < m_3$ and
is denoted by ${\bf N}$. It satisfies numerically the bound:
\be
\frac{m_1}{m_3} < \frac{m_2}{m_3} < 0.7
\label{nor}
\ee
\item Inverted hierarchy: characterized
by $m_3 < m_1 < m_2$ and is denoted by ${\bf I}$. It satisfies the bound:
\be
\frac{m_2}{m_3} > \frac{m_1}{m_3} > 1.3
\label{inv}
\ee
\item Degenerate hierarchy (meaning quasi- degeneracy): characterized
by $m_1\approx  m_2 \approx m_3$ and is denoted by ${\bf D}$. The corresponding numeric bound is
taken to be:
\be
0.7 < \frac{m_1}{m_3} < \frac{m_2}{m_3} < 1.3
\label{deg}
\ee
\end{itemize}
Moreover, we studied for each pattern the possibility of having a singular (non-invertible) mass matrix. The
viable singular mass matrix is
characterized by one of the masses ($m_1,
\mbox{and} \; m_3$) being equal to zero, as compatibility with the data prevents the
simultaneous vanishing of two masses and thus $m_2$ can not either vanish. The detailed investigation of singular
one-zero textures is presented in Section~5.

One can relate the one zero-textures analysis to the analysis of the one vanishing minor texture. This comes because Eq.~(\ref{diagM}) can be put, assuming $M_\nu$ is invertible, in an equivalent form:
 \bea V^{T} M_\n^{-1} V &=& \mbox{Diag}\left({1\over m_1}, {1\over m_2}, {1\over m_3}\right)\eea
By comparing the two forms, we see that we can write the inverse mass matrix in the same form as the mass matrix after having inverted
the masses and having conjugated the phases. Since a zero in $M_\n^{-1}$ is equivalent to a vanishing minor for $M_\n$, we get a mapping
between the one-zero textures and the one vanishing minor patterns in the following sense. Suppose a one-zero texture
at entry ($a,b$) led to an analytical relation in terms of the free parameters of the form:
\bea f(m_1,m_2,m_3,\th_x,\th_y,\th_z,\r, \s, \d) &=& 0 \eea then one can deduce that a vanishing minor at the same location ($a,b$) would
lead to an analytical relation of the form \bea f({1\over m_1},{1\over m_2},{1\over m_3},\th_x,\th_y,\th_z,-\r,-\s,-\d)\equiv g(m_1,m_2,m_3,\th_x,\th_y,\th_z,\r, \s, \d) &=& 0 \label{mapping-minor-zero}\eea

However, phenomenologically speaking, one can only qualitatively relate the correlations obtained in \cite{LashinChamoun2} with the correlations corresponding to the one-zero textures in this paper, even when using the same neutrino data. In fact, inverting the masses would change the
hierarchy type from normal into inverted and vice versa, whereas the degenerate hierarchy will keep its type. Now, if the $\Sigma$ constraint
 (equation \ref{nosdata}), say, was satisfied for a one-zero texture, then upon inverting the masses it might stop being satisfied in the corresponding vanishing minor pattern. Thus, we conclude that the experimental
bounds on the masses are not symmetric with respect to the mapping, and hence one can not deduce quantitatively the experimentally acceptable regions in the parameter space from those of  \cite{LashinChamoun2}. Lastly, this equivalence mapping is no more valid for the singular models, where one of the masses is equal to zero. Actually, the predictiveness of the singular model in \cite{LashinChamoun2} was weak, whereas
it is quite powerful in the singular one-zero texture models.

As to the symmetry $T_1$ introduced in Eqs.(\ref{sy1}-\ref{tr1}), it induces
equivalence between different textures as,
$(M_{\n\,12} = 0) \leftrightarrow (M_{\n\,13} =0) $ and $(M_{\n\,22} =0) \leftrightarrow (M_{\n\,33} =0) $.
However, this equivalence for $\t_y$ is a reflection about the first bisectrix,
i.e. it maps the $\t_y$ from the first octant to the second octant and vice versa. Similarly, the image points
of the map differ in $\d$ from their original points by a shift equal to $\pi$. This means that the accepted
points for a pattern imply for the equivalent pattern the same accepted points but after changing
the $\t_y$ and $\d$ correspondingly. Unfortunately the recent oscillation data presented in Table~(\ref{tab1}), has a range for $\t_y$ which is not perfectly symmetric about the first bisectrix especially for the 1-$\s$ and 2-$\s$ level values, thus making the $T_1$ symmetry of limited use in relating the equivalent patterns. However, at the 3-$\s$ level, the range of $\t_y$ is nearly symmetric  about the first bisectrix, and so the $T_1$-symmetry is in action and the phenomenology of any pattern can be deduced from its $T_1$-symmetric equivalent one. Hence, the equivalence between different patterns should be taken carefully especially for data determined at the 1- and 2-$\s$ level.

Thus, it is enough now to present four possible cases, instead of six ones, corresponding
to one zero texture. We stated, simple writing permitting, only the leading terms in powers
of $s_z$ of the analytical expressions in terms of
($\t_x ,\t_y , \t_z , \d, \r$ and $\s$). Moreover, since the mapping (Eq. \ref{mapping-minor-zero}) should be valid term by term
to all orders of the expansion in powers
of $s_z$, as the parameter $\t_z$ is not affected by the mapping, then we could indeed check our leading order formulae
by comparing them to those of \cite{LashinChamoun2}. Also, we have verified
in the patterns where the obtained formulae are
all exact (not expanded), i.e. ($M_{11}$) here and ($C_{11}$) in \cite{LashinChamoun2}, the validity of the mapping, which provided an additional crosscheck
for the formulae.

\section{Results of one zero textures }
In this section, we present the results of our numerical analysis for the four possible independent
textures based upon the approach described in the previous section. We give the coefficients $A's$ (Eq.~\ref{Ah})
defining each texture, and we produce also, in order to get some interpretation of the numerical results,
 the analytical expressions of the mass ratios up to leading order
in $s_z$, except in the last texture $(M_{11}=0)$ where we give the full analytical
expressions of the mass ratios and other experimental parameters. We emphasize here that our numerical analysis is based on the exact formulae and not on the approximate ones.

The large number of correlation figures is organized in plots, at the 3-$\s$-error level, by dividing
 each figure into left and right panels (halves) denoted accordingly by the letters L and R. Additional
labels (D,N and I) are attached to the plots to indicate the type of hierarchy (Degenerate, Normal and
Inverted, respectively). Any missing label D, N or I on the figures of
certain texture means the absence of the corresponding hierarchy type in this texture.

Tables (\ref{tab2}) and (\ref{tab3}) list, for the three types of hierarchy and the
three precision levels, the extremum values that the different parameters can take. The corresponding ranges
should get larger with higher-$\s$ precision levels. However, as in \cite{LashinChamoun2}, these bounds are evaluated by spanning the
parameter space with some given number (of order $(10^8-10^9)$) of points chosen randomly in the parameter
space. This way of random spanning is more efficient than a regular meshing with nested loops where we need,
with a fixed step of `modest' order of 1 degree, and one hundred points to cover the $\d m^2$ range, around $10^{12}$ points so that to cover the experimentally allowed space, and so requires, in order to be efficient, a `dynamic' step for a finer meshing in the regions
full of accepted points compensated by less spanning in the disallowed regions. We
do not have this problem with the random spanning. Moreover, the randomness of our spanning allowed us to check the stability of our results for different randomly chosen points when we ran the programs several times. Thus, the values in the tables are meant to give only a strong qualitative indication.

%%%%%%%%%%%%%%%%%%%%%%%%%%%%%%%%%%%%%%%%%%%%%%%%%%%%%%%%%%%%%%%%%%%%%%%%%%%%%%%
%%%%%%%%%%%%%%%%Case M33 =0
%%%%%%%%%%%%%%%%%%%%%%%%%%%%%%%%%%%%%%%%%%%%%%%%%%%%%%%%%%%%%%%%%%%%%%%%%%%%%%%
\subsection{ Pattern of vanishing $\mathbf {M_{\nu\, 33}}$}
In this pattern, the relevant expressions for $A_1$, $A_2$ and $A_3$ are
\bea
  A_1 &=& (-c_x c_y s_z + s_x s_y e^{-i\,\d})^2, \nn\\
 A_2 & = & (s_x c_y s_z  +  c_x s_y e^{-i\,\d})^2,\nn\\
 A_3 &=& c_y^2 c_z^2 ,
 \label{m33}
 \eea
leading to
\bea
 \frac{m_1}{m_3} &\approx& \frac{s_{2\s - 2\d}}{s_x^2  t_y^2s_{2\r - 2\s}} +
 O \left( s_z \right) \nn\\
 \frac{m_2}{m_3} &\approx& \frac{s_{2\d - 2\r}}{c_x^2 t_y^2  s_{2\r - 2\s}}
  +O \left( s_z \right)
\label{mr33}
\eea

\begin{figure}[hbtp]
\centering
\begin{minipage}[l]{0.5\textwidth}
\epsfxsize=8cm
\centerline{\epsfbox{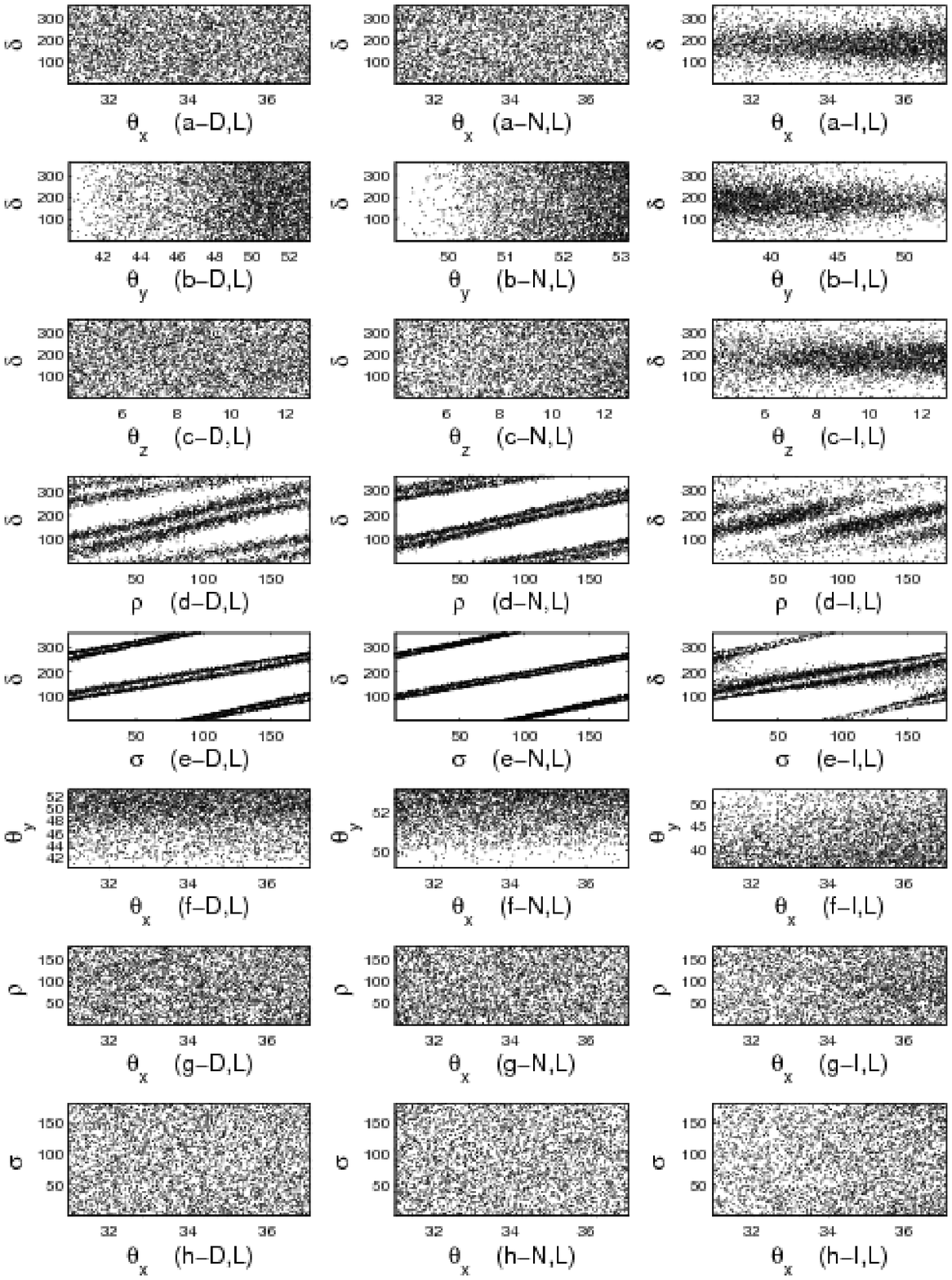}}
\end{minipage}%
\begin{minipage}[r]{0.5\textwidth}
\epsfxsize=8cm
\centerline{\epsfbox{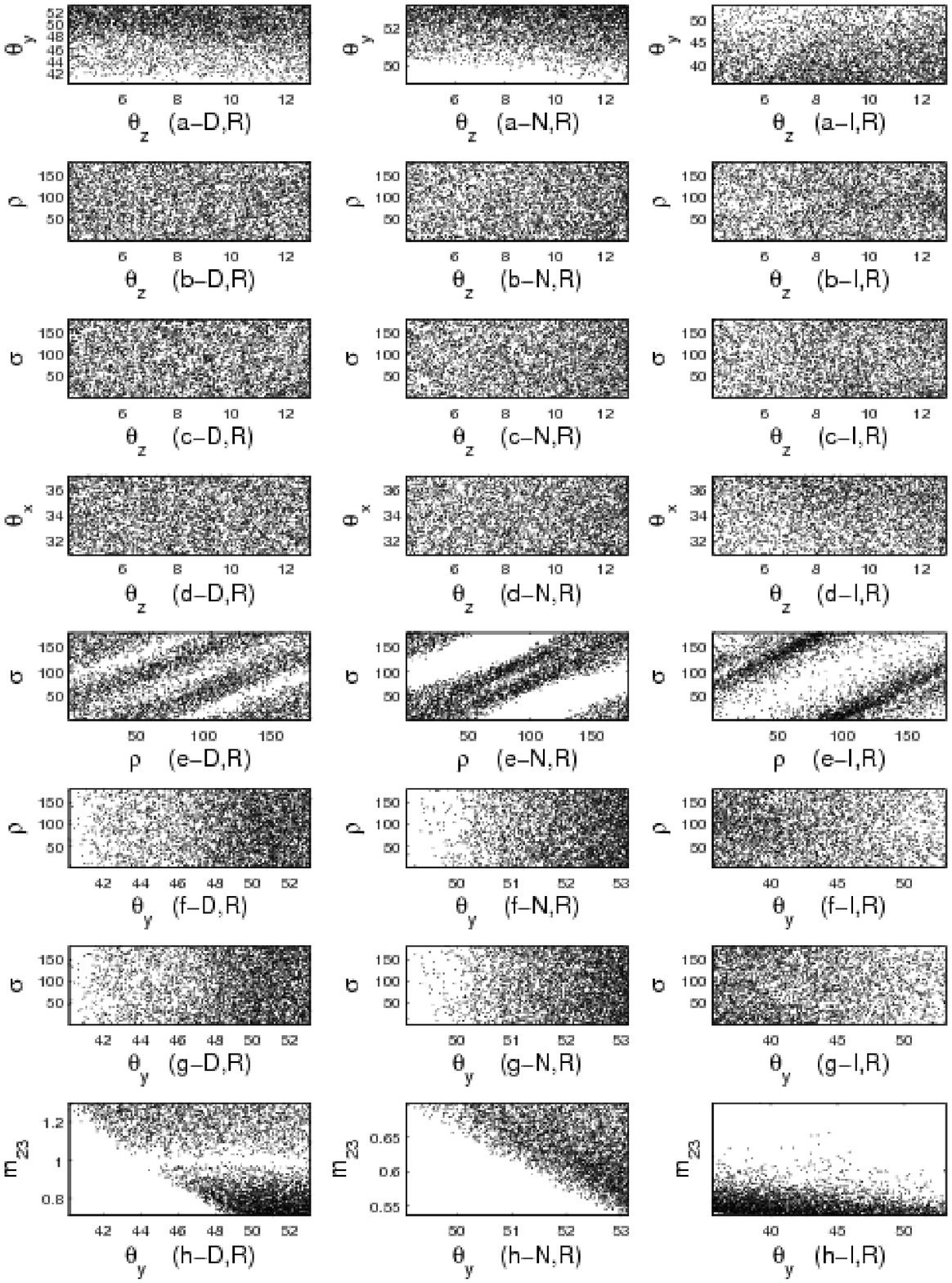}}
\end{minipage}
\vspace{0.5cm}
\caption{{\footnotesize Pattern $\mathbf M_{\nu\,33}=0:$ The left panel (the left three columns) presents correlations of $\delta$ against
mixing angles and Majorana phases ($\r$ and $\s$) and those of $\t_x$ against $\t_y$, $\r$ and $\s$.
The right panel (the right three columns) shows the correlations of $\t_z$ against $\t_y$, $\r$ ,
$\s$, and $\t_x$ and those of $\r$ against $\s$ and $\t_y$, and also the correlation of $\t_y$ versus
$\s$ and $m_{23}$.}}
\label{m33fig1}
\end{figure}

In Figure \ref{m33fig1}, left and right panels, we present all the possible fifteen pair correlations related to
 the three mixing angles and the three Majorana and Dirac phases $(\t_x, \t_y, \t_z, \d, \r, \s)$,
while the last plot in the right panel is reserved for the correlation of $(m_{23} \equiv{m_2\over m3})$ against $\t_y$.

\begin{figure}[hbtp]
\centering
\begin{minipage}[l]{0.5\textwidth}
\epsfxsize=8cm
\centerline{\epsfbox{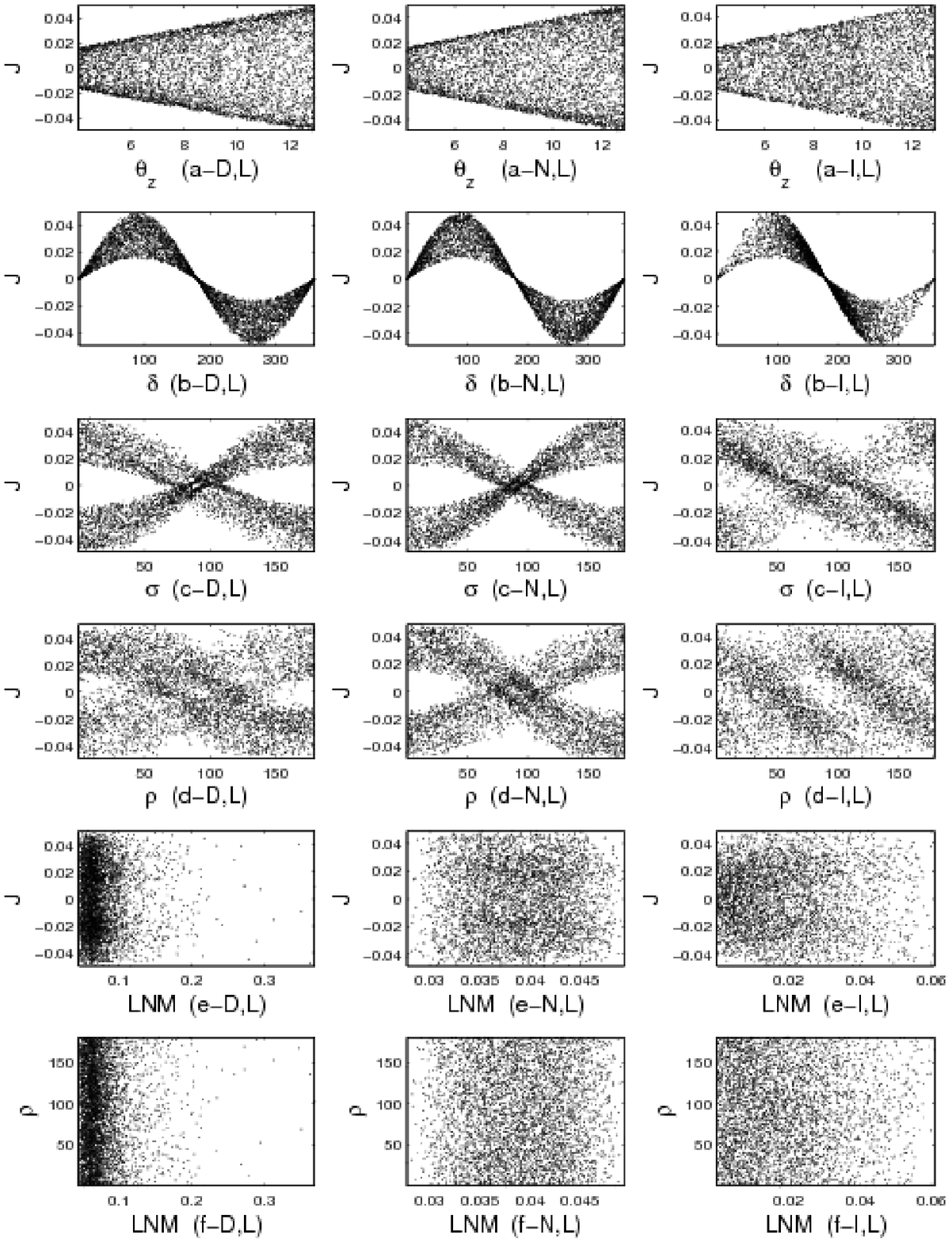}}
\end{minipage}%
\begin{minipage}[r]{0.5\textwidth}
\epsfxsize=8cm
\centerline{\epsfbox{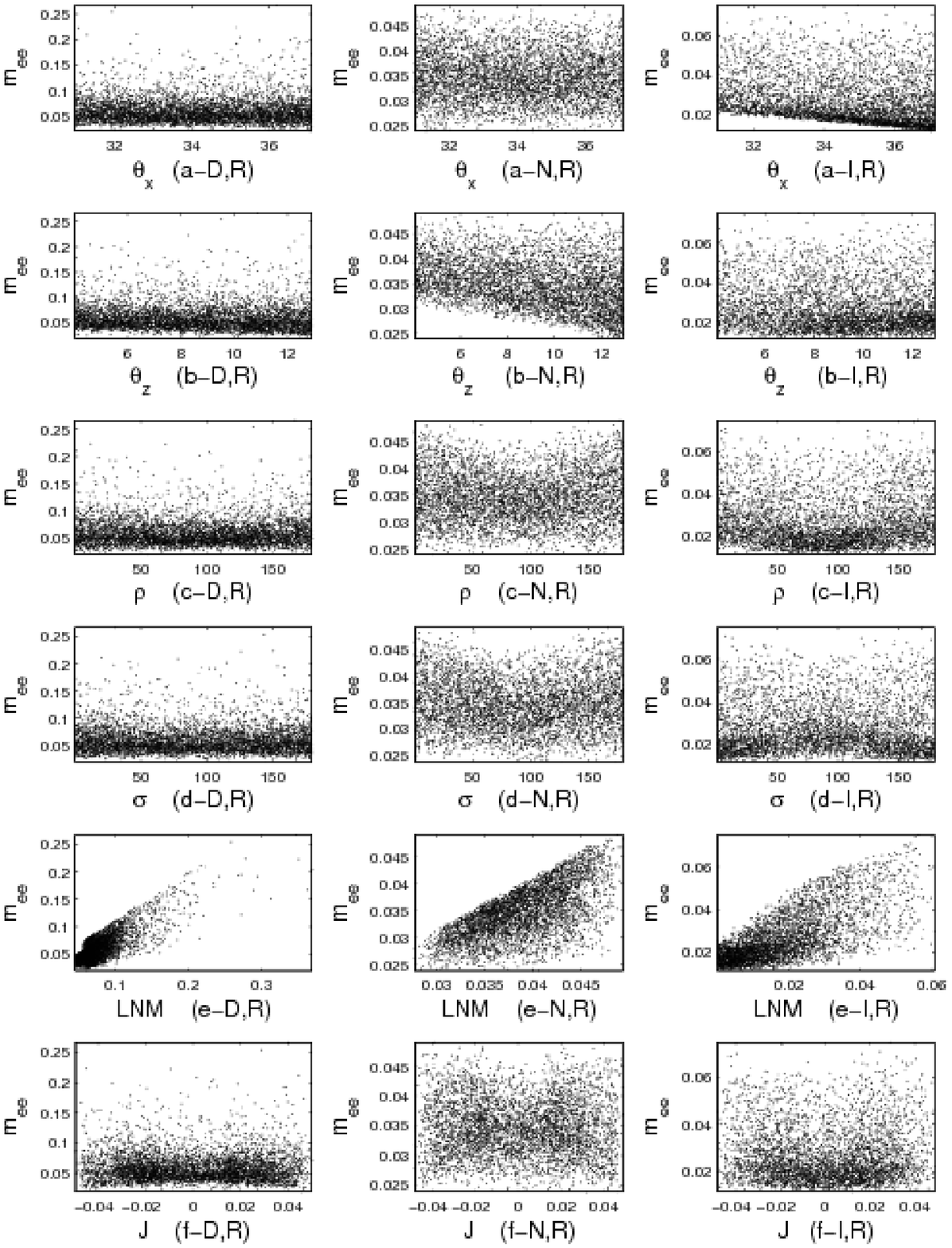}}
\end{minipage}
\vspace{0.5cm}
\caption{{\footnotesize Pattern $\mathbf M_{\nu\,33}=0:$ Left panel presents correlations of $J$ against
$\t_z$, $\d$,  $\s$ , $\r$, and lowest neutrino mass ({\bf LNM}), while the last one depicts the
correlation of LNM against $\r$. The right panel shows correlations of $m_{ee}$ against $\t_x$,
$\t_z$, $\r$, $\s$, {\bf LNM} and $J$. }}
\label{m33fig2}
\end{figure}

The left panel of Figure \ref{m33fig2} presents five correlations of $J$ against ($\t_z,\d, \s, \r$
and {\bf LNM}) and the correlation of $\r$ versus {\bf LNM}. As to the right panel
of this figure, it presents the  correlations of $\mee$ against $\t_x$, $\t_z$, $\r$, $\s$, $\mbox{LNM}$,  and $J$.

\begin{figure}[hbtp]
\centering
\epsfxsize=7cm
\epsfbox{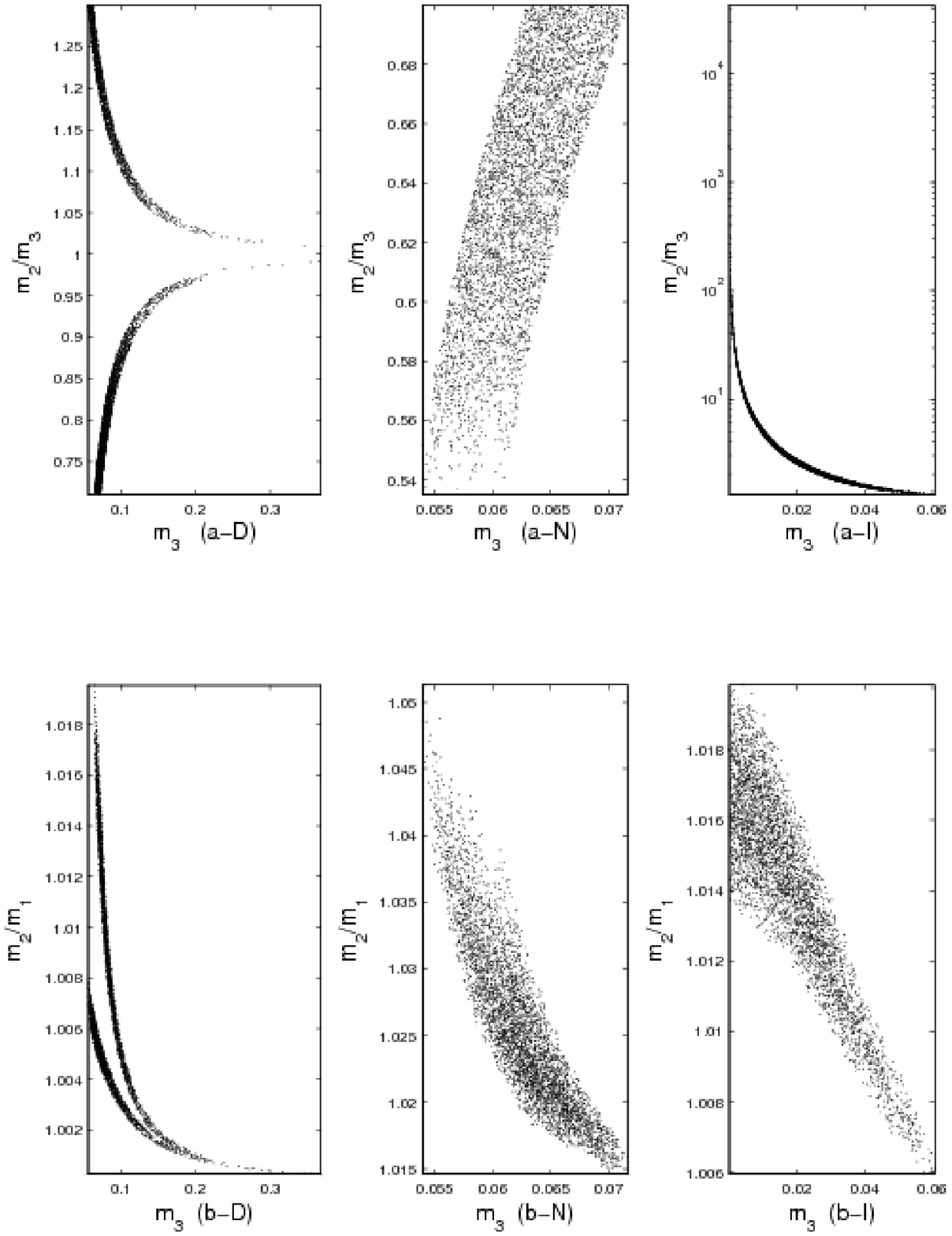}
\caption{{\footnotesize Pattern $\mathbf M_{\nu\,33}=0:$ correlations of mass ratios ${m_2\over m_3}$ and
${m_2\over m_1}$ against $m_3$.}}
\label{m33fig3}
\end{figure}

As to Figure \ref{m33fig3}, and in a similar way, it presents the correlation of $m_3$ against
$\frac{m_2}{m_3}$ and  $\frac{m_2}{m_1}$ for the three types of hierarchy. In all we have thirty types
of correlations for each hierarchy type.

We see in Figure~\ref{m33fig1}~(plots: a-L $\rightarrow$ c-L, as examples) that all the experimentally
allowed ranges of mixing angles, at $3 \s$
error levels, can be covered in this pattern except for normal and degenerate hierarchy types where $\t_y$ is restricted to be greater than $49^0$ ($40^0$) for normal and (degenerate) ones. This restriction on $\t_y$ is a characteristic of the normal hierarchy type in this pattern.
As to the Dirac CP-phase $\d$, it is not restricted at all as evident from the same plots. Likewise, the plots (g-L, h-L),
in Figure~\ref{m33fig1} show that the Majorana phases ($\r, \s$) are not constrained either.

The plots in Figure~\ref{m33fig1} show no obvious clear correlation except those concerning correlations
among phases.  The plots (d-L, e-L) show a strong linear correlation of $\d$ versus $(\r, \s)$ in the case of degenerate and normal hierarchy, whereas the inverted case does not reveal such a linearity.
We see also also in (plot e-R), a linear correlation between the Majorona phases ($\r, \s$), which is
rather obvious in the normal hierarchy type while quite blurred in the degenerate and inverted cases.
As to the correlation of $m_{23}$ against $\t_y$, the plot (h-D,R) shows that $m_2$ is greater than $m_3$
provided that the angle $\t_y$ is in the first octant for the degenerate hierarchy type.

The correlations ($J,\t_z$) and ($J,\d$) have each a specific geometrical
shape regardless of the hierarchy type as can be seen from  Figure~\ref{m33fig2}~(plots: a-L ,b-L).
 In fact, from Eq.~(\ref{jg}) we can see the correlation
($J,\d$) as a superposition of many sinusoidal graphs in $\d$ whose `positive' amplitudes are
determined by the acceptable mixing angles, whereas the ($J,\t_z$) correlation can be seen as being formed by the superposition of straight-lines in $s_z \sim \t_z$, for small $\t_z$, whose slopes can be positive or negative depending on the sign of $s_\d$.  The resulting shape for ($J,\t_z$) correlation is trapezoidal instead
of isosceles due to the exclusion of zero and its neighborhood for $\t_z$ according to the latest oscillation  data.

The left panel of Figure~\ref{m33fig2} (plots: c-L, d-L), reveals a correlation of  $J$ versus $(\r, \s)$
 which is a direct consequence of the `linear' correlations of  $\d$ against $(\r, \s)$ and of the `geometrical' correlation
 of ($J, \d$). The two correlations
 concerning the {\bf LNM} (plots: e-L, f-L)
disclose that as the {\bf LNM} increases the parameter space becomes more restricted. This seems to be a general trend in all the patterns, where the {\bf LNM} can reach in the degenerate case values higher than the other hierarchies.

The correlations of $\mee$ against $(\t_x, \t_z, \r, \s, \mbox{LNM}, J)$, as inferred from the right panel  of Figure~\ref{m33fig2}, show that the increase of $\mee$ would generally
constrain the allowed parameter space. We note also a general tendency of increasing $\mee$ with increasing
LNM in all cases of hierarchy (plots e-R). Another point concerning $\mee$ is that
it can not attain the zero-limit in all types of hierarchy, as is evident from the graphs or explicitly
from the corresponding covered range in Table~\ref{tab2}.

For the mass spectrum, we see from Figure~\ref{m33fig3} that the normal hierarchy is of
moderate type in that the mass ratios do not reach extremely high, nor low, values. In contrast, the inverted hierarchy could be a severe one passing to the limit of vanishing $m_3$. All hierarchy types are characterized by nearly equal values of $m_1$ and $m_2$. We also see that if $m_3$ is large enough then only the degenerate case with $m_1 \sim m_2$ can be compatible with data. We shall find later in section~$5$ that there is a singular  texture, with vanishing $m_3$ which can accommodate the data in the inverted hierarchy case. The possibility of vanishing $m_3$ could be seen from the coverable ranges of masses $m_3$ in Table~\ref{tab2}.
This table also shows that no normal hierarchy type of this pattern can be obtained at the 1-$\s$ precision level, that is in accordance with what we found as bound on $\t_y$ to be greater than $49^0$.

\subsection{  Pattern of vanishing $\mathbf {M_{\n\,22}}$}
In this pattern, the relevant expressions for $A_1$, $A_2$ and $A_3$
are
\bea
A_1 &=& (c_x  s_y  s_z + s_x  c_y e^{-i\,\d})^2,\nn \\
A_2 & = & (s_x s_y s_z - c_x c_y e^{-i\,\d})^2,\nn\\
A_3 &=& s_y^2 c_z^2 ,
\label{m22}
\eea
We get
\bea
\frac{m_1}{m_3} &\approx& \frac{t_y^2 s_{2\s - 2\d}}{s_x^2 s_{2\r - 2\s}} +
O \left( s_z \right)\nn \\
\frac{m_2}{m_3} &\approx& \frac{t_y^2 s_{2\d - 2\r}}{c_x^2  s_{2\r - 2\s}}
 +O \left( s_z \right)
 \label{mr22}
 \eea

Again, there is a singular such texture which can accommodate the data. As for the plots, and since
this pattern is related by $T_1$-symmetry to the pattern $(M_{\n\,33}=0)$, they can be deduced from those of the latter pattern but after changing $\t_y$ and $\d$ accordingly.
%%%%%%%%%%%%%%%%%%%%%%%%%%%%%%%%%%%%%%%%%%%%%%%%%%%%%%%%%%%%%%%%%%%%%%%%%%%%%%%%%
%%%%%%%%%%%%%%%%%%%%%%%%%%%%  CASE M32=0      %%%%%%%%%%%%%%%%%%%%%%%%%%%%%%%%%%%%%%
\subsection{  Pattern of vanishing $\mathbf {M_{\n\, 32}}$}
The relevant expressions for $A_1$, $A_2$ and $A_3$ for this model
are
\bea
A_1 &=& -\left(s_z c_x s_y + s_x c_y\,e^{-i\,\d} \right)
\left(-c_x c_y s_z  +  s_x s_y e^{-i\,\d}\right),\nn \\
A_2 & = & \left(s_x s_y s_z - c_x c_y\, e^{-i\,\d} \right)
\left(s_x c_y s_z + c_x s_y\,e^{-i\,\d}\right),\nn\\
A_3 &=& s_y c_y c_z^2.
\label{m32}
\eea
We get
\bea
\frac{m_1}{m_3} &\approx& \frac{ s_{2\s - 2\d}}{s_x^2 s_{2\s - 2\r}} +
O \left( s_z \right)\nn \\
\frac{m_2}{m_3} &\approx& \frac{s_{2\r - 2\d}}{c_x^2  s_{2\r - 2\s}}
 +O \left( s_z \right)
 \label{mr32}
 \eea

\begin{figure}[hbtp]
\centering
\begin{minipage}[l]{0.5\textwidth}
\epsfxsize=8cm
\centerline{\epsfbox{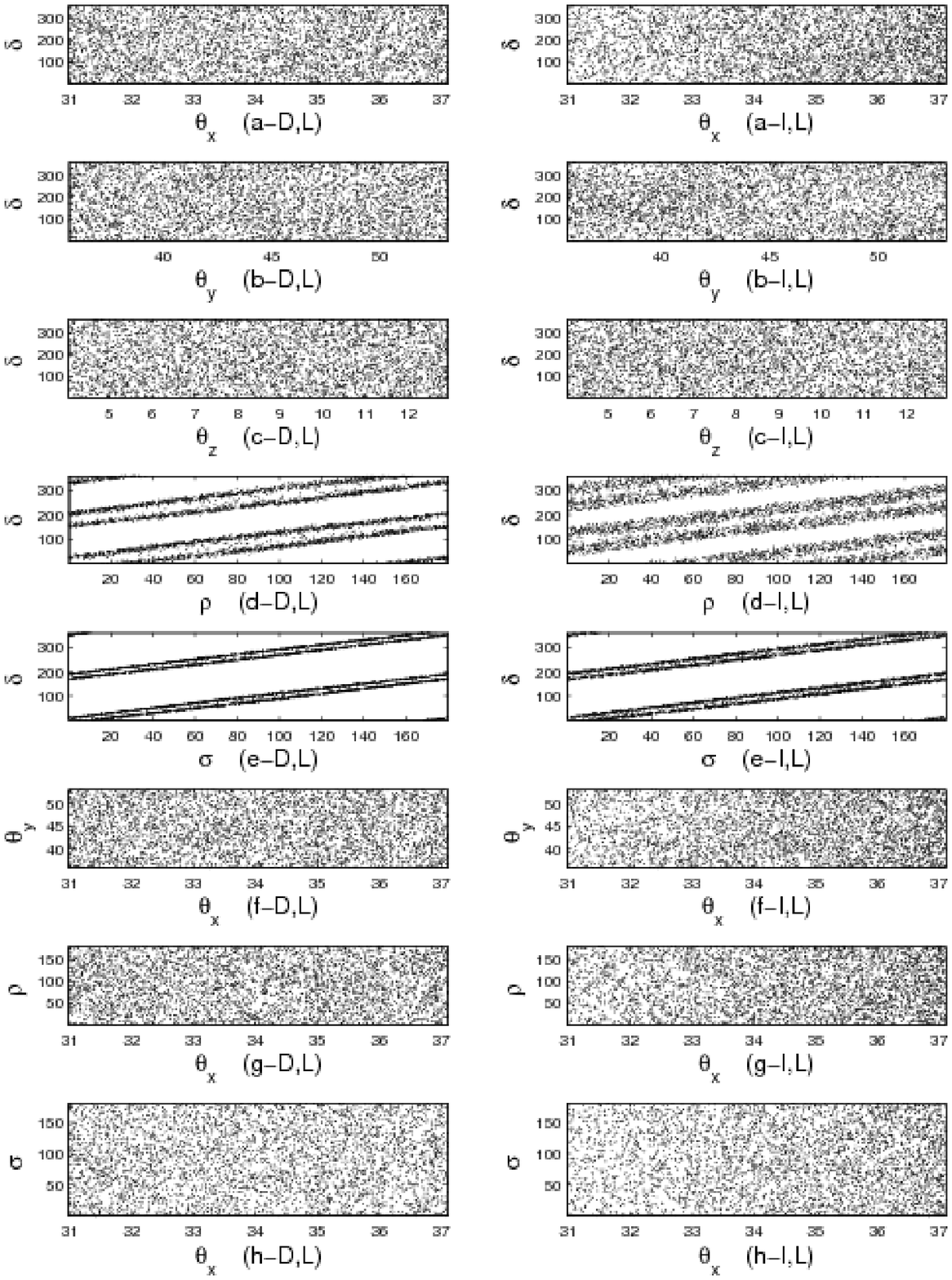}}
\end{minipage}%
\begin{minipage}[r]{0.5\textwidth}
\epsfxsize=8cm
\centerline{\epsfbox{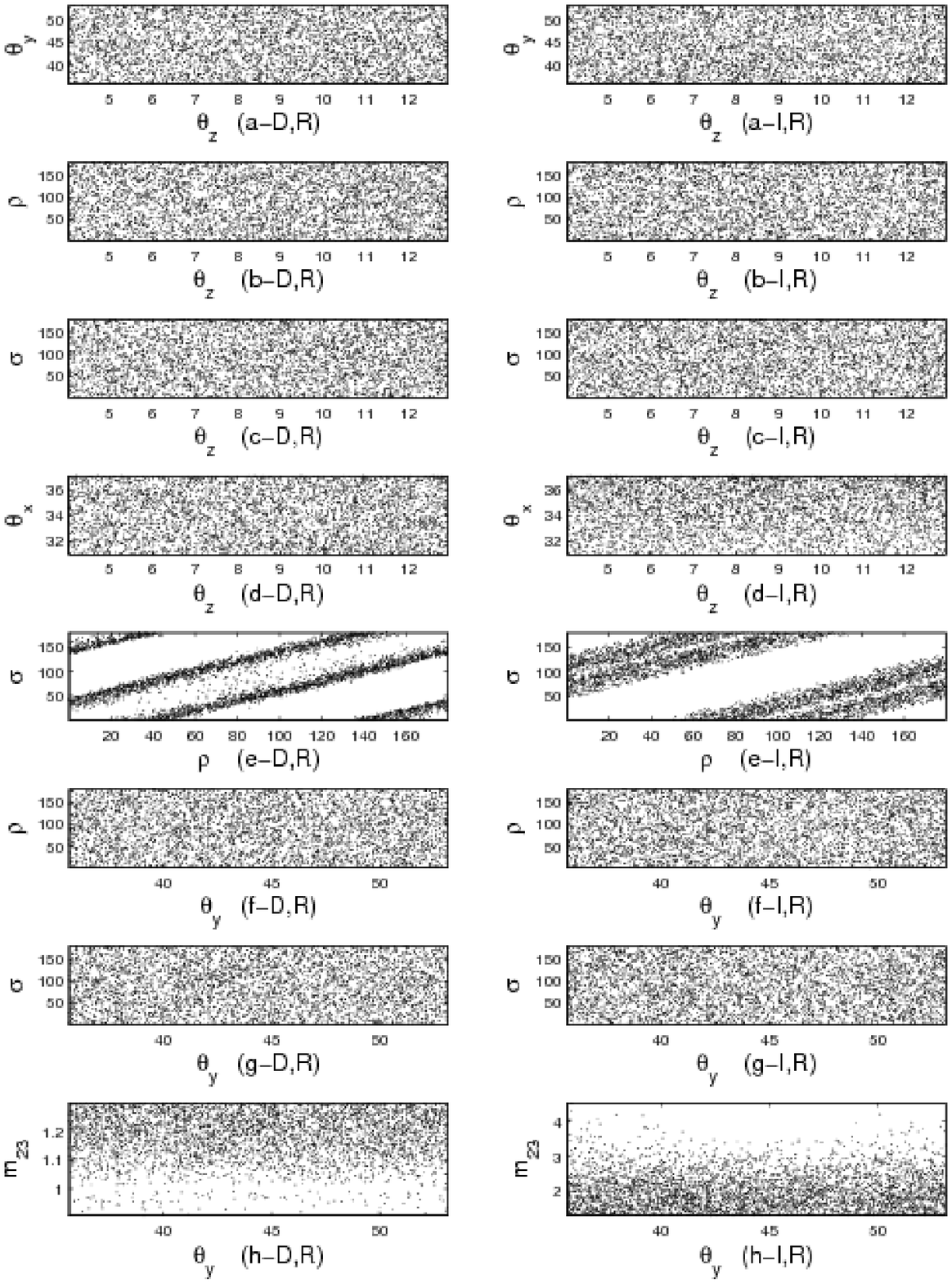}}
\end{minipage}
\vspace{0.5cm}
\caption{{\footnotesize Pattern $\mathbf M_{\n\,32}=0:$ Left panel presents correlations of $\delta$ against
mixing angles and Majorana phases ($\r$ and $\s$) and those of $\t_x$ against $\t_y$, $\r$ and $\s$.
while right panel shows the correlations of $\t_z$ against $\t_y$, $\r$ ,
$\s$, and $\t_x$ and those of $\r$ against $\s$ and $\t_y$, and also the correlation of $\t_y$ versus
$\s$ and $m_{23}$.}}
\label{m32fig1}
\end{figure}

\begin{figure}[hbtp]
\centering
\begin{minipage}[l]{0.5\textwidth}
\epsfxsize=8cm
\centerline{\epsfbox{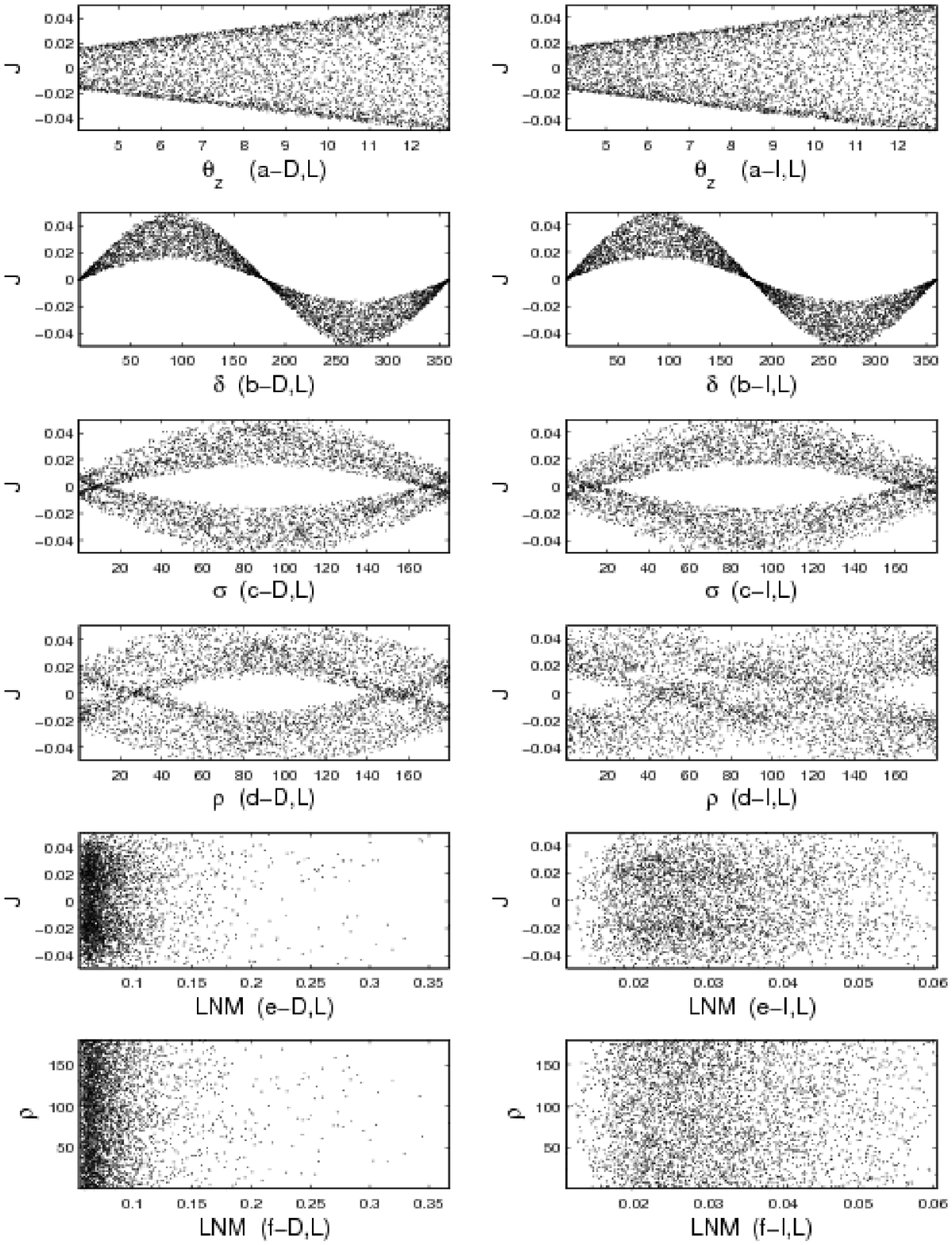}}
\end{minipage}%
\begin{minipage}[r]{0.5\textwidth}
\epsfxsize=8cm
\centerline{\epsfbox{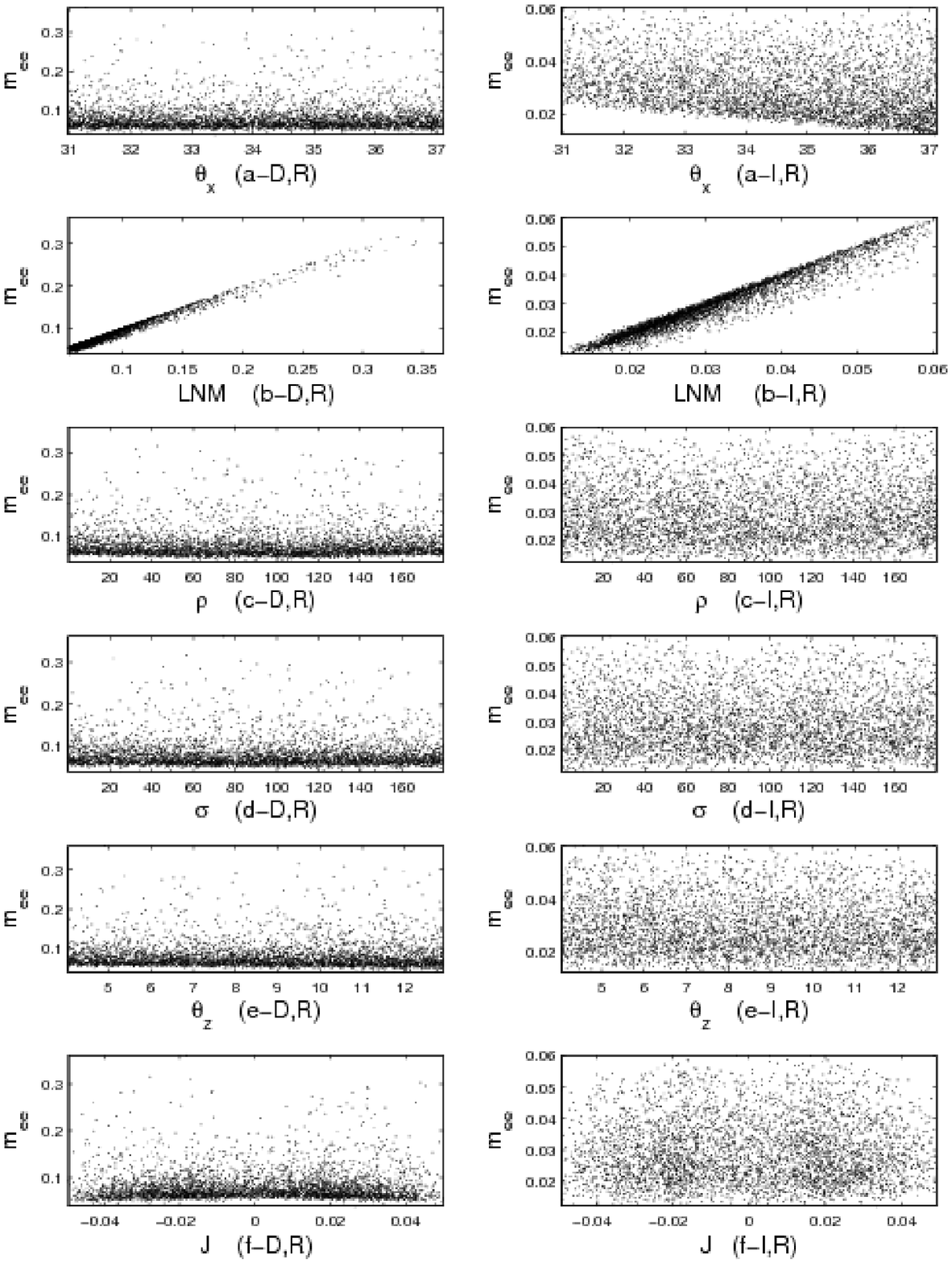}}
\end{minipage}
\vspace{0.5cm}
\caption{{\footnotesize Pattern $\mathbf M_{\n\,32}=0:$ Left panel presents correlations of $J$ against
$\t_z$, $\d$,  $\s$ , $\r$, and lowest neutrino mass ({\bf LNM}), while the last one depicts the
correlation of LNM against $\r$. The right panel shows correlations of $m_{ee}$ against $\t_x$,
LNM, $\r$, $\s$, $\t_z$ and $J$.}}
\label{m32fig2}
\end{figure}

\begin{figure}[hbtp]
\centering
\epsfxsize=7cm
\epsfbox{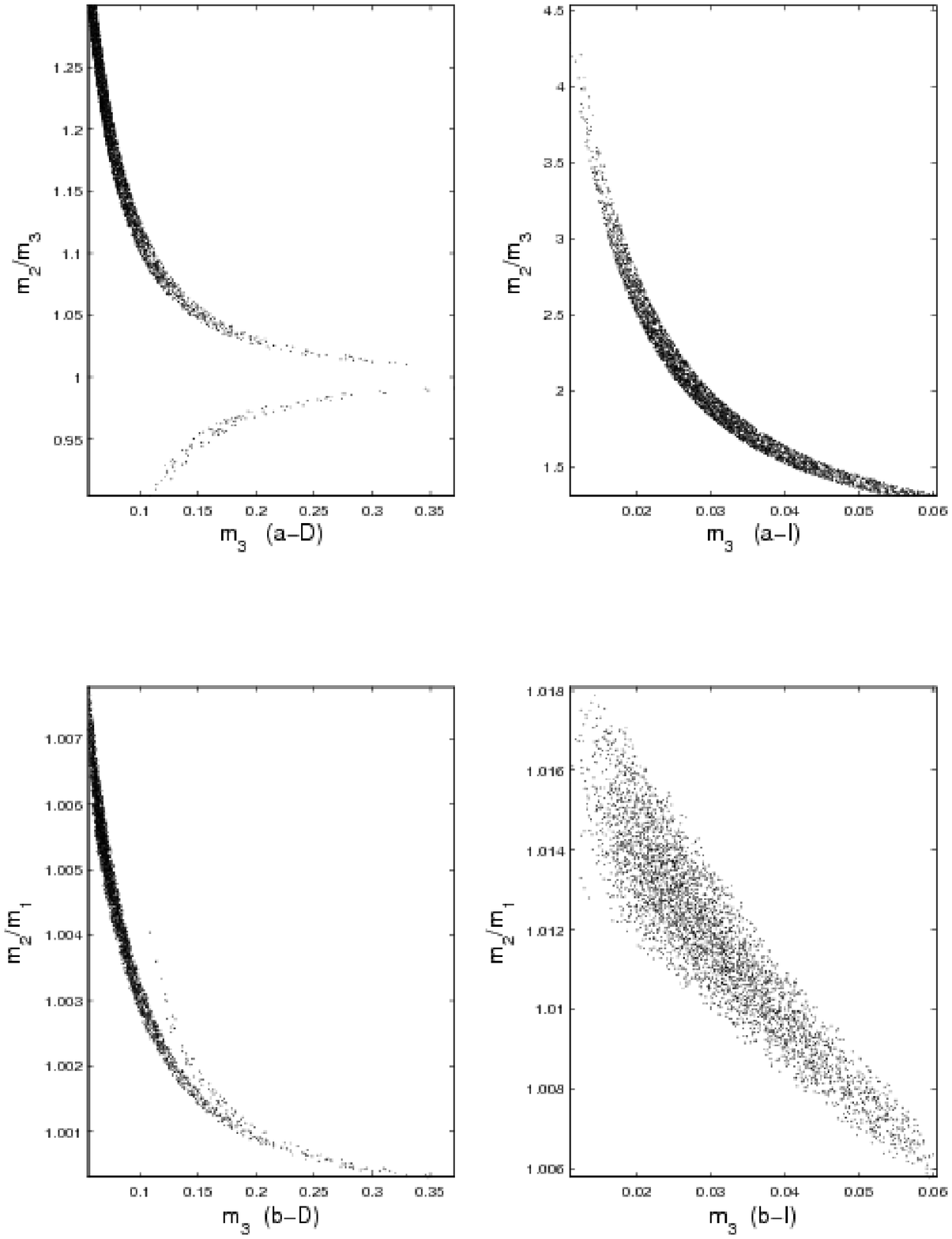}
\caption{{\footnotesize Pattern $\mathbf M_{\n\,32}:$ correlations of mass ratios ${m_2\over m_3}$ and
${m_2\over m_1}$ against $m_3$.}}
\label{m32fig3}
\end{figure}

We checked when we spanned the parameter space that no normal hierarchy could accommodate the data. Figures (\ref{m32fig1}, \ref{m32fig2}  and \ref{m32fig3}) show the corresponding correlation plots, with the same conventions as in the $(M_{\n\, 33}=0)$ case. We see that the
mixing angles and phase angles can cover their experimentally allowed regions. The pair correlations
 $(\d,\r)$,  $(\d,\s)$  and $(\r$, $\s)$ show linear behavior in both degenerate and inverted cases. The `sinusoidal' and `trapezoidal' shapes of the ($J,\d$) and ($J,\t_z$) correlations are invariably covered. Again, a lower bound on $m_{ee}$ restricts enormously the parameter space. The correlation-plots of $m_{ee}$, or alternatively
Table~\ref{tab1}, show that the limit $m_{ee} = 0$ can not be met in either one of the two acceptable hierarchy types.
Again, no clear correlation between ($m_{23}, \t_y$), nor between $\t_y$ and ($\r, \s$). As clear from Figure~\ref{m32fig2}((c-D,L) and (c-I,L)), there is an acute  correlation of $J$ versus $\s$, originating from the sharp linear correlation of $\d$ with $\s$ in both degenerate and inverted cases to be contrasted with the non acute one linking $J$ and $\r$ presented in Figure~\ref{m32fig2}((d-D,L) and (d-I,L)). As to the {\bf LNM} correlations, they show that it tends to have a low value, in that increasing it would reduce hugely the parameter space (this is illustrated
 in plots e-L and f-L of Figure~\ref{m32fig2} in the degenerate case, where the range of the LNM is larger than that in the inverted case and so a concentration of the accepted dots near a low value of the LNM is more obvious). The general tendency of increasing $\mee$ linearly with increasing
LNM in all possible  hierarchies (plots b-R of Figure~\ref{m32fig2}) can be easily recognized.

For the mass spectrum, the inverted hierarchy is not sharp, in that the ratio $m_2/m_3$ has an upper bound of order unity -$\approx 4.5$-
(Figure~\ref{m32fig3}, plot a-I ). We note also that no mass can approach too closely to zero. We see this in the inverted hierarchy either by looking at (Figure~\ref{m32fig3}, plot b-I) and noting that $\frac{m_2}{m_1}$ is not reaching very large values corresponding to very minute $m_1$, or by checking the coverable mass regions in Table~\ref{tab2}. There is no non-invertible such texture which can accommodate the current data.

\clearpage
\clearpage
%%%%%%%%%%%%%%%%%%%%%%%%%%%%%%%%%%%%%%%%%%

%%%%%%%%%%%%%%%%%%%%%%%%%%%%%%%%%%%%%%%%%%%%%%%%%%%%%%%%%%%%%%%%%%%%%%%%%%%%%%%%%%
%%%%%%%%%%%%%%%%%%%%%%%%%%%%  case M13 =0       %%%%%%%%%%%%%%%%%%%%%%%%%%%%%%%%%%%%%%
\subsection{  Pattern of vanishing $\mathbf {M_{\n\,13}}$}
The relevant expressions for $A_1$, $A_2$ and $A_3$ for this model
are
\bea
A_1 &=& c_z c_x (-c_x c_y s_z + s_x s_y\, e^{-i\,\d}),\nn \\
A_2 & = & -s_x c_z (s_x c_y s_z +  c_x s_y\, e^{-i\,\d}),\nn\\
A_3 &=& s_z c_z c_y.
\label{m13}
\eea
We obtain
\bea
\frac{m_1}{m_3} &\approx& \frac{ s_{2\s - \d}\, s_z}{t_y c_x s_x s_{2\r - 2\s}} +
O \left( s_z^2 \right), \\
\frac{m_2}{m_3} &\approx& \frac{s_{\d - 2\r}\,s_z}{t_y c_x s_x s_{2\s - 2\r}}
 +O \left( s_z^2 \right).
 \eea

\begin{figure}[hbtp]
\centering
\begin{minipage}[l]{0.5\textwidth}
\epsfxsize=8cm
\centerline{\epsfbox{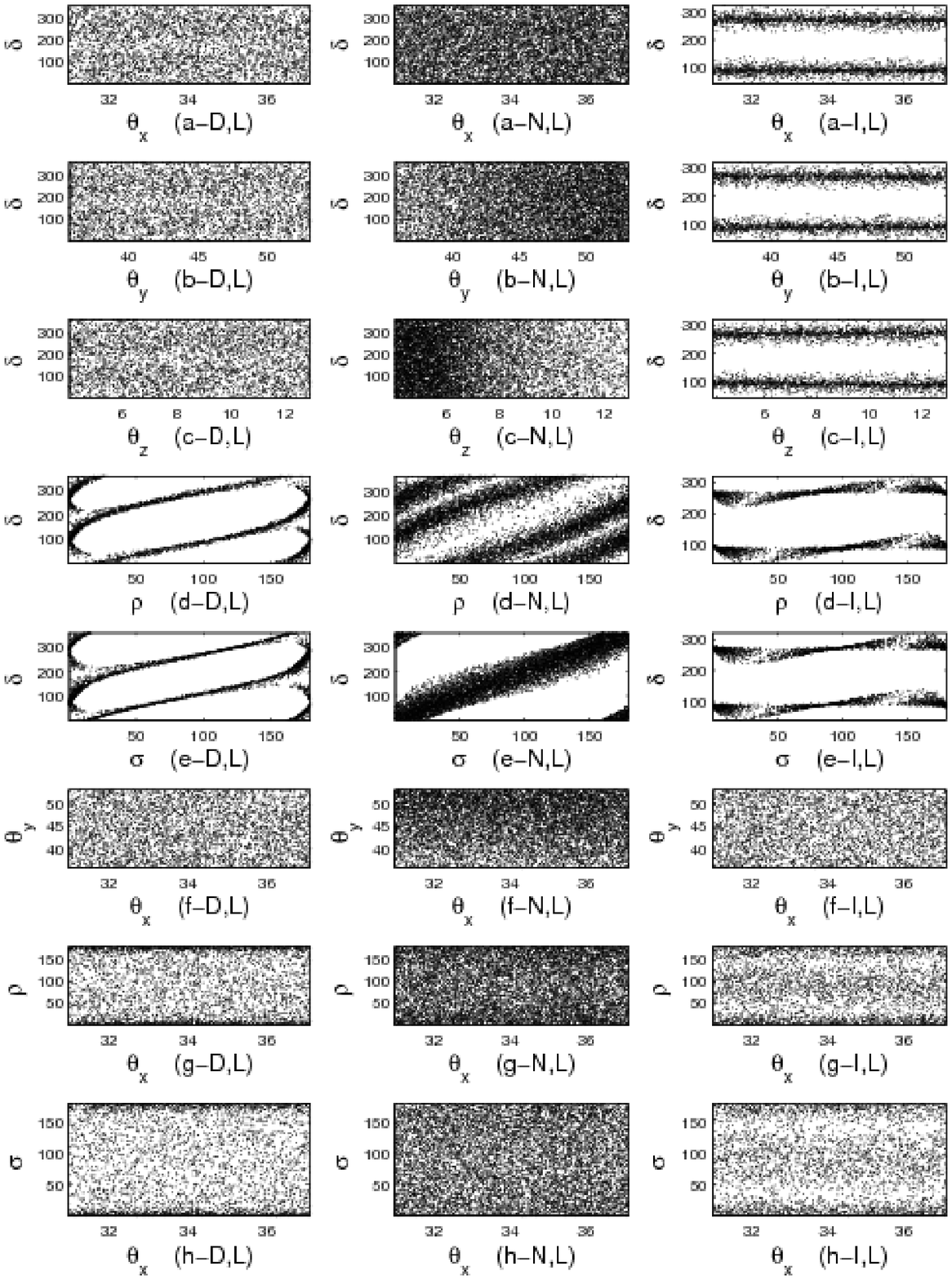}}
\end{minipage}%
\begin{minipage}[r]{0.5\textwidth}
\epsfxsize=8cm
\centerline{\epsfbox{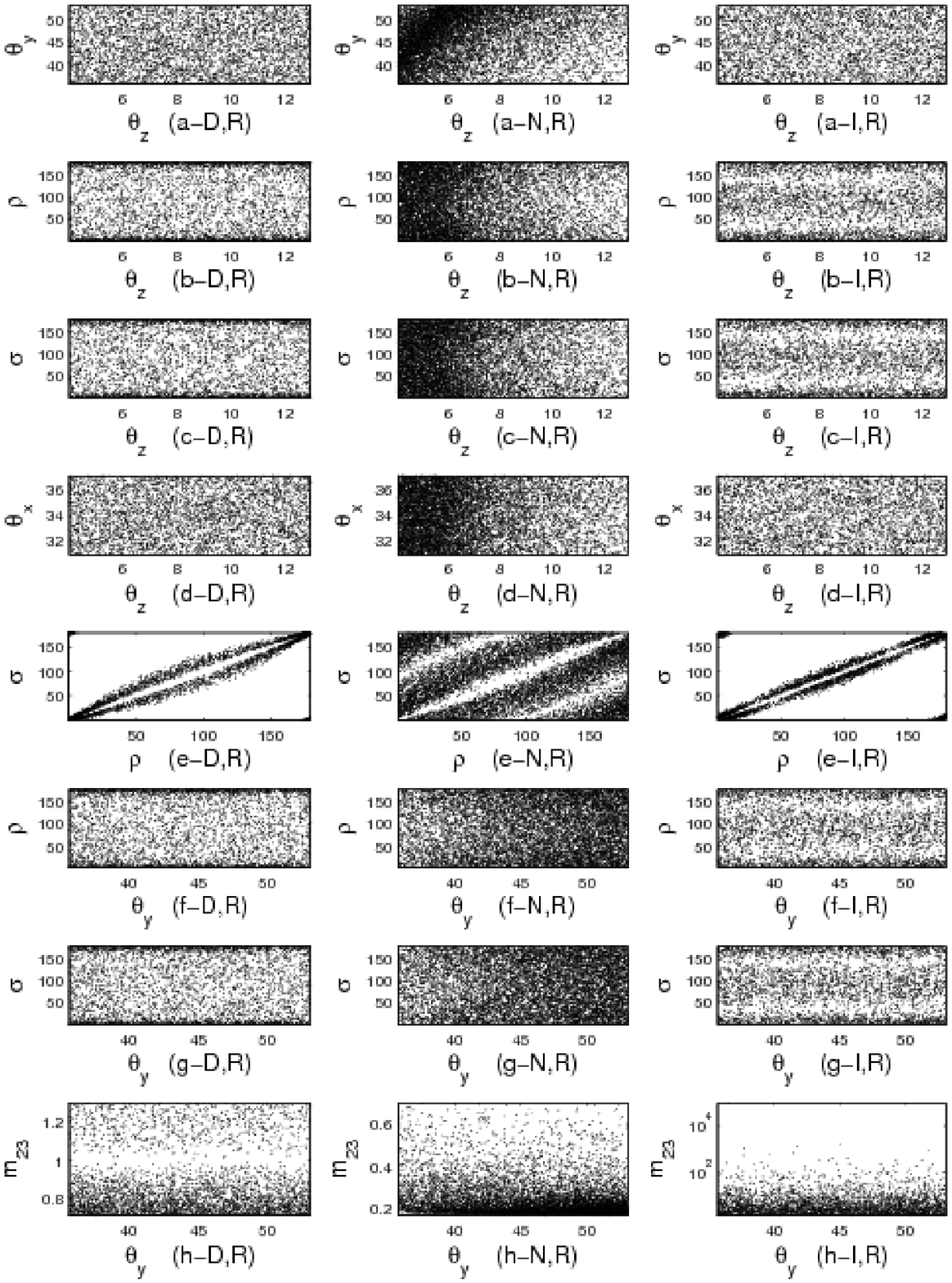}}
\end{minipage}
\vspace{0.5cm}
\caption{{\footnotesize Pattern $\mathbf M_{\n\,13}=0:$ Left panel presents correlations of $\delta$ against mixing angles and Majorana phases ($\r$ and $\s$) and those of $\t_x$ against $\t_y$, $\r$ and $\s$.
while right panel shows the correlations of $\t_z$ against $\t_y$, $\r$ ,
$\s$, and $\t_x$ and those of $\r$ against $\s$ and $\t_y$, and also the correlation of $\t_y$ versus $\s$ and $m_{23}$.}}
\label{m13fig1}
\end{figure}

\begin{figure}[hbtp]
\centering
\begin{minipage}[l]{0.5\textwidth}
\epsfxsize=8cm
\centerline{\epsfbox{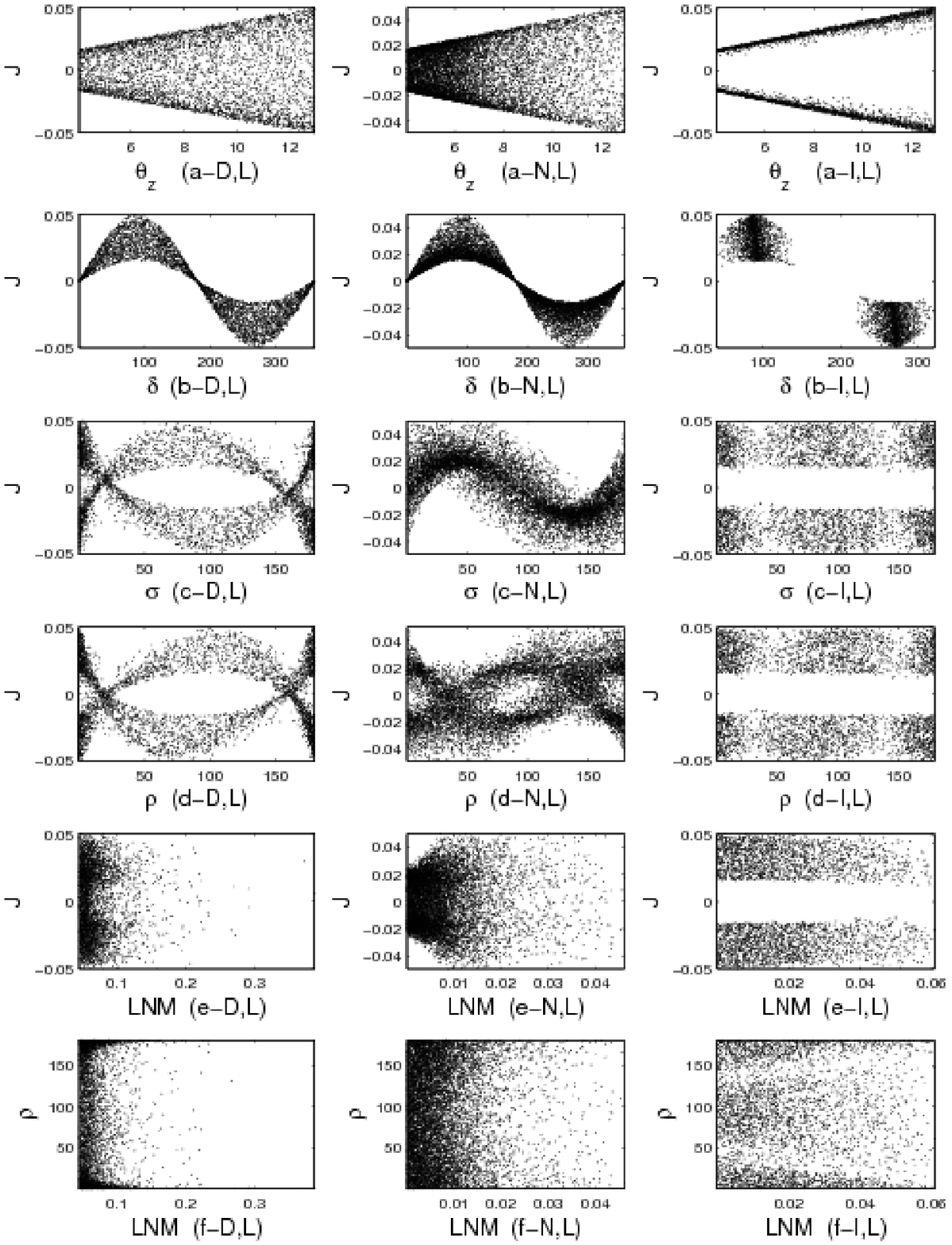}}
\end{minipage}%
\begin{minipage}[r]{0.5\textwidth}
\epsfxsize=8cm
\centerline{\epsfbox{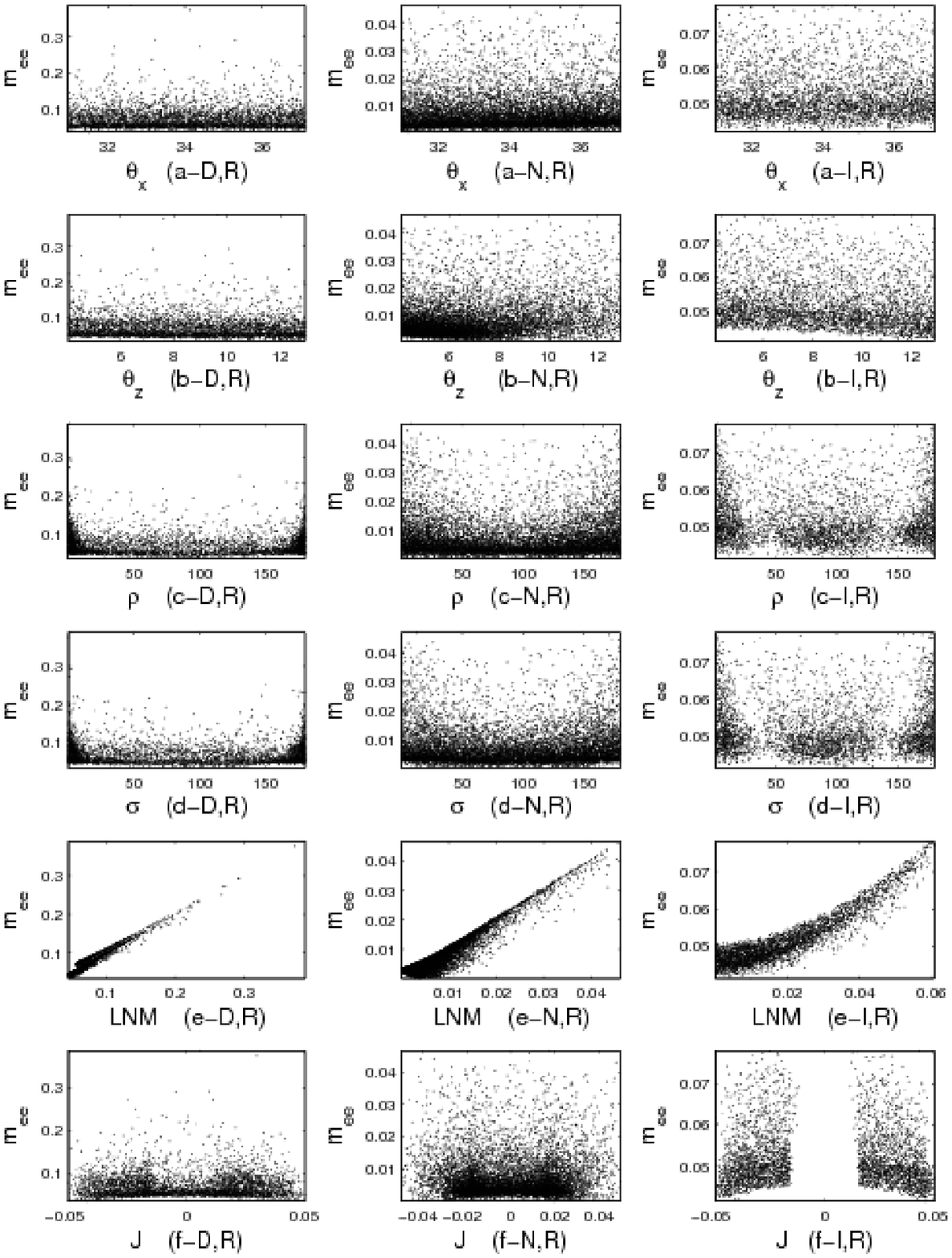}}
\end{minipage}
\vspace{0.5cm}
\caption{{\footnotesize Pattern $\mathbf M_{\n\,13}=0:$ Left panel presents correlations of $J$ against
$\t_z$, $\d$,  $\s$ , $\r$, and lowest neutrino mass ({\bf LNM}), while the last one depicts the correlation of LNM against $\r$. The right panel shows correlations of $m_{ee}$ against $\t_x$, $\t_z$, $\r$, $\s$, {\bf LNM} and $J$.}}
\label{m13fig2}
\end{figure}

\begin{figure}[hbtp]
\centering
\epsfxsize=7cm
\epsfbox{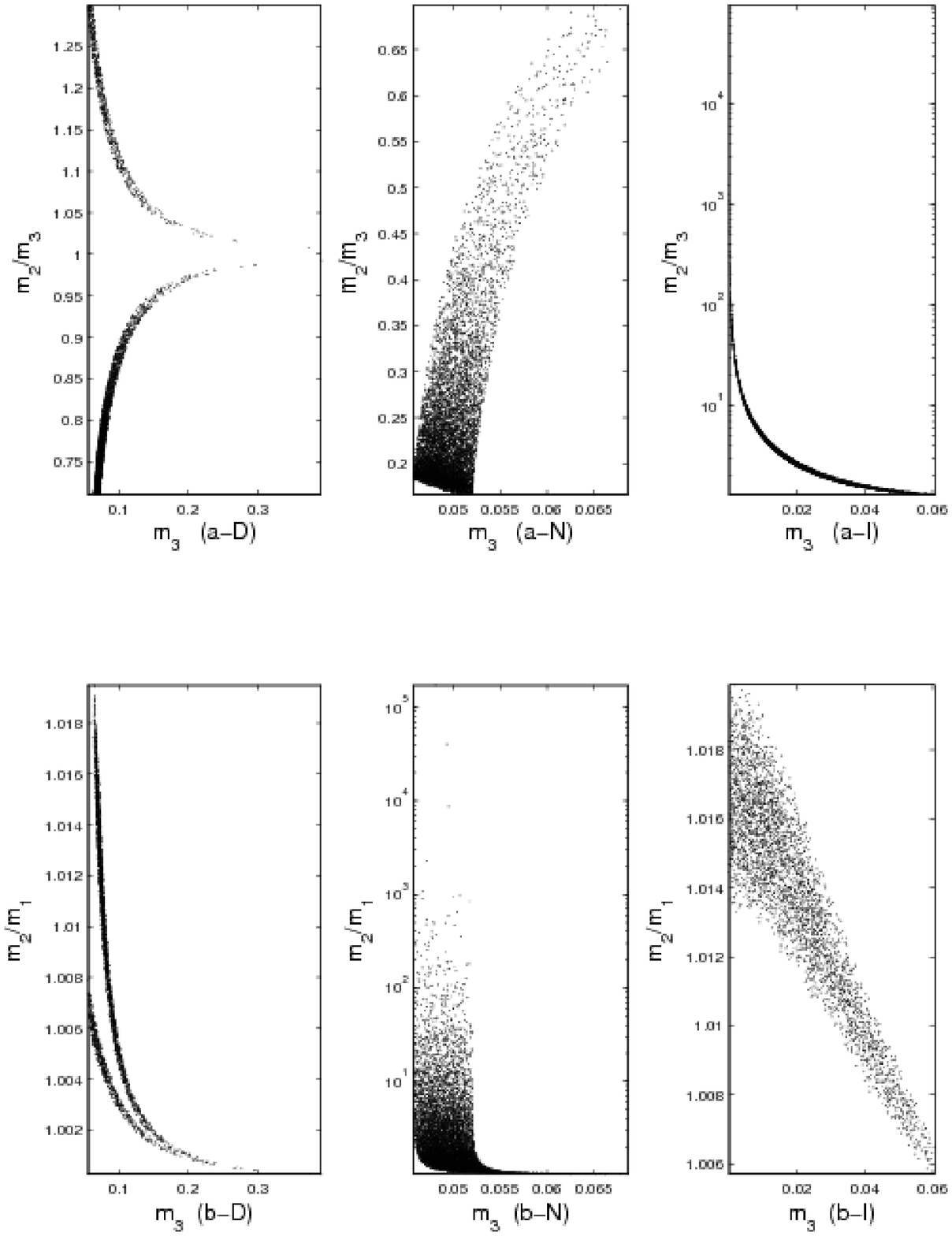}
\caption{{\footnotesize Pattern $\mathbf M_{\n\,13}=0:$ correlations of mass ratios ${m_2\over m_3}$ and
${m_2\over m_1}$ against $m_3$.}}
\label{m13fig3}
\end{figure}

We have also
\bea \label{Rt31} R_\nu &=& \frac{c_y^2 \left|s_{\d-2\,\s}^2-s_{\d-2\,\r}^2\right|}
{c_x^2 s_x^2 s_y^2 s_{2\r-2\s}^2}\,s_z^2+O \left(
s_z^3\right).
\eea
We plot the corresponding correlations in Figures (\ref{m13fig1}, \ref{m13fig2}  and \ref{m13fig3}) with the same conventions as before.
In contrast to the $(M_{\n\, 33}=0)$ case, we see here that the
mixing angles ($\t_x,\t_y,\t_z$) can cover all their allowable regions (Figure~\ref{m13fig1}, as examples,
plots: a-L $\rightarrow$ c-L) and in all hierarchy types.
The linear correlations of $\d$ versus $\r$ and $\s$ disappear in the inverted case, whereas they are
replaced by Lissajous-like patterns in the degenerate case (Figure~\ref{m13fig1}, plots: d-L, e-L). However,
there is a quasi linear correlation between $\r$ and $\s$ (Figure~\ref{m13fig1}, plot: e-R) in the degenerate and inverted cases. The special `sinusoidal' and `trapezoidal' shapes of $J$ versus $\d$ and $\t_z$ remain
(Figure~\ref{m13fig2}, plots: a-L, b-L), but we note that in the inverted case the sinusoidal shape is
severed and does not extend over the whole range of $\d$.  This would single out three disallowed regions for $\d$, one is from $0^0$ to $40.72^0$ and another from $140^0$ to $221.9^0$ while the third from $320.81^0$ to $360^0$ approximately. Again no clear correlation involves $\mee$
(Figure~\ref{m13fig2}, right panel). However, setting a lower bound on this parameter would contract enormously the parameter space, as can be seen from the upper bound of $m_{ee}$: ($< 0.05$ eV) in the normal hierarchy type, followed by ($< 0.08$ eV) in the inverted case, and then ($< 0.4$ eV) in the degenerate case. Apart from the usual correlations of $J$ versus $\r$ and
$\s$ (Figure~\ref{m13fig2} plots: c-L, d-L), initiated by the correlation of $\d$ with $\r$ and $\s$,
the other plots  does not show clear correlations. We see from Table~\ref{tab2} that the limit $m_{ee} = 0$
is not attainable in this pattern, although a tiny value for $\mee$ of $O(10^{-5}\,\mbox{eV})$ can be
achieved in the normal case. Again, the general tendency of increasing $\mee$  with increasing
LNM in all possible  hierarchies (Figure~\ref{m13fig2}, plots e-R) can be easily recognized.

For the mass spectrum, the plot b-I in Figure~\ref{m13fig3} tells us that the experimental data can be
accommodated in the inverted hierarchy type only when the two masses $m_1$ and $m_2$ are approximately equal. However, the mass ratio-parameter $m_2/m_3$ (plot a-I) indicates a strong hierarchy. Similarly, the
normal type hierarchy case (plots a-N and b-N) reveals a strong  hierarchy as shown by the mass ratios
 ${m_2\over m_1}$ in (plot b-N), while the mass ratios ${m_2\over m_3}$ are of order $O(1)$ as
 shown in (plot a-N). We see also that in contrast to the pattern of vanishing $M_{\n\,33}$, the limit $m_1 = 0$ can also be reached. In fact, there is a non-invertible such texture which can accommodate the current data, and this happens when $m_1=0$ or $m_3=0$.

%%%%%%%%%%%%%%%%%%%%%%%%%%%%%%%%%%%%%%%%%%%%%%%%%%%%%%%%%
\begin{landscape}
\begin{table}[h]
 \begin{center}
\scalebox{0.99}{
{\tiny
 \begin{tabular}{c|c|c|c|c|c|c|c|c|c|c|c|c}

 %%%%%%%%%%%%%%%%%%%%%%%%%%%%%%%%%%%%%%%%%%%%%%%%%%%%%%%%%%%%%%%%%%%%%%%%%%%%%%%%%%%%%%%%%%%%%%%%%
 %%%%%%%%%%%%%%%%%%%%%%%%%%%%%%Model  M_33 %%%%%%%%%%%%%%%%%%%%%%%%%%%%%%%%%%%%%%%%%%%%%%%
 %%%%%%%%%%%%%%%%%%%%%%%%%%%%%%%%%%%%%%%%%%%%%%%%%%%%%%%%%%%%%%%%%%%%%%%%%%%%%%%%%
 \hline
\multicolumn{13}{c}{\mbox{Model} $M_{\n\,33}=0$} \\
\hline
 \mbox{quantity} & $\th_x$ & $\th_y$& $\th_z$ & $m_1$ & $m_2$& $m_3$ & $\r$ & $\sig$ & $\d$ & $\me$
 & $\mee$ & $J$\\
 \hline
 \multicolumn{13}{c}{\mbox{Degenerate  Hierarchy}} \\
 \cline{1-13}
 $1\, \sig$ &$32.96 - 35.00$ & $40.40 - 44.99$ & $ 7.71 - 10.30$ & $0.0747 - 0.2879$ & $0.0752 - 0.2880$ &
  $0.0579 - 0.2840$& $0.0839-  179.90$ & $0.1234 - 179.94$&$0.0285 - 359.94$ & $0.0745 - 0.2878$
  &$0.0367 - 0.2752 $ & $-0.0401 - 0.0403$ \\
 \hline
 $2\, \sig$ & $31.95 - 36.09$ & $40.44 - 50.77$& $6.29 - 11.68$ & $0.0464 - 0.3161$ & $0.0473 - 0.3162$ &
 $0.0569 - 0.3133$&$ 0.0933 -  179.89$ & $0.1043 - 180$ & $0.1110 - 359.77$& $0.0473 - 0.3161$ &
 $0.0240 - 0.2972$ & $-0.0452 - 0.0455$  \\
 \hline
 $3\, \sig$ &$30.98 - 37.10$ &$40.17 -53.13$ & $4.05 -12.92$ & $0.0452- 0.3718$ & $0.0461- 0.3719$ &
  $0.0553 - 0.3683$& $0.0197 - 179.99 $ & $0.0057 - 179.94$ & $0.0790 -359.99$ &
  $0.0457-  0.3718$ & $0.0194 - 0.2668$ & $-0.0498 -  0.0502$ \\
 \hline
 %%%%%%%%%%%%%%%%%%%%%%%%%%%%%%%%%%%%%%%%%%%%%%%%%%%%%%%%%%%%%%%%%
 \multicolumn{13}{c}{\mbox{Normal  Hierarchy}} \\
 \cline{1-13}
 $1\, \sig$ &$\times $& $\times$ &$\times$ &$\times$ &$\times$ &
  $\times$& $\times$ &$\times$ & $\times$ &$\times$ &
  $\times$ & $\times$  \\
 \hline
 $2\, \sig$ & $31.95 - 36.09$ &$ 49.32- 50.77$ & $6.29 - 11.68 $ &$0.0372 -0.0484$ &$0.0382 - 0.0492$
 &$ 0.0602 - 0.0703$ & $0.0084 - 179.93$ & $ 0.0048 - 179.94$ & $0.1524 - 359.73$& $0.0386 - 0.0496$  &
 $0.0335 - 0.0492$ & $-0.0449 -  0.0445$  \\
 \hline
 $3\, \sig$ &$30.98- 37.11$ & $49.08 -53.13$ & $4.05 -12.92$& $0.0277-0.0494$ & $0.0290 - 0.0501$ &
  $ 0.0539 - 0.0716$& $0.0224 - 179.97$ & $0.0073 -179.94$ &$0.0213 - 359.91$ & $0.0297 - 0.0502$ &
  $ 0.0239 -  0.0490$ & $-0.0485 -  0.0490$ \\
 \hline
 %%%%%%%%%%%%%%%%%%%%%%%%%%%%%%%%%%%%%%%%%%%%%%%%%%%%%%%%%%%%%%%%%%%%%%%%
 \multicolumn{13}{c}{\mbox{Inverted  Hierarchy}} \\
 \cline{1-13}
 $1\, \sig$ &$32.96 - 35.00$ & $38.65 - 45.00$ & $7.71 - 10.30$ & $0.0472 -0.0762$ &
 $0.0479 - 0.0766$ & $3.05\times 10^{-7} - 0.0583$ & $0.0630 - 179.99$ & $0.0280 - 179.97$ & $0.7166 - 359.73$ &
  $0.0467 - 0.0760$& $ 0.0155 - 0.0746$ & $  -0.0398 - 0.0402$ \\
 \hline
 $2\, \sig$ &$31.95  - 36.09$ & $36.87- 50.77$ & $6.29 -11.68$ & $0.0461-  0.0785$ &
 $0.0470 -  0.0790$ & $1.078\times 10^{-5}- 0.0604$ & $0.0355 -179.93$ & $0.0057- 179.98$ & $0.2343 -358.68$ &$ 0.0455 - 0.0781$& $ 0.0135 - 0.0711$ & $ -0.0449 - 0.0453$  \\
 \hline
 $3\, \sig$ & $30.98 -37.11$ & $35.67 - 53.12$ & $4.05 -12.92$ & $0.0450- 0.0794$ &
 $ 0.0458 - 0.0799$ & $1.11\times 10^{-6} -0.0607$ & $0.0414 -179.98$ & $0.0105 -180$ & $0.2134 -359.15$ &
  $0.0442 -  0.0791$ & $ 0.0114 - 0.0749$& $ -0.0482 - 0.0487$  \\
 \hline
 %%%%%%%%%%%%%%%%%%%%%%%%%%%%%%%%%%%%%%%%%%%%%%%%%%%%%%%%%%%%%%%%%%%%%%%%%%%%%%%%%%%%%%%%%%%%%%%%%
 %%%%%%%%%%%%%%%%%%%%%%%%%%%%%%Model  M_22 %%%%%%%%%%%%%%%%%%%%%%%%%%%%%%%%%%%%%%%%%%%%%%%
 %%%%%%%%%%%%%%%%%%%%%%%%%%%%%%%%%%%%%%%%%%%%%%%%%%%%%%%%%%%%%%%%%%%%%%%%%%%%%%%%%
 \hline
\multicolumn{13}{c}{\mbox{Model} $M_{\n\,22}=0$} \\
\hline
 \mbox{quantity} & $\th_x$ & $\th_y$& $\th_z$ & $m_1$ & $m_2$& $m_3$ & $\r$ & $\sig$ & $\d$ & $\me$
 & $\mee$ & $J$\\
 \hline
 \multicolumn{13}{c}{\mbox{Degenerate  Hierarchy}} \\
 \cline{1-13}
 $1\, \sig$ &$ 32.96 -35.00$ & $38.65 -45.00$ & $7.71 -10.30$ & $0.0471- 0.3933$ & $0.0480 -  0.3933$ &
  $0.0580 -0.3903$& $0.0552 - 179.99$ & $0.0481 -179.99$&$0.2248 -359.99$ & $0.0480 - 0.3932$
  &$0.0268 - 0.2978$ & $-0.0398  - 0.0396$ \\
 \hline
 $2\, \sig$ & $31.95 - 36.09$ & $36.87 - 49.65$& $6.29 -11.68$ & $0.0462 -0.3898$ & $0.0471 -0.3899$ &
 $0.0567- 0.3885$&$0.0153- 179.83$ & $0.0457 - 179.99$ & $0.0996 -359.97$& $0.0471 - 0.3897$ &
 $0.0229 -0.3399$ & $-0.0449 - 0.0457$  \\
 \hline
 $3\, \sig$ &$30.98 - 37.11$ &$35.68 - 50.06$ & $4.05 - 12.92$ & $0.0453 -0.3827$ & $0.0462 -0.3828$ &
  $0.0552 - 0.3859$& $0.0038 -  179.89$ & $0.0431 - 179.97$ & $0.1258 -359.93$ &
  $ 0.0459 - 0.3828$ & $ 0.0203 - 0.3596$ & $-0.0496 - 0.0489$ \\
 \hline
 %%%%%%%%%%%%%%%%%%%%%%%%%%%%%%%%%%%%%%%%%%%%%%%%%%%%%%%%%%%%%%%%%
 \multicolumn{13}{c}{\mbox{Normal  Hierarchy}} \\
 \cline{1-13}
 $1\, \sig$ &$32.96 -35$& $38.65 - 40.50$ &$7.71 -10.30$ &$0.0361 - 0.0474$ &$0.0371 -0.0482$ &
  $0.0602 - 0.0689$& $0.2889 -  179.91$ &$0.1453 - 179.94$ & $0.1594 - 359.78$ &$ 0.0374 - 0.0484$ &
  $0.0333 - 0.0477$ & $-0.0394 - 0.0396$  \\
 \hline
 $2\, \sig$ & $31.95 -36.09$ &$36.87 - 40.67$ & $6.29 - 11.68$ &$0.0292 -0.0483$ &$0.0304 - 0.0491$
 &$0.0555 - 0.0702$ & $0.0029 -179.9$ & $ 0.0798 -179.99$ & $0.1001 - 359.93$& $0.0307- 0.0495$ &
 $0.0251 - 0.0488$ & $-0.0443 - 0.0447$  \\
 \hline
 $3\, \sig$ &$ 30.98-  37.11$ & $35.67 - 40.79$ & $4.05 -12.91$& $0.0250 -0.0492$ & $0.0265 -0.0500$ &
  $0.0522 - 0.0715$& $0.0087- 179.96$ & $0.0066  - 179.96$ &$0.0250 - 359.88$ & $ 0.0269 - 0.0505$ &
  $0.0209 - 0.0485$ & $-0.0484 - 0.0493$ \\
 \hline
 %%%%%%%%%%%%%%%%%%%%%%%%%%%%%%%%%%%%%%%%%%%%%%%%%%%%%%%%%%%%%%%%%%%%%%%%
 \multicolumn{13}{c}{\mbox{Inverted  Hierarchy}} \\
 \cline{1-13}
 $1\, \sig$ & $32.96 - 35$  &$38.65 - 45$ &$7.71- 10.30$ &$0.0473 - 0.0765$ &$0.0481 - 0.0770$&
 $0.0023 - 0.0587$ &$0.0479 - 179.96$ &$0.0171 -179.99$ &$0.0452 - 359.93$ &$0.0468 - 0.0762$
 &$0.0157- 0.0662$ &$-0.0398 - 0.0402$  \\
 \hline
 $2\, \sig$ &$31.95- 36.09$ & $36.89 -50.76$ & $6.29 - 11.68$ & $0.0462 - 0.0782$ &
 $0.0470 - 0.0786$ & $2.016\times 10^{-5}- 0.0599$ & $0.0320 -179.99$ & $0.0736 - 179.97$ &$0.0399-360$ &
  $0.0456 - 0.0780$& $ 0.0135 - 0.0741$ & $-0.0448- 0.0441$  \\
 \hline
 $3\, \sig$ & $30.99 -37.11$ & $35.71 - 53.13$ & $4.05 - 12.92$ & $0.0450 -  0.0784$ &
 $0.0458 - 0.0789$ & $1.33 \times 10^{-5}- 0.0602$ & $0.0537- 180$ & $0.0159 - 179.94$ &$0.0638 - 359.99$ &
  $0.0442 - 0.0783$ & $0.0117 - 0.0744$& $-0.0486 - 0.0492$  \\
 \hline
 %%%%%%%%%%%%%%%%%%%%%%%%%%%%%%%%%%%%%%%%%%%%%%%%%%%%%%%%%%%%%%%%%%%%%%%%%%%%%%%%%%%%%%%%%%%%%%%%%
 %%%%%%%%%%%%%%%%%%%%%%%%%%%%%%Model  M_32 %%%%%%%%%%%%%%%%%%%%%%%%%%%%%%%%%%%%%%%%%%%%%%%
 %%%%%%%%%%%%%%%%%%%%%%%%%%%%%%%%%%%%%%%%%%%%%%%%%%%%%%%%%%%%%%%%%%%%%%%%%%%%%%%%%
 \hline
\multicolumn{13}{c}{\mbox{Model} $M_{\n\,32}=0$} \\
\hline
 \mbox{quantity} & $\th_x$ & $\th_y$& $\th_z$ & $m_1$ & $m_2$& $m_3$ & $\r$ & $\sig$ & $\d$ & $\me$
 & $\mee$ & $J$\\
 \hline
 \multicolumn{13}{c}{\mbox{Degenerate  Hierarchy}} \\
 \cline{1-13}
 $1\, \sig$ &$32.96 -35$ & $38.65- 45$ & $7.71 -10.30$ & $0.0746 -0.3955$ &$0.0752 -0.3956$ &
  $0.0579 - 0.3926$& $0.0037 - 179.97$ & $0.0248 - 179.92$ &$0.1188 - 359.96$ & $0.0745 - 0.3954$
  &$0.0483 - 0.3617$ & $ -0.0395 -0.0404$ \\
 \hline
 $2\, \sig$ & $31.95 -36.09$ & $36.88 - 50.76$& $6.29 - 11.68$ & $0.0732 - 0.3873$ &$0.0738 - 0.3874$ &
 $0.0568 - 0.3841$&$ 0.1507 - 179.93$ & $ 0.0139 -179.96$ & $0.0203  - 360$& $0.0730 - 0.3873$ &
 $0.0457 - 0.3518$ & $-0.0445 - 0.0451$  \\
 \hline
 $3\, \sig$ &$30.98 -  37.11$ & $35.67 - 53.12$ & $4.05 -12.92$& $0.0713 -0.3673$ & $0.0719 -0.3674$ &
  $0.0553 -0.3708$& $0.0350 - 179.98$ & $ 0.0539 - 179.9$ &$0.0052 - 359.88$ & $0.0710 - 0.3674$ &
  $  0.0389 - 0.3609$ & $ -0.0496 -  0.0501$ \\
 \hline
 %%%%%%%%%%%%%%%%%%%%%%%%%%%%%%%%%%%%%%%%%%%%%%%%%%%%%%%%%%%%%%%%%
 \multicolumn{13}{c}{\mbox{Normal  Hierarchy}} \\
 \cline{1-13}
 $1\, \sig$ &$\times$& $\times$ &$\times$ &$\times$ &$\times$ &
  $\times$& $\times$ &$\times$ & $\times$ &$\times$ &
  $\times$ & $\times$  \\
 \hline
 $2\, \sig$ &$\times$& $\times$ &$\times$ &$\times$ &$\times$ &
  $\times$& $\times$ &$\times$ & $\times$ &$\times$ &
  $\times$ & $\times$  \\
 \hline
 $3\, \sig$ &$\times$& $\times$ &$\times$ &$\times$ &$\times$ &
  $\times$& $\times$ &$\times$ & $\times$ &$\times$ &
  $\times$ & $\times$  \\
 \hline
 %%%%%%%%%%%%%%%%%%%%%%%%%%%%%%%%%%%%%%%%%%%%%%%%%%%%%%%%%%%%%%%%%%%%%%%%
 \multicolumn{13}{c}{\mbox{Inverted  Hierarchy}} \\
 \cline{1-13}
 $1\, \sig$ &$32.96 - 35.00$ & $38.65 -   45.00$ & $7.71 - 10.30$ & $0.0500 - 0.0769$ &
 $0.0508 - 0.0773$ & $ 0.0156 - 0.0590$ & $ 0.0034 -179.86$ & $0.0529 - 179.84$ & $0.1501 - 360$ &
  $0.0496  - 0.0766$& $ 0.0166 -  0.0580$ & $-0.0397 - 0.0403$  \\
 \hline
 $2\, \sig$ &$31.95 - 36.09$ & $ 36.87 - 50.77$ & $6.29 -  11.68$ & $0.0484 -  0.0784$ &
 $0.0491 - 0.0789$ & $ 0.0128 -0.0602$ & $ 0.0441 - 179.97$ & $0.0091 -179.97$ & $0.1001 - 359.97$ &
  $0.0480 - 0.0781$& $  0.0144 - 0.0587$ & $-0.0455 - 0.0451$  \\
 \hline
 $3\, \sig$ & $30.98 - 37.11$ & $35.67 -53.13$ & $4.05 -12.92$ & $0.0463 -0.0790$ &
 $0.0471 - 0.0794$ & $0.0110 - 0.0605$ & $0.0900 - 179.88$ & $ 0.0075 -179.98$ & $0.1826 -359.76$ &
  $0.0459 -0.0790$ & $0.0122 - 0.0602$& $-0.0491 - 0.0491$  \\
 \hline

 \end{tabular}
 }}
 \end{center}
  \caption{\small \label{tab2} The various prediction for the patterns of
  one-zero texture $(M_{\n\,33}=0)$ , $(M_{\n\,22}=0)$ and $(M_{\n\,32}=0)$. All the angles (masses) are
  evaluated in degrees ($eV$).}
 \end{table}
\end{landscape}

%%%%%%%%%%%%%%%%%%%%%%%%%%%%%%%%%%%%%%%%%%%%%%%%%%%%%%%%%%%%%%%%%%%%%%%%%%%%%%
%%%%%%%%%%%%%%%%%%Second TAble%%%%%%%%%%%%%%%%%%%%%%%%%%%%%%%%%%%%%%%%%%%
%%%%%%%%%%%%%%%%%%Second Table%%%%%%%%%%%%%%%%%%%%%%%%%%%%%%%%%%%%%%%%%%%
\begin{landscape}
\begin{table}[h]
 \begin{center}
\scalebox{0.8}{
{\tiny
 \begin{tabular}{c|c|c|c|c|c|c|c|c|c|c|c|c}
 \hline
\multicolumn{13}{c}{\mbox{Model} $M_{\n\,13}=0$} \\
\hline
 \mbox{quantity} & $\th_x$ & $\th_y$& $\th_z$ & $m_1$ & $m_2$& $m_3$ & $\r$ & $\sig$ & $\d$ & $\me$
 & $\mee$ & $J$\\
 \hline
 \multicolumn{13}{c}{\mbox{Degenerate  Hierarchy}} \\
 \cline{1-13}
 $1\, \sig$ &$32.96 -35.00$ & $38.65-45.00$ & $7.71 -10.30$ & $0.0470 -0.3938$& $0.04783 -0.3939$ &
  $0.0579 -0.3968$ & $0.0057 -179.99$&$ 0.0012 -179.99$ & $0.1688- 359.94$
  &$0.0480 -0.3939$ & $ 0.0397-0.3939$ & $-0.0402 - 0.0398$\\
 \hline
 $2\, \sig$ & $31.95 - 36.09$ & $36.87 -50.76$& $6.29 -11.68$ & $0.0461 -0.3802$ & $0.0470 - 0.3803$ &
 $0.0575- 0.3771$& $0.0108 -180$ & $ 0.0013 -179.99$ & $0.1427 - 359.95$& $0.0471 - 0.3802$ &
 $0.0364 - 0.3562$ & $-0.0454 -0.0456$  \\
 \hline
 $3\, \sig$ &$30.98 -37.11$ &$35.673- 53.13$ & $4.05 -12.92$ & $0.0451-0.3878$ & $0.0458 -0.3880$ &
  $0.0555 - 0.3908$ & $0.0026 -180$ & $0.0057 - 180$ & $0.1566 - 359.82$ &
  $0.0457 - 0.3880$ & $0.0341 - 0.3879$ & $-0.0503 -0.0504$ \\
 \hline
 %%%%%%%%%%%%%%%%%%%%%%%%%%%%%%%%%%%%%%%%%%%%%%%%%%%%%%%%%%%%%%%%%
 \multicolumn{13}{c}{\mbox{Normal  Hierarchy}} \\
 \cline{1-13}
 $1\, \sig$ &$32.96 - 35.00$& $38.66 - 45.00$ &$7.71 -10.30$ &$0.0045 - 0.0468$ &$0.0099 -0.0476$ &
  $0.0483 - 0.0681$& $0.0588 - 179.96$ &$0.0137 - 179.99$ & $0.1474 -359.89$ &$0.0094 - 0.0478$ &
  $0.0003 - 0.0460$ & $-0.0398 -0.0404$  \\
 \hline
 $2\, \sig$ & $31.95 - 36.09$ &$36.87- 50.77$ & $6.29 - 11.68$ &$0.0003 - 0.0477$ &$ 0.0085 - 0.0484$
 &$ 0.0470 - 0.0693$ & $0.0434 -179.95$ & $ 0.0263 - 179.94$ & $0.0879 - 359.91$& $0.0071 - 0.0484$ &
 $0.0001 -  0.0482$ & $-0.0447 - 0.0453$  \\
 \hline
 $3\, \sig$ &$30.98 - 37.11$ & $35.69 - 53.13$ & $4.05 - 12.92$& $4.94\times 10^{-8} -0.0462$ &
 $0.0084 -0.0471 $&
  $0.0458 - 0.0686$& $0.0037 - 179.99$ & $0.0007  -179.98$ &$0.0396 -359.99$ & $0.0056 -0.0468$ &
  $5.65\times 10^{-5} -0.0466$ & $-0.0495 - 0.0496$ \\
 \hline
 %%%%%%%%%%%%%%%%%%%%%%%%%%%%%%%%%%%%%%%%%%%%%%%%%%%%%%%%%%%%%%%%%%%%%%%%
 \multicolumn{13}{c}{\mbox{Inverted  Hierarchy}} \\
 \cline{1-13}
 $1\, \sig$ & $32.96 - 35.00$  &$38.65- 45.00$ &$7.71-10.30$ &$0.0471 - 0.0765$ &$0.0480 - 0.0770$
 &$1.04\times 10^{-5} -0.0588$&$0.0096 -179.99$ &$0.0024 -179.99$ &$40.27-135.4 \cup 222.6-319.73$ &$0.0467 -0.0763$
 &$0.0450 -0.0758$ &$-0.0403 -0.0404$  \\
 \hline
 $2\, \sig$ & $31.95 -36.09$  &$36.87 -50.76$ &$6.29 -11.68$ &$0.0462 -0.0783$ &$0.0470 -0.0787$&
 $2.49\times 10^{-6}- 0.0602$&$0.0341 -179.99$ &$0.0029 - 180$ &$38.40-138.7 \cup 220 - 317.5$ &$0.0456 -  0.0779$
 &$ 0.0436 - 0.0776$ &$-0.0456 -0.0457$  \\
 \hline
 $3\, \sig$ & $30.98 - 37.11$  &$35.67- 53.12$ &$4.06 -  12.92$ &$0.0450 - 0.0791$ &$0.0459 -0.0796$&
 $5.38\times 10^{-7} -  0.0606$&$0.0040 -180$ &$0.0490 - 179.99$ &$40.72 -140 \cup 221.9 - 320.81$ &$0.0443 -0.0790$
 &$0.0414 - 0.0781$ &$-0.0501 - 0.0505$  \\
 \hline
 %%%%%%%%%%%%%%%%%%%%%%%%%%%%%%%%%%%%%%%%%%%%%%%%%%%%%%%%%%%%%%%%%%%%%%%%%%%%%%%%%%%%%%%%%%%%%%%%%
 %%%%%%%%%%%%%%%%%%%%%%%%%%%%%%Model  M_12 =0 %%%%%%%%%%%%%%%%%%%%%%%%%%%%%%%%%%%%%%%%%%%%%%%
 %%%%%%%%%%%%%%%%%%%%%%%%%%%%%%%%%%%%%%%%%%%%%%%%%%%%%%%%%%%%%%%%%%%%%%%%%%%%%%%%%
 \hline
\multicolumn{13}{c}{\mbox{Model} $M_{\n\,12}=0$} \\
\hline
 \mbox{quantity} & $\th_x$ & $\th_y$& $\th_z$ & $m_1$ & $m_2$& $m_3$ & $\r$ & $\sig$ & $\d$ & $\me$
 & $\mee$ & $J$\\
 \hline
 \multicolumn{13}{c}{\mbox{Degenerate  Hierarchy}} \\
 \cline{1-13}
 $1\, \sig$ &$32.96  -35.00$ & $38.65 - 45.00$ & $ 7.71 - 10.30$ & $0.0470 - 0.3567$ & $0.0478 -0.3568$ &$0.0579 - 0.3601$& $0.0062 - 179.99$ & $0.0094 -180$&$0.0241 -359.67$ & $0.0479 - 0.3569$
  &$0.0416  - 0.3502$ & $-0.0395 - 0.0399$ \\
 \hline
 $2\, \sig$ & $31.95  - 36.09$ & $36.88 -50.77$& $6.29 -11.68$ & $0.0461- 0.3818$ & $ 0.0470 - 0.3819$ &
 $0.0575 - 0.3849$& $0.0044 -180$ & $ 0.0111- 179.99$ & $0.1679 - 359.56$& $0.0468 - 0.3819$ &
 $0.0389 -0.3819$ & $-0.0453- 0.0453$  \\
 \hline
 $3\, \sig$ &$30.98- 37.11$ &$35.68 - 53.13$ & $4.05 -12.92$ & $ 0.0450 -0.3442$ & $0.0457 - 0.3443$ &
 $0.0555 -0.3481$& $0.0253 -180$ & $0.0574 -179.99$ & $0.2258- 359.84$ &
 $ 0.0458 -0.3443$ & $  0.0343 -0.3331$ & $-0.0494 - 0.0503$ \\
 \hline
 %%%%%%%%%%%%%%%%%%%%%%%%%%%%%%%%%%%%%%%%%%%%%%%%%%%%%%%%%%%%%%%%%
 \multicolumn{13}{c}{\mbox{Normal  Hierarchy}} \\
 \cline{1-13}
 $1\, \sig$ &$32.96 -35.00$& $38.65 -45.00$ &$7.71 -10.30$ &$0.0025 - 0.0454$ &$0.0090 -0.0463$ &
  $0.0480 -  0.0675$& $ 0.0028 -179.98$ &$0.0886 -179.96$ & $0.0682 -359.89$ &$ 0.0085 - 0.0465$ &
  $5.82\times 10^{-5} -0.0462$ & $-0.0398 - 0.0400$  \\
 \hline
 $2\, \sig$ & $ 31.95- 36.09$ &$ 36.87 - 50.77$ & $6.29 - 11.68$ &$2.14\times 10^{-6} - 0.0477$&
 $0.00849- 0.0485$&$0.0470 - 0.0699$ & $0.0001-180$ & $ 0.0219 - 179.92$ & $0.0273 -  359.97$&
 $0.0070 - 0.0487$ &$5.78 \times 10^{-5} - 0.0484$ & $-0.0442 - 0.0455$  \\
 \hline
 $3\, \sig$ &$30.98 -37.11$ & $35.67 -53.10$ & $4.05 -12.89$& $2.14\times 10^{-7}-  0.0473$ &
 $0.0083 - 0.0481$ &$0.0458 - 0.0700$& $0.0181-180$ & $0.0267 - 179.99$ &$0.1138 -359.84$ &
 $0.0055 - 0.0481$ &$0.0001 - 0.0479$ & $-0.0492 - 0.0495$ \\
 \hline
 %%%%%%%%%%%%%%%%%%%%%%%%%%%%%%%%%%%%%%%%%%%%%%%%%%%%%%%%%%%%%%%%%%%%%%%%
 \multicolumn{13}{c}{\mbox{Inverted  Hierarchy}} \\
 \cline{1-13}
 $1\, \sig$ & $ 32.96 -35.00$  &$38.65 -45.00$ &$7.71 - 10.30$ &$0.0471 -0.0767$ &$0.0479 - 0.0772$ &
 $2.01\times 10^{-5} -0.0590$ &$0.0408 -180$ &$0.0045 - 179.99$ &$41.69 -138.90   \cup 224.1 - 316.64$ &$0.0467 - 0.0765$ &$ 0.0454 - 0.0758$ &$-0.0402 - 0.0402$  \\
 \hline
 $2\, \sig$ &$31.95 -36.09$ & $36.87- 50.77$ & $6.29 - 11.68$ & $0.0462 - 0.0773$ &
 $0.0470  - 0.0777$ & $ 6.16\times 10^{-6} - 0.0589$ & $0.0004  - 179.97$ & $0.0303 -179.98$ &
 $41.789 - 138.7 \cup  220.7 -  314.96$ &
  $0.0456 - 0.0769$& $ 0.0437 - 0.0768$ & $-0.0456 - 0.0458$  \\
 \hline
 $3\, \sig$ & $30.98- 37.11$ & $35.68- 53.13$ & $4.05 - 12.92$ & $ 0.0450 - 0.0798$ &
 $0.0458- 0.0802$ & $ 6.25 \times 10^{-6}- 0.0611$ &$0.0036 -179.99$ &
 $0.0123 - 179.98$ & $43.773 - 136.00 \cup   223.7 -316.82 $ &
  $0.0442 - 0.0794$ & $ 0.0417 - 0.0785$& $-0.0503 - 0.0506$  \\
 \hline
 %%%%%%%%%%%%%%%%%%%%%%%%%%%%%%%%%%%%%%%%%%%%%%%%%%%%%%%%%%%%%%%%%%%%%%%%%%%%%%%%%%%%%%%%%%%%%%%%%
 %%%%%%%%%%%%%%%%%%%%%%%%%%%%%%Model  M_11 %%%%%%%%%%%%%%%%%%%%%%%%%%%%%%%%%%%%%%%%%%%%%%%
 %%%%%%%%%%%%%%%%%%%%%%%%%%%%%%%%%%%%%%%%%%%%%%%%%%%%%%%%%%%%%%%%%%%%%%%%%%%%%%%%%
 \hline
\multicolumn{13}{c}{\mbox{Model} $M_{\nu 11} = 0$} \\
\hline
 \mbox{quantity} & $\th_x$ & $\th_y$& $\th_z$ & $m_1$ & $m_2$& $m_3$ & $\r$ & $\sig$ & $\d$ & $\me$
 & $\mee$ & $J$\\
 \hline
 \multicolumn{13}{c}{\mbox{Degenerate  Hierarchy}} \\
 \cline{1-13}
 $1\, \sig$ & $\times$  &$\times$ &$\times$ &$\times$ &$\times$ &$\times$
 &$\times$ &$\times$ &$\times$ &$\times$ &$\times$ &$\times$  \\
 \hline
 $2\, \sig$ & $\times$  &$\times$ &$\times$ &$\times$ &$\times$ &$\times$
 &$\times$ &$\times$ &$\times$ &$\times$ &$\times$ &$\times$  \\
 \hline
 $3\, \sig$ & $\times$  &$\times$ &$\times$ &$\times$ &$\times$ &$\times$
 &$\times$ &$\times$ &$\times$ &$\times$ &$\times$ &$\times$  \\
 \hline
 %%%%%%%%%%%%%%%%%%%%%%%%%%%%%%%%%%%%%%%%%%%%%%%%%%%%%%%%%%%%%%%%%
 \multicolumn{13}{c}{\mbox{Normal  Hierarchy}} \\
 \cline{1-13}
 $1\, \sig$ &$32.96-35.00$& $38.65 -45.00$ &$7.71 -10.30$ &$0.0015-0.0080$ &$0.0080 - 0.0118$ &
  $0.0480- 0.0506$& $1.57-88.82 \cup 92.38- 179.71$ &$ 3.06-89.83\cup 92.46178.32$ & $0.0296 -359.97$ &$0.0085 -0.0128$ &
  $0$ & $ -0.0403 -0.0404$  \\
 \hline
 $2\, \sig$ & $31.95- 36.09$ &$36.88 -50.77$ & $6.29 -11.68$ &$ 0.0005 - 0.0099$ &$0.0085 -0.0132$
 &$0.0470 -0.0517$ & $0.559-89.3 \cup 92.25 -179.35$ & $ 2.71-179.12$ & $0.3021 - 359.9$& $0.0077 - 0.0147$ &
 $0 $ & $ -0.0458 - 0.0453$  \\
 \hline
 $3\, \sig$ &$30.98 - 37.11$ & $35.68 -53.13$ & $4.06 -12.92$& $6.31\times 10^{-7} - 0.0115$ &
 $0.0084 - 0.0146$ &$0.0458 -0.0532$& $0.458- 88.52 \cup 92.16-179.46$ & $4.20-178.11$ &$0.0546- 359.99$ &
 $0.0065 - 0.0167$ &$0$ & $ -0.0498 -0.0503$ \\
 \hline
 %%%%%%%%%%%%%%%%%%%%%%%%%%%%%%%%%%%%%%%%%%%%%%%%%%%%%%%%%%%%%%%%%%%%%%%%
 \multicolumn{13}{c}{\mbox{Inverted  Hierarchy}} \\
 \cline{1-13}
 $1\, \sig$ & $\times$  &$\times$ &$\times$ &$\times$ &$\times$ &$\times$
 &$\times$ &$\times$ &$\times$ &$\times$ &$\times$ &$\times$  \\
 \hline
 $2\, \sig$ & $\times$  &$\times$ &$\times$ &$\times$ &$\times$ &$\times$
 &$\times$ &$\times$ &$\times$ &$\times$ &$\times$ &$\times$  \\
 \hline
 $3\, \sig$ & $\times$  &$\times$ &$\times$ &$\times$ &$\times$ &$\times$
 &$\times$ &$\times$ &$\times$ &$\times$ &$\times$ &$\times$  \\
\hline
 \end{tabular}
 }}
 \end{center}
  \caption{\small \label{tab3} The various prediction for the patterns  of
  one-zero texture $(M_{\n\,13}=0)$ , $(M_{\n\,12}=0)$ and $(M_{\n\,11}=0)$. All the angles (masses) are
  evaluated in degrees ($eV$).}
 \end{table}
\end{landscape}
\subsection{  Pattern of vanishing  $\mathbf { M_{\nu 21}}$}
The relevant expressions for $A_1$, $A_2$ and $A_3$ for this model
are
\bea
A_1 &=& - c_z c_x (c_x s_y s_z +  s_x c_y\, e^{-i\,\d}),\nn \\
A_2 & = & -c_z s_x (s_x s_y s_z -  c_x c_y\, e^{-i\,\d}),\nn\\
A_3 &=& s_z c_z s_y.
\label{M21}
\eea
We get
\bea
\frac{m_1}{m_3} &\approx& \frac{ s_{2\s - \d}\, s_z \,t_y}{c_x s_x s_{2\s - 2\r}} +
O \left( s_z^2 \right), \nn\\
\frac{m_2}{m_3} &\approx& \frac{s_{2\r - \d}\,s_z\,t_y}{c_x s_x s_{2\s - 2\r}}
 +O \left( s_z^2\right).
 \label{mt21}
 \eea
with
\bea \label{Rt21} R_\nu &\approx& \frac{t_y^2 \left|s_{\d-2\,\s}^2-s_{\d-2\,\r}^2\right|}
{c_x^2 s_x^2  s_{2\r-2\s}^2}\,s_z^2+O \left(s_z^3\right).
\eea

The phenomenological analysis of this pattern can be deduced from that of vanishing $M_{\n\;13}$ by applying the symmetry $T_1$.

Also, and as in the pattern of vanishing $M_{\n\;13}$, there is a non-invertible such texture which can
accommodate the current data.

%%%%%%%%%%%%%%%%%%%%%%%%%%%%%%%%%%%%%%%%%%%%%%%%%%%%%%%%%%%%%%%%%%%%%%%%%%%%%%%%%

%%%%%%%%%%%%%%%%%%%%%%%%%%%%%%%%%%%%%%%%%%%%%%%%%%%%%%%%%%%%%%%%%%%%%%%%%%%%%%%%%
%%%%%%%%%%%%%%%%%%%%%%%%%%%%  CASE M11 =0    %%%%%%%%%%%%%%%%%%%%%%%%%%%%%%%%%%%%%%

\subsection{ Pattern of vanishing  $\mathbf {M_{\nu\, 11}}$}
In this pattern, the relevant expressions for $A_1$, $A_2$ and $A_3$ are
\bea
A_1=c_z^2 c_x^2, & A_2 = c_z^2 s_x^2,
& A_3= s_z^2.
\label{M11}
\eea
The analytical expressions for all relevant computed parameters are simple and
independent of $\d$.
The mass ratios take the forms:
 \bea
\frac{m_1}{m_3} &=&  \frac{t_z^2\,s_{2\s} }{c_x^2 s_{2\r-2\s} },\nn \\
\frac{m_2}{m_3} &=& \frac{t_z^2\,s_{2\r} }{s_x^2 s_{2\s-2\r} }.
\label{mt11}
\eea
Using the randomly generated  $\d m^2$ within its acceptable range as found in Table~(\ref{tab1}),
$m_3$ can be calculated to be
\be
m_3 = \sqrt{\d m^2}\,
{\left|s_{2\s-2\r}\right| \over {t_z^2\;\sqrt{\left|\left({s_{2\r}^2 \over s_x^4}-
{s_{2\s}^2\over c_x^4}\right)\right|}}},
\label{m3t11}
\ee
Using this expression of $m_3$, one can get the corresponding expression  of
$\D m^2$ as:
\be
\D m^2 = m_3^2 \left|1-{1\over 2}\,{t_z^4\over s_{2\s-2\r}^2}\left({s_{2\s}^2\over c_x^4}
 + {s_{2\r}^2\over s_x^4} \right)\right|.
\label{dmatmt11}
\ee

The non oscillation parameters $\langle m\rangle_e $, $\langle m\rangle_{ee} $
and $\Sigma$ are given as
\bea
\langle m\rangle_e &=& m_3\;\sqrt{\left[ {c_z^2\,t_z^4\over s_{2\r-2\s}^2}\;
\left({s_{2\s}^2\over c_x^2}+ {s_{2\r}^2\over s_x^2 }\right) +
s_z^2\right]},\nonumber\\
\langle m\rangle_{ee}&=& m_3\;\left| {t_z^2\, s_{2\s}\,c_z^2\over s_{2\r-2\s}}\, e^{2\,i\r}
+ {t_z^2 s_{2\r}\,c_z^2\over s_{2\s-2\r}}\,e^{2\,i\s} + s_z^2\right|,\nonumber \\
\Sigma &=& m_3\;\left({t_z^2\over c_x^2}
{s_{2\s}\over s_{2\r-2\s}} + {t_z^2\over s_x^2}
{s_{2\r}\over s_{2\s-2\r}} + 1\right).
\label{nost11}
\eea
where $m_3$ is given in Eq.(\ref{m3t11}). Finally the parameter $R_\nu$
has the form,
\bea
R_\nu &=& {t_z^4\over s_{2\r-2\s}^2}\;{\left|{s_{2\r}^2 \over s_x^4}-
{s_{2\s}^2\over c_x^4}\right| \over \left| 1-{1\over 2}\,{t_z^4\over s_{2\s-2\r}^2}\left({s_{2\s}^2\over c_x^4}
 + {s_{2\r}^2\over s_x^4} \right)\right|}.
\label{Rt11}
\eea

\begin{figure}[hbtp]
\centering
\begin{minipage}[l]{0.5\textwidth}
\epsfxsize=8cm
\centerline{\epsfbox{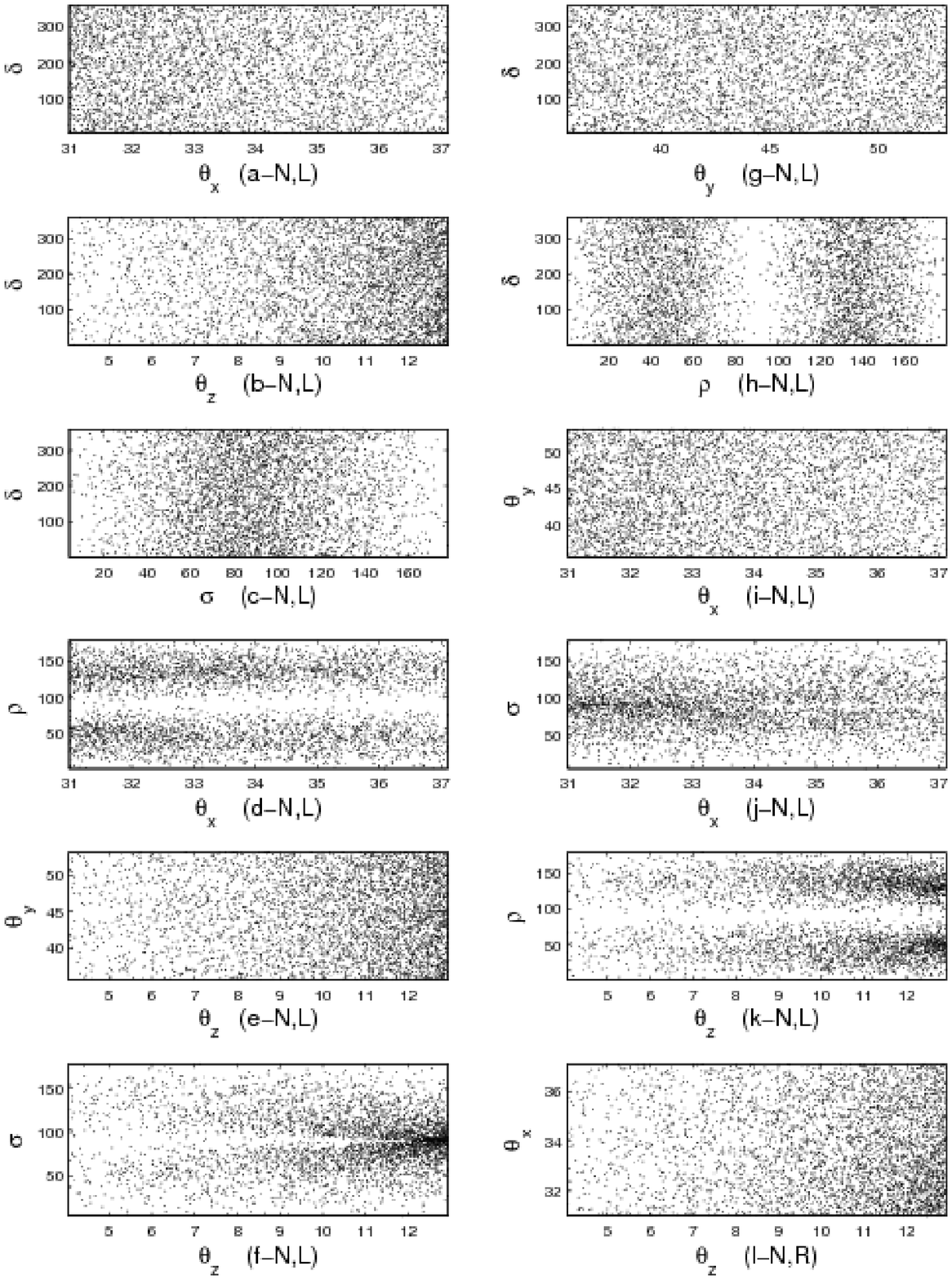}}
\end{minipage}%
\begin{minipage}[r]{0.5\textwidth}
\epsfxsize=8cm
\centerline{\epsfbox{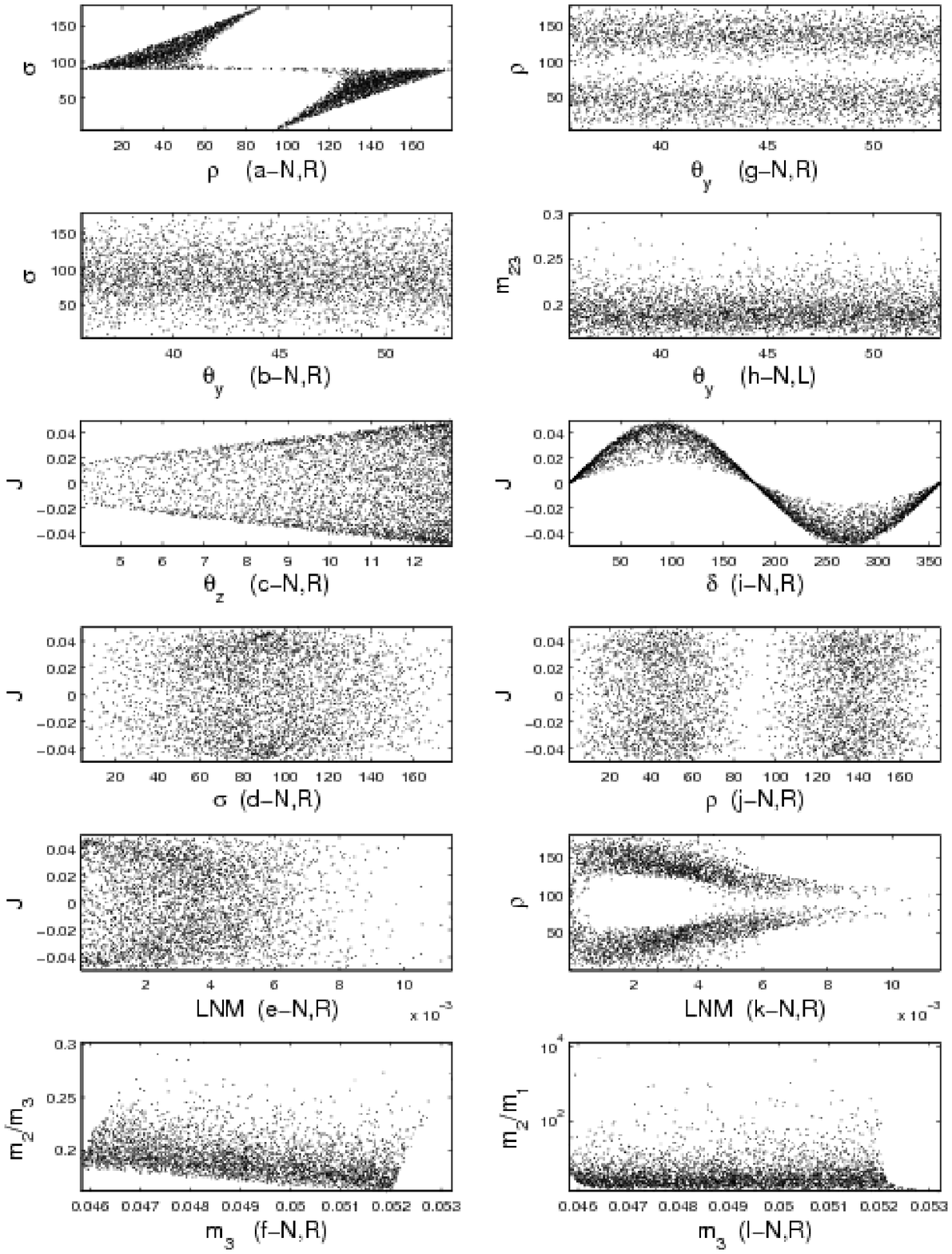}}
\end{minipage}
\vspace{0.5cm}
\caption{{\footnotesize Pattern $\mathbf M_{\n\,11}=0:$ Left panel (the left two columns) presents all pair correlations concerning
mixing angles and CP-phases except those of $(\s, \r)$, $(\s, \t_y)$and $(\r, \t_y)$. while right panel (right two columns) displays the previously excluded ones besides $(m_{23}={m_2\over m_3}, \t_y)$ correlation  and those of $J$ against $\t_z$ ,$\d$, $\s$, an $\r$ besides those of lowest neutrino mass $(\mbox{LNM})$ against $\r$ and $J$. finally those of $m_3$ against
${m_2 \over m_3}$ and ${m_2 \over m_1}$.}}
\label{t11fig1}
\end{figure}

As mentioned earlier, we have crosschecked our calculations by applying the mapping (Eq.~\ref{mapping-minor-zero}) and getting
 exactly the formulae corresponding to the pattern ($C_{11}$) in \cite{LashinChamoun2}.
 This pattern shows only normal-type hierarchy, and the corresponding plots are shown in
Figures (\ref{t11fig1}). We see in this Figure (plots: a-L,  b-L and g-L) that the mixing angles ($\t_x,\t_y,\t_z$) and the Dirac phase angle $\d$ cover all their allowable regions. Figure~(\ref{t11fig1}, l-N,R) and Table~(\ref{tab3}) show that this texture allows a vanishing value of $m_1$ only at the 3-$\s$ error level, as we shall see later when studying the singular textures.
The non vanishing of $m_1$, at the 1-$\s$ and 2-$\s$ error levels, has far reaching
consequences on Majorana phases in that it excludes  the tight region neighborhood around $\s = \frac{\pi}{2}$ which is clearly evident from the disallowed region $[89.83^0,92.46^0]$ for the phase $\s$  at the 1-$\s$ error level as deducted from Table~(\ref{tab3}).
Similarly, the non vanishing of $m_2$  implies the same consequence but for  $\r$, which can be easily
understood through mass ratios given in Eq.~\ref{mt11} (look at the two plots: d-L, h-L). In the same
way, the  region of $\r-\s$ equal to a multiple of  $\frac{\pi}{ 2}$ would be excluded due to the nonvanishing of $m_3$ (look at the two strips in the plot: a-R).

Furthermore, setting the mass ratio, $\left({m_2\over m_1} = {s_{2\r}\over s_{2\s} t_x^2}\right) $ to be
larger than $1$ and taking into account that $t_x$ is less than one, for the phenomenologically accepted $\t_x$,
would force small non-vanishing lower bounds for $\r$ as can be seen in Table~\ref{tab2}.

Plots (c-L, h-L) show no strong correlation between ($\d, \r$), nor between ($\d, \s$), whence no
clear correlation between $J$ versus $\r$, or between $J$ versus $\s$ (plots: d-R, j-R). The absence
of correlation concerning delta is expected since $\d$ drops out from all expression defining the pattern.
There is a strong `kite-shaped' correlation between $\s$ against $\r$ (plot: a-R) showing that
$\r$ being in the first quarter forces $\s$ to be in the second quarter, and vice versa.
In this pattern $\mee$, as given by Eq.~(\ref{mee}),  vanishes as a direct consequence of the relation defining the pattern, namely  $M_{\n\;11}=0$. The vanishing of $\mee$ clearly leaves its imprint on all resulting
correlation through its functional dependence on mixing and phase angles as given in Eq.~(\ref{nost11}).
The LNM correlation with $\r$ (plot k-R) again excludes the region around
$\r = \frac{\pi}{2}$.

The mass spectrum in (plot: f-R) shows a strong normal hierarchy  with
$m_2$ quite larger than  $m_1$ (plot: l-R) and we can approach the limit
$m_1 = 0$. The singular limit ($m_1=0$) is absent  at the 1-$\s$ and 2-$\s$ error levels
as evident from Table~(\ref{tab3}).

\clearpage
%%%%%%%%%%%%%%%%%%%%%%%%%%%%%%%%%%%%%%%%%%%%%%%%%%%%%%%%%%%%%%%%%%%%%%%%%%%%%%%%%%%%%%%%%
%%%%%%%%%%%%%%%%%%%%%%%%%%%%%%%%%%%%%%%%%%%%%%%%%%%%%%%%%%%%%%%%%%%%%%%%%%%%%%%%%%%%%%%%%%

\section{Singular patterns}
The viable singular patterns can be respectively divided into two classes, the first class is characterized by vanishing $m_1$ while the second class
is defined by letting $m_3$ equal to zero. Table ~\ref{tab4} is the analog of Tables (~\ref{tab2},~\ref{tab3}) for the experimentally acceptable
  singular patterns.

\subsection{Vanishing $m_1$ singular patterns}
    We have the following normal-hierarchy mass spectrum
 \bea
 m_1 =0, & m_2= \sqrt{\d m^2}, & m_3 = \sqrt{\D m^2 + {\d m^2\over 2}}
 \approx \sqrt{ \D m^2},
 \label{m1spec}
 \eea
 which constrains the mass ratio ${m_2\over m_3}$ to be
 \be
 {m_2\over m_3} = \sqrt{\frac{R_\n}{1+{R_\n\over 2}}}\approx \sqrt{R_\n},
 \label{m23con}
 \ee
and the mass ratio value can be deduced from that of $R_\n$ given in Table~(\ref{tab1}).
The vanishing of $m_1$ together with imposing one zero texture implies,
\be
A_2\, m_2 \, e^{2\,i\,\s} + A_3\, m_3 = 0,
\label{m1zero}
\ee
 leading to
 \bea
 {m_2\over m_3} = \left|{A_3\over A_2}\right|, &&\s = {1\over 2}\;\mbox{Arg}
 \left(-{A_3\, m_3\over A_2\, m_2}\right).
 \label{m1zerocon}
 \eea
 The Majorana phase $\r$ becomes unphysical in this class, since $m_1$ vanishes, and can be dropped out.

 One can use the mass constraint Eq.~\ref{m23con} to check the viability of the singular patterns. We find that we have three unviable cases with vanishing $m_1$, whose inability to accommodate data can be revealed easily from the mass ratio formula ${m_2\over m_3}$.
 In fact, this mass ratio for the textures $M_{\n\;22}=0$ ($M_{\n\;33}=0$, and $M_{\n\;23}=0$), up to leading order in $s_z$, is respectively ${t_y^2\over c_x^2}$ (${1\over t_y^2\, c_x^2}$ and ${1\over c_x^2}$), which is larger than $\sqrt{{R_\n}}$ for all allowed range of
 $\t_x$, $\t_y$ and $\t_z$. The relevant mass ratio formulae for both viable and unviable patterns are listed in Table~\ref{singmass}.

\subsubsection{Pattern of vanishing $m_1$ and $M_{\n\;11}$}
\begin{figure}[hbtp]
\epsfxsize=8cm
\centerline{\epsfbox{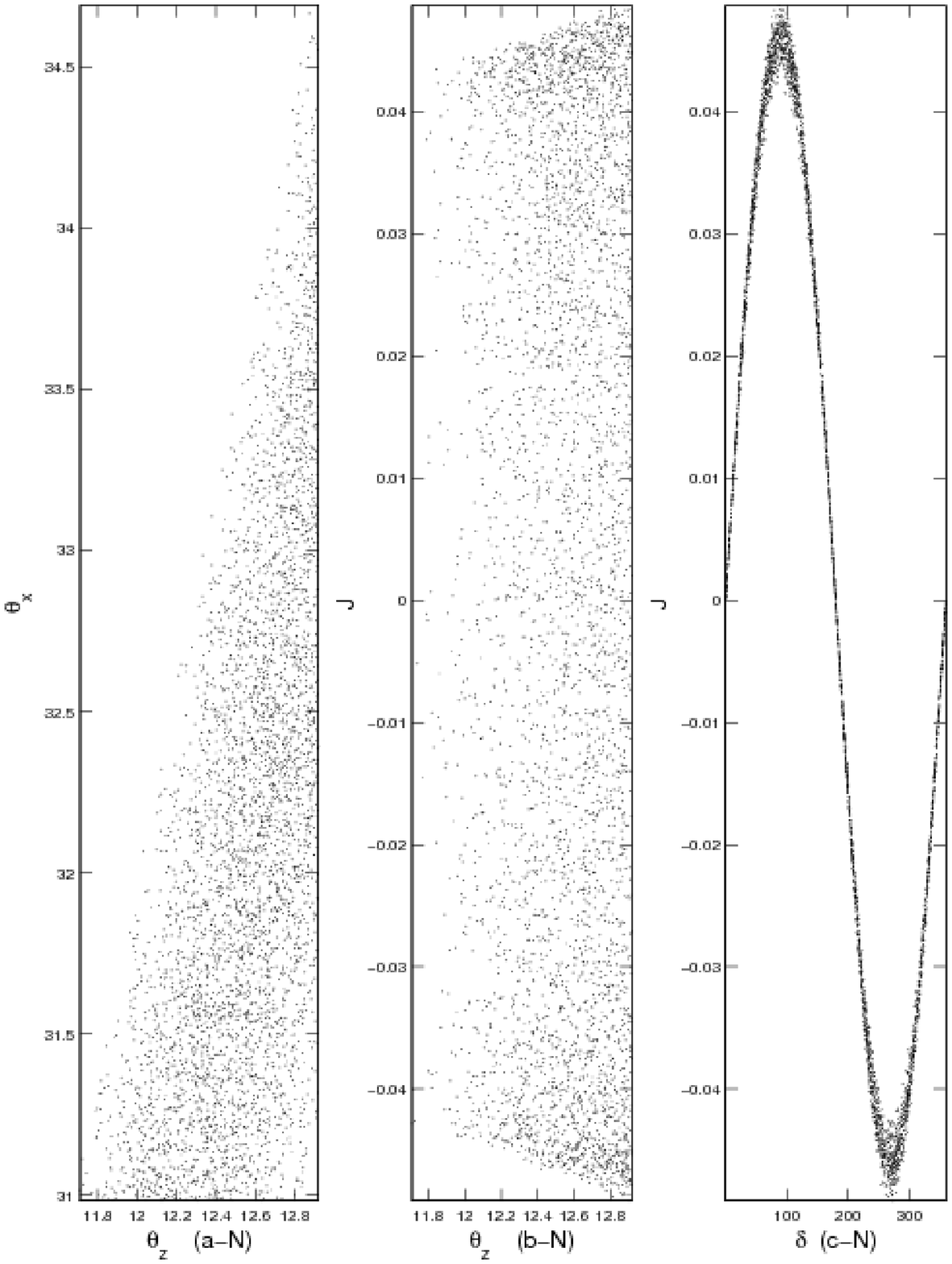}}
\caption{{\footnotesize Pattern $(\mathbf M_{\n\,11}=0, m_1=0):$ Relevant correlations concerning $(\t_z, \t_x)$, $(J, \t_z)$ and $(J, \d)$ at
the 3-$\s$ precision level.}}
\label{m1tex11}
\end{figure}

In this pattern the relevant mass ratio is
\be
{m_2 \over m_3} = {t_z^2 \over s_x^2}.
\label{m111}
\ee
In this case there is no restriction on $\t_y$ and $\d$. As to the Majorana phase
$\s$, as given by Eq.\ref{m1zerocon}, it is restricted to be ${\pi\over 2}$ since both of $A_3$ and $A_2$ are
positive real (Eq.~\ref{M11}). Eq.~\ref{m111} imposes the constraint on $\t_z$ and $\t_x$ given by
${t_z^2\over s_x^2}=\sqrt{R_\n \over 1 +{R_\n\over 2}}$. This constraint can be satisfied only at the 3-$\s$ level. The correlations in Figure~\ref{m1tex11} indicates that $\t_z$ is restricted to be in a narrow range from $11.71^0$ to $12.91^0$, while $\t_x$ is from $30.98^0$ to $34.69^0$.

\subsubsection{Pattern of vanishing $m_1$ and $M_{\n\;13}$}
\begin{figure}[hbtp]
\centering
\begin{minipage}[l]{0.5\textwidth}
\epsfxsize=8cm
\centerline{\epsfbox{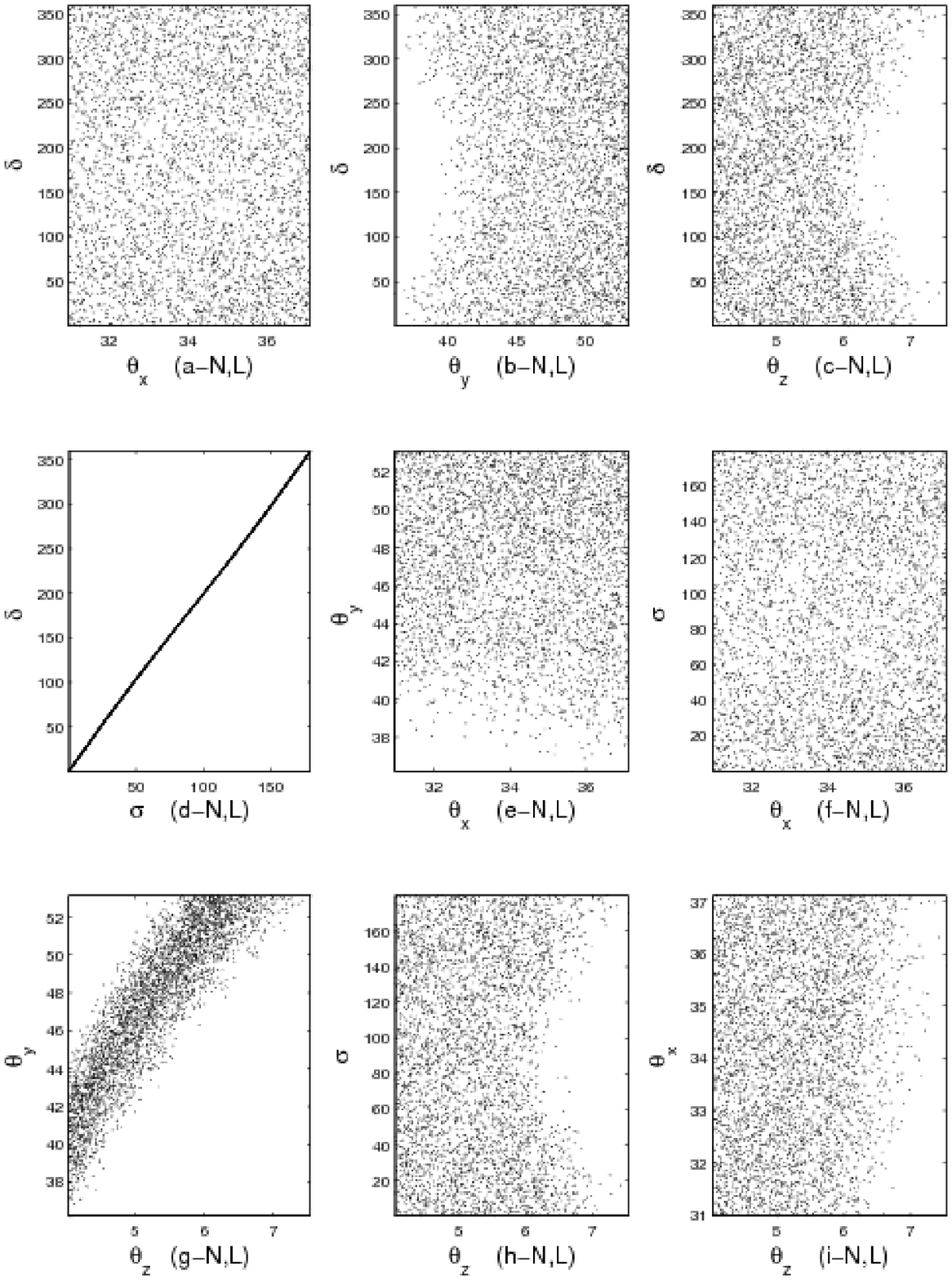}}
\end{minipage}%
\begin{minipage}[r]{0.5\textwidth}
\epsfxsize=8cm
\centerline{\epsfbox{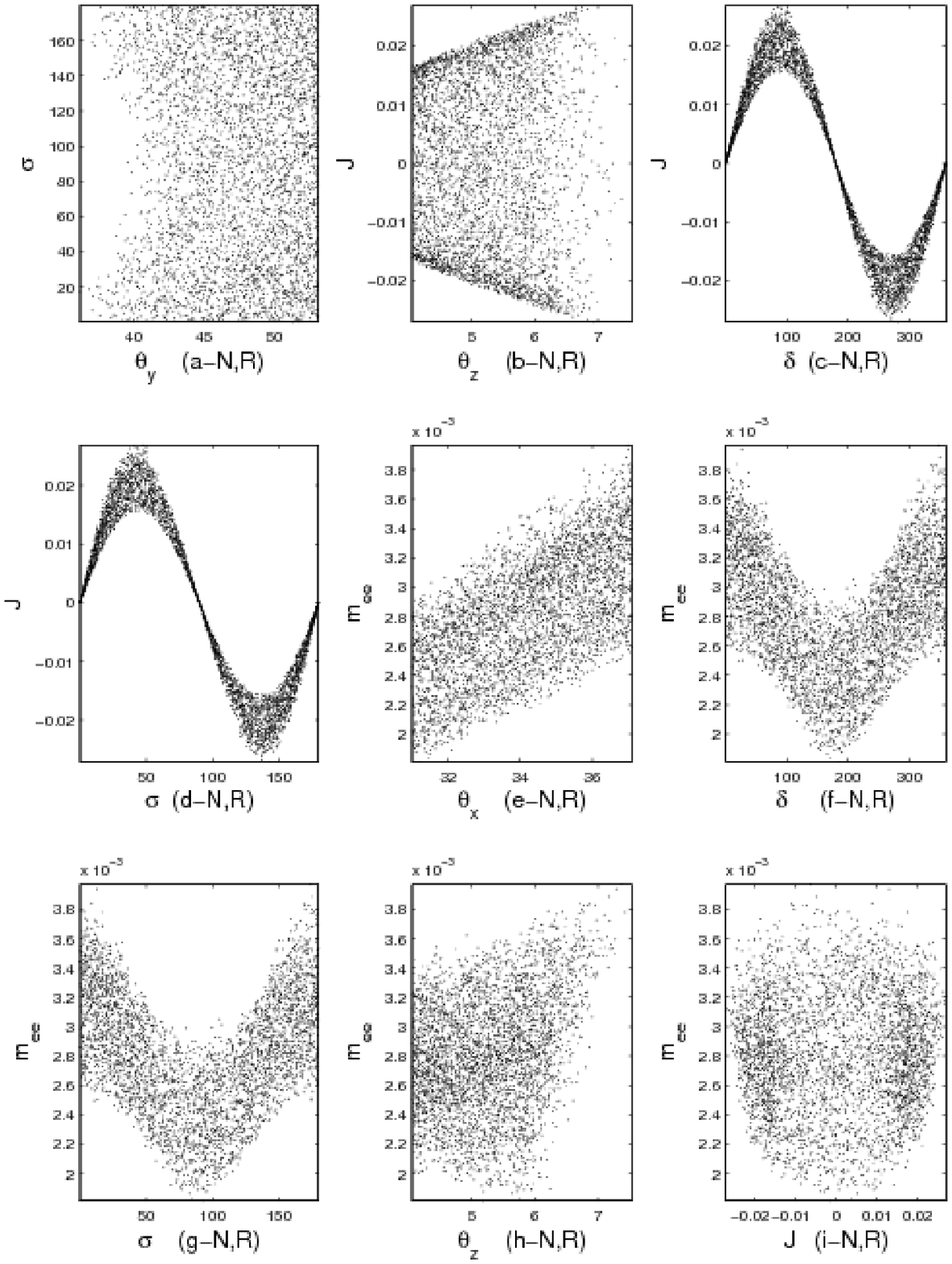}}
\end{minipage}
\vspace{0.5cm}
\caption{{\footnotesize Pattern $(\mathbf M_{\n\,13}=0, m_1=0):$ The left panel plots and the top  plot
of the right panel present all pair correlations concerning mixing angles and relevant CP-phases.
The remaining right panel plots display the correlations of $J$ against $\t_z$ ,$\d$, and $\s$, besides
those of $\mee$  against $\t_x$, $\d$, $\s$, $\t_z$ and $J$. All data correspond to 3-$\s$ precision level.}}
\label{m1t13fig1}
\end{figure}

In this pattern the relevant mass ratio  and the Majorana phase $\s$ are given, expanded in terms of $s_z$, as
\bea
{m_2 \over m_3} &\approx& {s_z \over t_y s_x\,c_x}\,\left(1- {c_\d\,t_x\,s_z\over t_y}\right)+ O(s_z^3).\nn\\
\s &\approx& {\d\over 2}  + O(s_z).
\label{m113}
\eea
In this case, the mass ratio in Eq.~\ref{m113} imposes a constraint on the mixing angles and phases that can not be satisfied at the 1-$\s$ error level, whereas it can be met at the other levels, and this manifests itself in the clear pair correlations of Figure~\ref{m1t13fig1} (plots: (b-N,L),(c-N,L),(e,N-L),(g,N-L), (h,N-L) and (i,N-L)). Table~(\ref{tab4}) reveals the restrictions on the mixing angles $\t_z, \t_y$ and the phase $\d$ at the various $\s$ levels. The notable constraints, at the 2-$\s$ level, are the restriction on
$\t_z$ and $\t_y$  to be respectively around $\t_z \approx 6.5^0$ and $\t_y \approx 49.5^0$. The valid range for $\d$, at the 2-$\s$ error level, is  $[0.015^0,86.24^0]\cup [267.2^0,360^0]$ in order to keep ${m_2\over m_3}$ consistent with Eq.~(\ref{m23con}). At the 3-$\s$ error level, we find that only $\t_z$ has the  restricted range $[4.05,7.54^0]$, while $\t_x$, $\t_y$ and $\d$ almost cover their acceptable range. In Figure~\ref{m1t13fig1}(c-N,L), it is clear that as $\t_z$ increases, the phase angle $\d$ tends to be  around $0^0$ or $360^0$ and this helps in  understanding the resulting correlation between $J$ and $\t_z$ concentrating around a vanishing $J$ for relatively large $\t_z$. The corresponding range for $\s$ is affected  and can be determined from the $\d$-range through the linear correlation in Eq.~(\ref{m113}).
The correlations $(J,\d)$ and $(J,\t_z)$ can be explained as before (see, for example, subsection~4.1, Figure~\ref{m33fig2}, plots a-L and b-L, and the interpretation therein), albeit with care regarding the mentioned restricted ranges for $\t_z$ and $\d$ and their mutual correlation.

The linear correlation between $\d$ and $\s$ (Eq.~\ref{m113}) is clearly manifested in Figure~\ref{m1t13fig1}(d-N,L). The correlation $(J,\s)$, in
Figure~\ref{m1t13fig1} (plot d-N,R), can be understood to be induced by the $(\d,\s)$ linear correlation and by the $(\d,J)$ `sinusoidal' correlation.
The correlation between $\mee$ and $\s$ can be understood through the relation defining $\mee$ in
Eq.~\ref{mee}, which under approximation to leading order in $s_z$ reads as
\be
\mee \approx m_3\,\sqrt{R_\n}\,s_x^2\left(1+ { s_z^2 \over s_x^2\,\sqrt{R_\n}} \cos{2\,\s}\right)
+ O(s_z^4).
\label{mee12}
\ee
Thus the resulting curve can be viewed as a superposition of many sinusoidal graphs in 3-$\s$ precision level, whose `positive' amplitudes are determined by the acceptable mixing angles, and shifted by constant amount. Again the correlation $(\mee, \d)$ is implied through $(\d, \s)$ correlation.

\subsubsection{Pattern of vanishing $m_1$ and $M_{\n\;12}$}
The phenomenology of the
pattern $(m_1=0, M_{\n\;12}=0)$ can be deduced from that of the pattern ($m_1=M_{\n13}=0$) by applying $T_1$ symmetry,
but after appropriate change of $\t_y$ and $\d$.

\subsection{Vanishing $m_3$ singular patterns}

The inverted-hierarchy mass spectrum here is
 \bea
 m_1 =\sqrt{\D m^2 - {\d m^2\over 2}}, & m_2= \sqrt{\D m^2 + {\d m^2\over 2}}, & m_3 = 0,
 \label{m3spec}
 \eea
 This will constrain the mass ratio ${m_2\over m_1}$ to be
 \be
 {m_2\over m_1} = \sqrt{{1+{R_\n\over 2}}\over{1-{R_\n\over 2}}} \approx 1 + {R_\n\over 2} \gtrsim 1.
 \label{m21con}
 \ee
The vanishing of $m_3$ together with imposing one zero texture implies,
\be
A_1\, m_1 \, e^{2\,i\,\r} + A_2\, m_2 \, e^{2\,i\,\s} = 0,
\label{m3zero}
\ee
 leading to
 \bea
 {m_2\over m_1} = \left|{A_1\over A_2}\right|, &&\r-\s = {1\over 2}\;\mbox{Arg}
 \left(-{A_2\, m_2\over A_1\, m_1}\right).
 \label{m3zerocon}
 \eea
 The only relevant physical combination of Majorana phases in this class of singular patterns is the difference  $\r-\s$.

 Again, the constraints in Eq.~\ref{m21con} helps to test the validity of the
 singular pattern. We have here two unviable cases $M_{\n\;11}=0$, and  $M_{\n\;23}=0$. The corresponding  mass ratios ${m_2 \over m_1}$ are  respectively  ${1\over t_x^2} $ and $ t_x^2 + O(s_z)$. Both ratio are ruled out because in the first case it is larger than $1 + \frac{R_\n}{2}$ while for the second is smaller than $1$ for the allowable data. We list
also the relevant mass ratio formulae for both viable and unviable patterns in the right part of Table~\ref{singmass}.

\subsubsection{Pattern of vanishing $m_3$ and $M_{\n\;13}$}
\begin{figure}[hbtp]
\centering
\begin{minipage}[l]{0.5\textwidth}
\epsfxsize=8cm
\centerline{\epsfbox{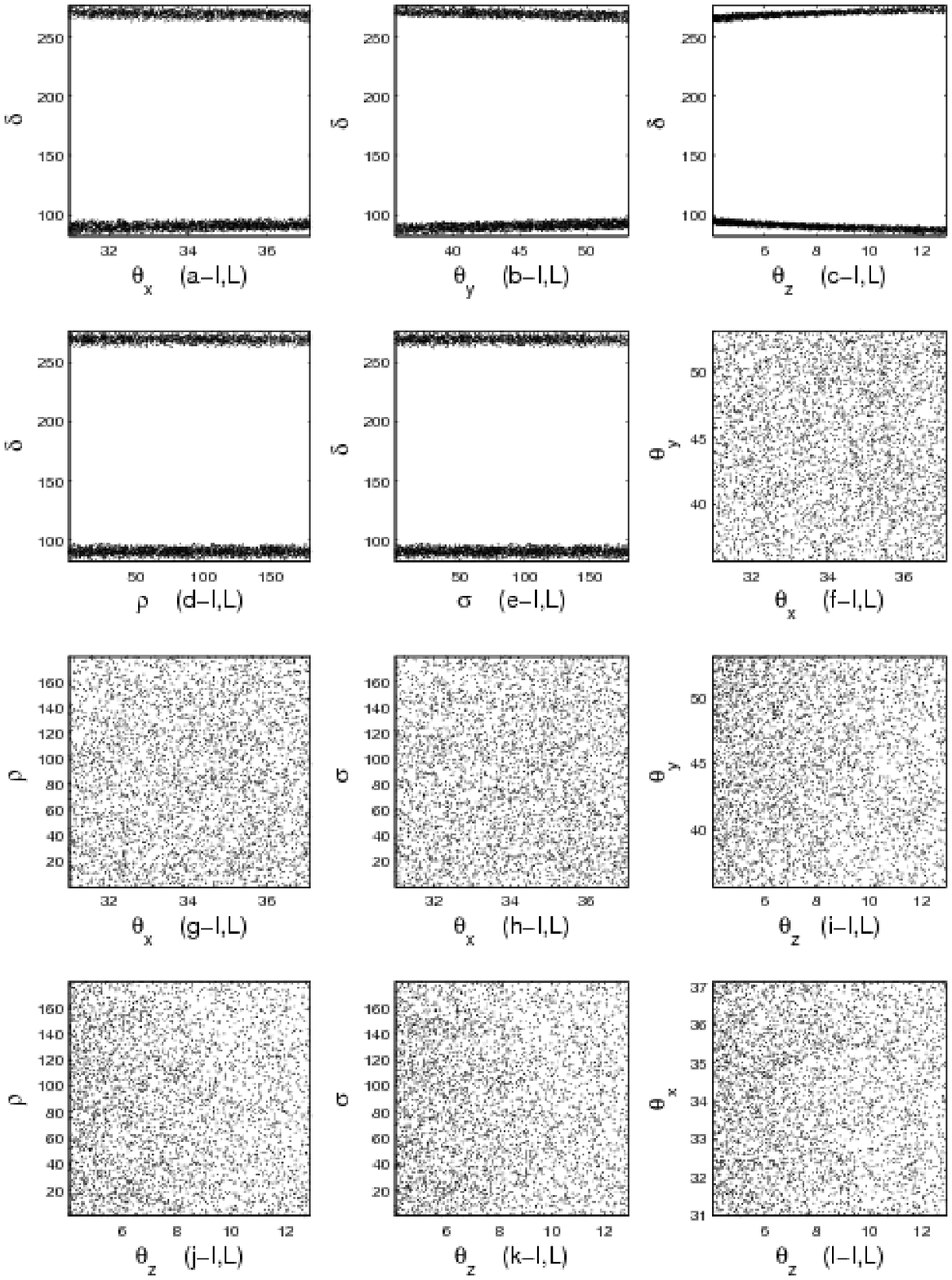}}
\end{minipage}%
\begin{minipage}[r]{0.5\textwidth}
\epsfxsize=8cm
\centerline{\epsfbox{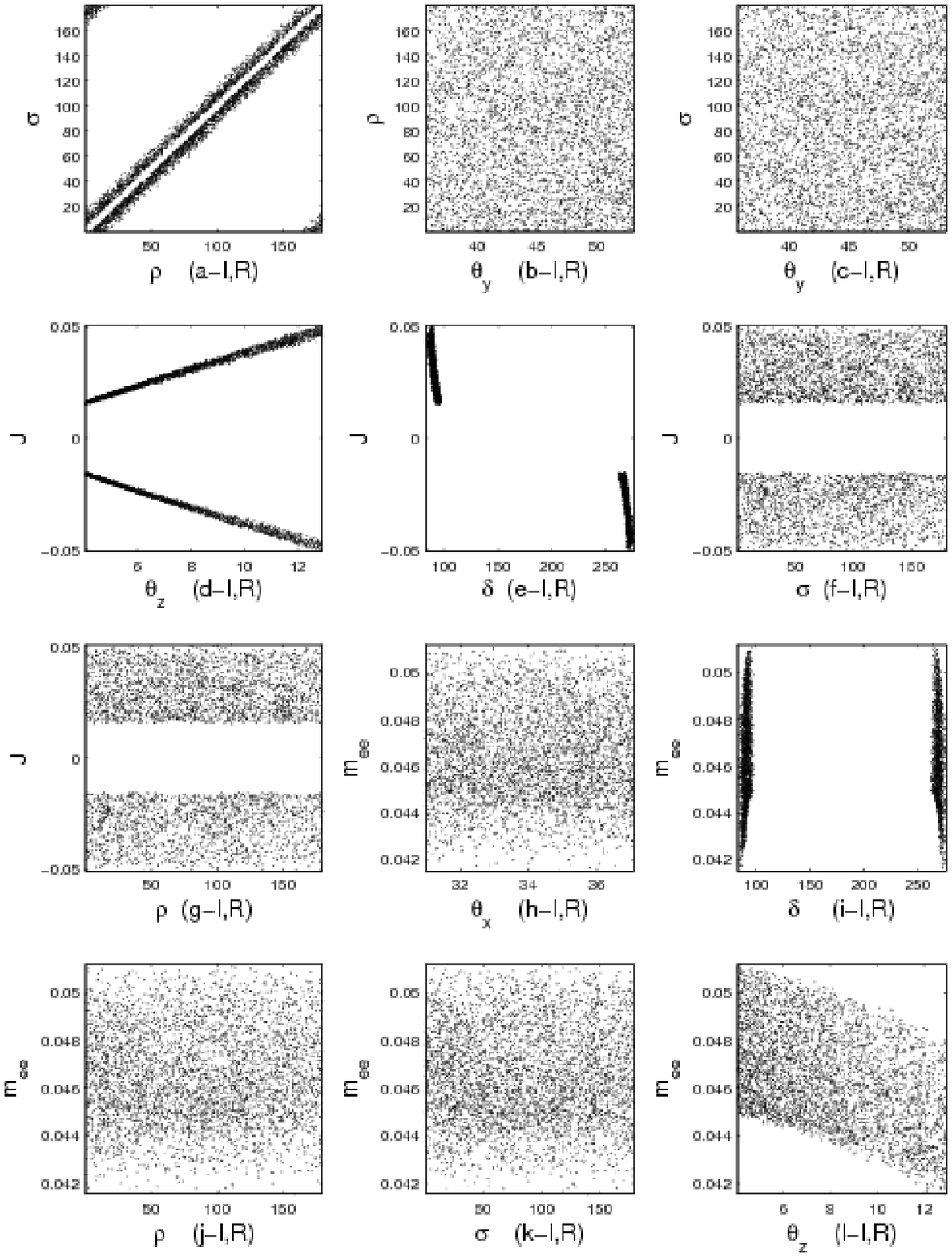}}
\end{minipage}
\vspace{0.5cm}
\caption{{\footnotesize Pattern $(\mathbf M_{\n\,13}=0, m_3=0):$ The left panel plots and the top row  plot of the right panel  present all pair correlations concerning mixing angles and relevant CP-phases. The rest of the right panel displays the correlations of $J$ against $\t_z$ ,$\d$, $\s$ and $\r$, besides those of $\mee$  against $\t_x$, $\d$, $\r$, $\s$ and  $\t_z$.}}
\label{m3t13fig1}
\end{figure}

In this pattern, we get
\bea
{m_2 \over m_1} &\approx& 1- {c_\d\,s_z\over t_y\,s_x\,c_x}  -{s_z^2\,\left(c_\d^2\,\left(c_x^2 - 3 \,s_x^2\right) - c_{2x}\right) \over 2 c_x^2\,s_x^2\,t_y^2}+ O(s_z^3).\nn\\
\r &\approx& \s + {s_\d\,s_z\over 2\,c_x\,s_x\,t_y} + O(s_z^2).
\label{m313}
\eea
In this case, the compatibility of the mass ratio in Eq.~\ref{m313} with the constraint given by Eq.~(\ref{m21con}) imposes restrictions on the
mixing angles and phases, which is made evident in the clear pair correlations in Figure~\ref{m3t13fig1} [plots:
(a-I,L) $\rightarrow$ (e,I-L)].
The mixing angles cover all their allowed range at the 3-$\s$ level.
The Dirac phase $\d$ is restricted to be  around  ${\pi\over 2}$ or ${3\,\pi\over 2}$ in order to keep ${m_2\over m_1}$ greater than one as evident from Figure~\ref{m3t13fig1} (plot: (c-I,L)). The restriction on $\d$ results from an interplay between terms of various orders namely $O(s_z)$ and $O(s_z^2)$ which leads to the specified range
$[82.22^0, 97.39^0] \cup [262.2^0, 276.46^0]$. The correlations $(J,\d)$ and $(J,\t_z)$ appearing in Figure~\ref{m3t13fig1} (plots:(d-I,R) and (e-I,R) respectively) can again be explained as before (see subsubsection 5.1.2), but with care about the mentioned restricted range for $\d$.

The linear correlation between $\r$ and $\s$ is clearly manifested in Figure~\ref{m3t13fig1}(plot: (a-I,R))
and  is in agreement with the corresponding relation in Eq.~\ref{m313}. The relation in Eq.~\ref{m313}
implies the existence of two lines whose intercepts with the $\r$-axis are determined by the term of order $s_z$ with sign controlled by $s_\d$.
The restricted range of $\d$ leads to the blank strip between the two lines.
The correlation between $\mee$ and $\t_z$, as depicted in Figure~\ref{m3t13fig1}(plot: (l-I,R)), can be
understood through the relation defining $\mee$ in Eq.~\ref{mee}, which under approximation to leading
order in $s_z$ reads as
\be
\mee \approx m_1\left(1 - s_z^2 \right) + O(s_z^4).
\label{mee313}
\ee
Thus the resulting curve can be viewed as a superposition of many parabolae graphs in the $(\mee, \t_z)$ plane.
No correlation for $\mee$  versus $\t_x$, $\d$, $\r$ and $\s$.

\clearpage
%%%%%%%%%%%%%%%%%%%%%%%%%%%%%%%%%%%%%%%%%%%%%%%%%%%%%%%%%%%%%%%%%%%%%%%%%%%%%%%%%%%%%%%%%%%%%

\subsubsection{Pattern of vanishing $m_3$ and $M_{\n\;12}$}
The phenomenology of the pattern
$(m_3=0, M_{\n\;12}=0)$ can be inferred from the corresponding one in the case $(m_3=0,M_{\n\;13}=0)$ since they are related
by `$T_1$' symmetries, but after appropriate changes of $\t_y$ and $\d$.

%%%%%%%%%%%%%%%%%%%%%%%%%%%%%%%%%%%%%%%%%%%%%%%%%%%%%%%%%%%%%%%%%%%%%%%%%%%%%%%%%%%%%%%%%%%%%%%
\subsubsection{Pattern of vanishing $m_3$ and $M_{\n\;33}$}
\begin{figure}[hbtp]
\centering
\begin{minipage}[l]{0.5\textwidth}
\epsfxsize=8cm
\centerline{\epsfbox{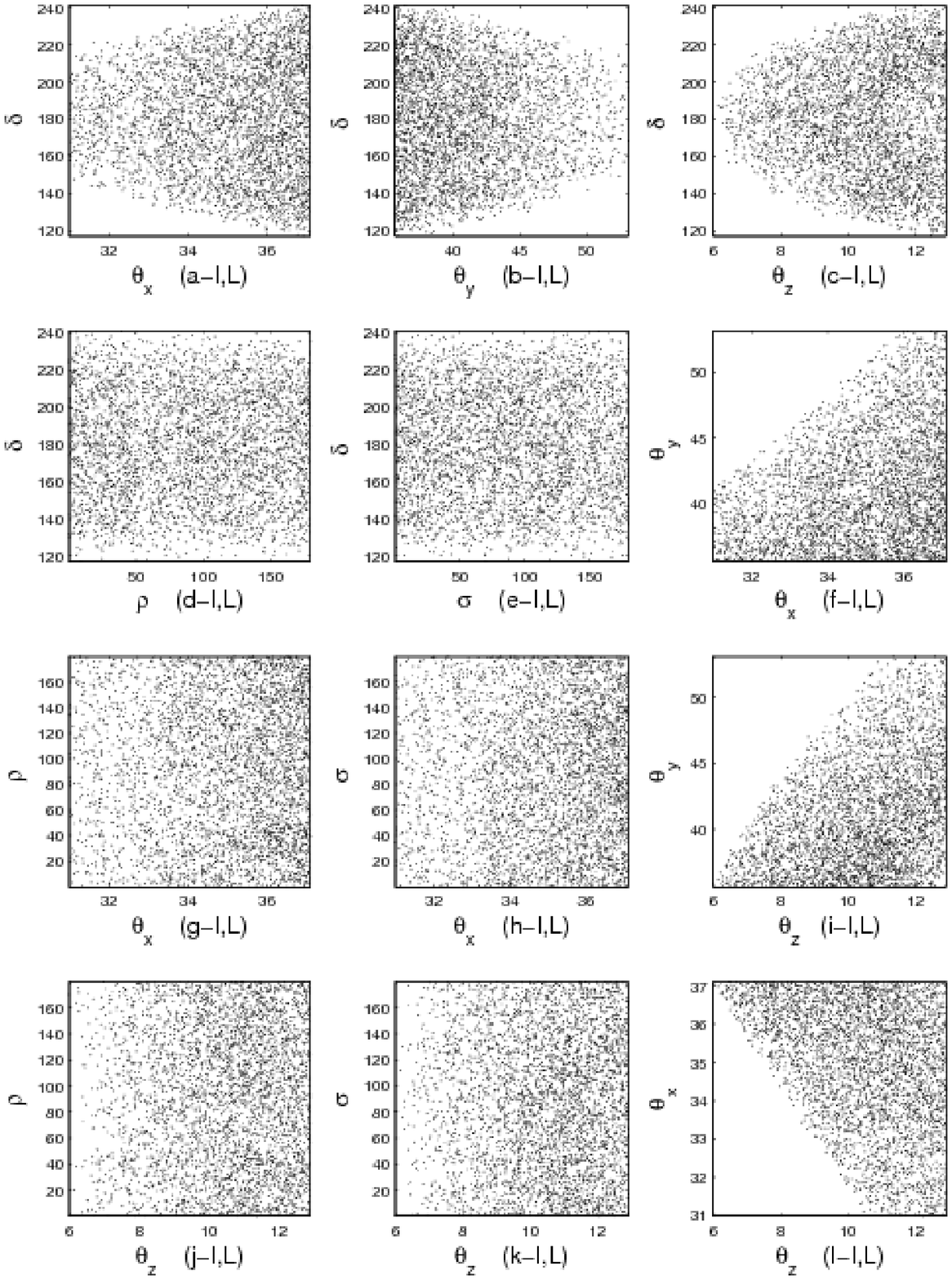}}
\end{minipage}%
\begin{minipage}[r]{0.5\textwidth}
\epsfxsize=8cm
\centerline{\epsfbox{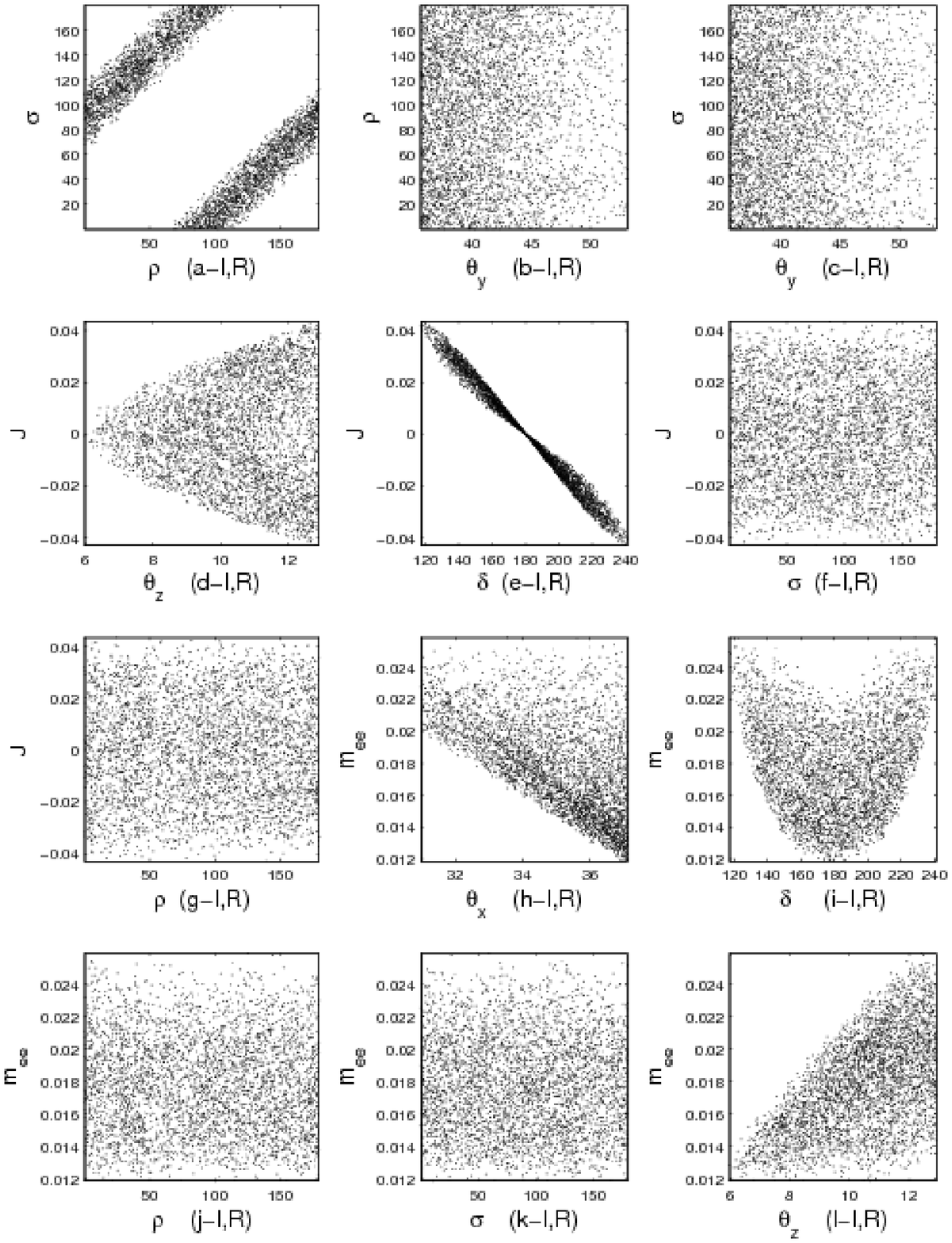}}
\end{minipage}
\vspace{0.5cm}
\caption{{\footnotesize Pattern $(\mathbf M_{\n\,33}=0, m_3=0):$ The left panel and the top row plot of the right panel
present all pair correlations concerning mixing angles and relevant CP-phases.
The other plots of the right panel display the correlations of $J$ against $\t_z$ ,$\d$, $\s$ and
$\r$, besides those of $\mee$  against $\t_x$, $\d$, $\r$, $\s$ and $\t_z$.}}
\label{m3t33fig1}
\end{figure}

In this pattern the relevant mass ratio  and the relation among  Majorana phases are given, expanded in terms of
$s_z$, as
\bea
{m_2 \over m_1} &\approx& t_x^2\,\left(1- {2\,c_\d\,s_z\over t_y\,s_x\,c_x}\right) + O(s_z^2).\nn\\
\r &\approx& \s + {\pi\over 2} + {s_\d\,s_z\over t_y\, s_x\, c_x} + O(s_z^2).
\label{m333}
\eea
Here, the equality of the mass ratio in Eq.~(\ref{m333}) with that given in Eq.~(\ref{m21con}) imposes a constraint on the mixing angles and phases, which is displayed in the clear pair correlations in Figure~\ref{m3t33fig1} (plots:(a-I,L) $\rightarrow$ (e,I-L). The Dirac phase $\d$ is restricted to be in the second and third quadrant ($\d \in [117.13^0 , 241.19^0]$) to keep$ {m_2\over m_1}$ greater than one. This behavior can be easily demonstrated from the mass ratio formula in Eq.~(\ref{m333}) which indicates that $c_\d$ should be negative in order to enhance the leading order result, ($t_x^2$), to be greater than one.   As $\t_z$ decreases, $\d$ tends to be restricted around $180^0$  as evident from Figure~\ref{m3t33fig1}(plot: (c-I,L)).  The remaining mixing angles almost cover all their
allowed range at the 3-$\s$ level, but as $\t_z$ increases $\t_x$ and $\t_y$ tends to be unrestricted as
evident from Figure~\ref{m3t33fig1} (plots: (i-I,L), (l-I,L)). The correlations $(J,\d)$ and $(J,\t_z)$
appearing in Figure~\ref{m3t33fig1} (plots:(d-I,R) and (e-I,R) respectively) can be explained as before (see subsubsection 5.1.2), paying attention to the mentioned restricted range for both $\d$ and $\t_z$ and their mutual correlation.

The linear correlation between $\r$ and $\s$ is clearly manifested in Figure~\ref{m3t33fig1} (plot: (a-I,R))
and agrees with the corresponding relation in Eq.~\ref{m333} and the higher order terms $O(s_z)$ can account for the strip appearance instead of a sharp line.
The correlation between $\mee$ and $\t_x$, as depicted in Figure~\ref{m3t33fig1} (plot: (h-I,R)), can be
understood through the relation defining $\mee$ in Eq.~\ref{mee}, which leads to
\be
\mee \approx m_1\,\cos{(2\,\t_x)}\,c_z^2,
\label{mee333}
\ee
In deriving the above equation, we took into account the relation between $\r$  and $\s$ as presented in
Eq.~\ref{m333}. In the allowed range for $\t_z$, the factor $c_z^2$ does not vary much and thus $m_{ee}$ can be seen depending mainly on the variable $\t_x$.
The resulting figure, Figure~\ref{m3t33fig1} (plot:(h-I,R), can be viewed then as a cosine curve, for the allowed range of $(\t_x, \t_z)$, whose amplitude is modulated by the allowed value of $m_1$. The other correlations involving $\mee$ presented in
Figure~\ref{m3t33fig1} (right panel) can be explained easily  through the formula in Eq.~(\ref{mee333}) together with the correlations involving $\t_x$ and $\t_z$ with other mixing angles and phases.

\subsubsection{Pattern of vanishing $m_3$ and $M_{\n\;22}$}
The phenomenology of the pattern $(m_3=0, M_{\n\;22}=0)$ can be deduced from those of the pattern
$(m_3=0,M_{\n\;33}=0)$  after appropriate changes of $\t_y$ and $\d$ that endowed by `$T_1$' symmetries.

%%%%%%%%%%%%%%%%%%Foruth TAble%%%%%%%%%%%%%%%%%%%%%%%%%%%%%%%%%%%%%%%%%%%
%%%%%%%%%%%%%%%%%%Fourth Table%%%%%%%%%%%%%%%%%%%%%%%%%%%%%%%%%%%%%%%%%%%
\begin{landscape}
\begin{table}[h]
 \begin{center}
{\tiny
 \begin{tabular}{c|c|c|c|c|c|c|c|c|c}
 \hline
\multicolumn{10}{c}{\mbox{Model} $(m_1=0, M_{\n\,11}=0)$} \\
\hline
 \mbox{quantity} & $\th_x$ & $\th_y$& $\th_z$ & $\r$ & $\sig$ & $\d$ & $\me$
 & $\mee$ & $J$\\
 \hline
 \multicolumn{10}{c}{\mbox{Normal  Hierarchy}} \\
 \cline{1-10}
 $1\, \sig$ & $\times$  &$\times$ &$\times$ &$\times$ &$\times$ &$\times$
 &$\times$ &$\times$ & $\times$  \\
 \hline
 $2\, \sig$ & $\times$  &$\times$ &$\times$ &$\times$ &$\times$ &$\times$
 &$\times$ &$\times$ & $\times$  \\
 \hline
 $3\, \sig$ & $30.98 -34.69$  &$33.56 - 53.13$ &$11.71 - 12.92$ &$ 0 - 180$ &$90$ &$0.0548 -359.94$
 &$0.0108 - 0.0126$ &$0$ &$-0.0492 - 0.0487 $ \\
 %%%%%%%%%%%%%%%%%%%%%%%%%%%%%%%%%%%%%%%%%%%%%%%%%%%%%%%%%%%%%%%%%%%%%%%%%%%%%%%%%%%%%%%%%%%%%%%%%
 %%%%%%%%%%%%%%%%%%%%%%%%%%%%%%Model  M_12 =0 %%%%%%%%%%%%%%%%%%%%%%%%%%%%%%%%%%%%%%%%%%%%%%%
 %%%%%%%%%%%%%%%%%%%%%%%%%%%%%%%%%%%%%%%%%%%%%%%%%%%%%%%%%%%%%%%%%%%%%%%%%%%%%%%%%
 \hline
\multicolumn{10}{c}{\mbox{Model} $(m_1=0, M_{\n\,12}=0)$} \\
\hline
 \mbox{quantity} & $\th_x$ & $\th_y$& $\th_z$ & $\r$ & $\sig$ & $\d$ & $\me$
 & $\mee$ & $J$\\
 \hline
 \multicolumn{10}{c}{\mbox{Normal  Hierarchy}} \\
 \cline{1-10}
 $1\, \sig$ &$\times$ & $\times$ & $\times$ &$\times$ & $\times$&$\times$ & $\times$ &$\times$ &
 $\times$\\
 \hline
 $2\, \sig$ & $31.95 - 36.09$ & $36.87 - 40.81$& $6.29- 7.32$ & $0 - 180$ & $ [0.0034,86] \cup [96, 179.99]$ &
 $12.29 - 353.17$& $0.0068 - 0.0079$ & $0.0023 - 0.0038$ & $-0.0263 - 0.0264$  \\
 \hline
 $3\, \sig$ &$30.98 - 37.11$ &$35.67 - 53.02$ & $4.05 -7.78$ & $0 - 180$ & $0.0174 - 179.97$ &
 $0.1588 - 359.95$ & $0.0055- 0.0082$ & $0.0018 - 0.0040$ & $-0.0279 - 0.0279$ \\
 %%%%%%%%%%%%%%%%%%%%%%%%%%%%%%%%%%%%%%%%%%%%%%%%%%%%%%%%%%%%%%%%%%%%%%%%%%%%%%%%%%%%%%%%%%%%%%%%%
 %%%%%%%%%%%%%%%%%%%%%%%%%%%%%%Model  M_13\bigcap %%%%%%%%%%%%%%%%%%%%%%%%%%%%%%%%%%%%%%%%%%%%%%%
 %%%%%%%%%%%%%%%%%%%%%%%%%%%%%%%%%%%%%%%%%%%%%%%%%%%%%%%%%%%%%%%%%%%%%%%%%%%%%%%%%
 \hline
\multicolumn{10}{c}{\mbox{Pattern} $(m_1=0, M_{\nu 13} = 0)$} \\
\hline
 \mbox{quantity} & $\th_x$ & $\th_y$& $\th_z$ & $\r$ & $\sig$ & $\d$ & $\me$
 & $\mee$ & $J$\\
 \hline
 \multicolumn{10}{c}{\mbox{Normal  Hierarchy}} \\
 \cline{1-10}
 $1\, \sig$ & $\times$  &$\times$ &$\times$ &$\times$ &$\times$ &
 $\times$&$\times$ &$\times$ &$\times$  \\
 \hline
 $2\, \sig$ & $32.16 - 36.09$  &$48.85 - 50.77$ &$6.29  - 6.74$ &$0 -180$ &$[0.0071,41.3]\cup[135.5, 180]$ &$[0.0151, 86.24] \cup [267.2 , 360]$
 &$0.007 - 0.0076$ &$0.0030 -0.0037$ &$ -0.0252 - 0.0251$ \\
 \hline
 $3\, \sig$ & $30.98 - 37.11$  &$36.20 - 53.13$ &$4.05 -7.54$ &$0 - 180$ &$ 0.0098 - 179.97$&
 $0.0206 - 359.94$&$0.0055 - 0.0081$ &$0.0018 - 0.0040$ &$-0.0271 - 0.0269$  \\
 \hline
%%%%%%%%%%%%%%%%%%%%%%%%%%%%%%%%%%%%%%%%%%%%%%%%%%%%%%%%%%%%%%%%%%%%%%%%%%%%%%%%%
 \hline
\multicolumn{10}{c}{\mbox{Pattern} $(m_3=0, M_{\nu 22} = 0)$} \\
\hline
 \mbox{quantity} & $\th_x$ & $\th_y$& $\th_z$ & $\r$ & $\sig$ & $\d$ & $\me$
 & $\mee$ & $J$\\
 \hline
 \multicolumn{10}{c}{\mbox{Inverted Hierarchy}} \\
 \cline{1-10}
 $1\, \sig$ & $\times$  &$\times$ &$\times$ &$\times$ &$\times$ &$\times$
 &$\times$ &$\times$ & $\times$  \\
 \hline
 $2\, \sig$ & $31.95 - 36.09$  &$38.76 - 50.77$ &$ 7.65 - 11.68$ &$0.0290 -180$ &$ 0.0096 -179.99$ &
 $ [0.0051,48.73]\cup [310.2,360]$&$ 0.0455 - 0.0501$ &$0.0135 - 0.0219$ &$-0.0347 - 0.0334$ \\
 \hline
 $3\, \sig$ & $30.99 - 37.11$  &$35.77 -53.13$ &$6.30 -12.92$ &$0.0213 -180$ &$0.0299 - 179.96$&
 $[0.0253,61.85 ]\cup [ 300, 360]$&$0.0441 - 0.0511$ &$ 0.0118 - 0.0255$ &$ -0.0432 - 0.0431$  \\
 \hline
 %%%%%%%%%%%%%%%%%%%%%%%%%%%%%%%%%%%%%%%%%%%%%%%%%%%%%%%%%%%%%%%%%
 %%%%%%%%%%%%%%%%%%%%%%%%%%%%%%%%%%%%%%%%%%%%%%%%%%%%%%%%%%%%%%%%%%%%%%%%%%%%%%%%%
 \multicolumn{10}{c}{\mbox{Pattern} $(m_3=0, M_{\nu 33} = 0)$} \\
\hline
 \mbox{quantity} & $\th_x$ & $\th_y$& $\th_z$ & $\r$ & $\sig$ & $\d$ & $\me$
 & $\mee$ & $J$\\
 \hline
 \multicolumn{10}{c}{\mbox{Inverted  Hierarchy}} \\
 \cline{1-10}
 $1\, \sig$ & $32.96 -35$  &$38.65 - 44.64$ &$8.35 -10.3$ &$0.0176 -179.96$ &$0.1035 -179.97$ &
 $144.15 -215.86$&$0.0466 -0.0490$ &$0.0154 -0.0196$ & $-0.0232 - 0.0230$ \\
 \hline
 $2\, \sig$ & $ 31.96 - 36.09$&$36.87 -50.68$ &$7.04 -11.68$ &$0.0829 - 179.97$ &$ 0.0185 -179.99$ &
 $126.46 -233.22$&$0.0455 - 0.0502$&$0.0136 - 0.0233$ &$-0.0356 -0.0356$ \\
 \hline
 $3\, \sig$ & $30.98 -  37.11$ &$35.67 - 53.12$ &$5.98 - 12.92$ &$ 0.0870 -180$ &
 $0.0263 -179.93$&$117.13 - 241.19$&$0.0441 - 0.0511$ &$0.0119 - 0.0259$ &$ -0.0429 - 0.0436$  \\
 \hline
 %%%%%%%%%%%%%%%%%%%%%%%%%%%%%%%%%%%%%%%%%%%%%%%%%%%%%%%%%%%%%%%%%
 %%%%%%%%%%%%%%%%%%%%%%%%%%%%%%%%%%%%%%%%%%%%%%%%%%%%%%%%%%%%%%%%%%%%%%%%%%%%%%%%%
 \hline
\multicolumn{10}{c}{\mbox{Pattern} $(m_3=0, M_{\nu 12} = 0)$} \\
\hline
 \mbox{quantity} & $\th_x$ & $\th_y$& $\th_z$ & $\r$ & $\sig$ & $\d$ & $\me$
 & $\mee$ & $J$\\
 \hline
 \multicolumn{10}{c}{\mbox{Inverted  Hierarchy}} \\
 \cline{1-10}
 $1\, \sig$ & $32.96 - 35$  &$38.65- 45$ &$7.71 -10.30$ &$0.0036 -179.97$ &$0.0259 -179.99$ &
 $ [88.17,91.88 ]\cup [268.3 , 271.95]$&$0.0466 -0.0491$ &$0.0453 - 0.0484$ &$\pm 0.0296- \pm 0.0405$  \\
 \hline
 $2\, \sig$ & $ 31.95 -36.09$&$36.88  -50.77$ &$6.29-11.68$ &$0.0384 -179.98$ &$0.0140 -179.95$ &
 $ [86.15,94.4 ]\cup [265.1 , 273.56]$&$0.0455 - 0.0502$ &$0.0435 - 0.0496$ &$\pm 0.0238 - \pm 0.0457$ \\
 \hline
 $3\, \sig$ & $ 30.98- 37.11$  &$35.67- 53.13$ &$4.05 -12.92$ &$0.0055 -179.97$ &
 $0.0227 - 179.95$&$[82.22,96.63]\cup [263.2 ,278.53]$&$0.0441-0.0514$ &$0.0415 -0.0511$ &$\pm 0.0151 - \pm 0.0506$  \\
 \hline
 %%%%%%%%%%%%%%%%%%%%%%%%%%%%%%%%%%%%%%%%%%%%%%%%%%%%%%%%%%%%%%%%%
 %%%%%%%%%%%%%%%%%%%%%%%%%%%%%%%%%%%%%%%%%%%%%%%%%%%%%%%%%%%%%%%%%%%%%%%%%%%%%%%%%
 \hline
\multicolumn{10}{c}{\mbox{Pattern} $(m_3=0, M_{\nu 13} = 0)$} \\
\hline
 \mbox{quantity} & $\th_x$ & $\th_y$& $\th_z$ & $\r$ & $\sig$ & $\d$ & $\me$
 & $\mee$ & $J$\\
 \hline
 \multicolumn{10}{c}{\mbox{Inverted  Hierarchy}} \\
 \cline{1-10}
 $1\, \sig$ & $32.96 - 35$&$38.65 - 45$ &$7.71 -10.30$ &$0.0589- 179.98$ &$0.0046 -179.94$ &
 $[86.42, 90.47]\cup [ 269.6  ,273.57]$&$ 0.0467 - 0.0491$ &$0.0448 -0.0482$ &$\pm 0.0297 - \pm 0.0406$  \\
 \hline
 $2\, \sig$ & $31.95-  36.09$ &$36.87 -50.77$ &$6.29 - 11.68$ &$0.021 - 179.99$ &$0.0016 -179.94$ &
 $[ 84.63, 93.57]\cup [ 266.7  ,275.62]$&$0.0455 - 0.0502$ &$0.0432 - 0.0497$ &$\pm 0.0238 - \pm 0.0460$ \\
 \hline
 $3\, \sig$ & $30.98 -37.11$  &$35.67 -53.12$ &$4.05- 12.92$ &$0.0554 -179.95$ &$0.0285 -179.93$&
 $[ 82.22, 97.39]\cup [ 262.2  ,276.46]$&$0.0441 - 0.0514$ &$0.0415 -0.0512$ &$\pm 0.0152 - \pm 0.0509$  \\
 \hline
 %%%%%%%%%%%%%%%%%%%%%%%%%%%%%%%%%%%%%%%%%%%%%%%%%%%%%%%%%%%%%%%%%
 \end{tabular}
 }
 \end{center}
  \caption{\small \label{tab4} The various predictions for the viable singular
  one-zero textures. All the angles (masses) are evaluated in degrees ($eV$).}
 \end{table}
\end{landscape}

%%%%%%%%%%%%%%%%%%%%%%%%%%%%%%%%%%%

%%%%%%%%%%%%%%%%%Third Table%%%%%%%%%%%%%%%%%%%
\begin{landscape}
\begin{table}[hbtp]
\begin{center}
\begin{tabular}{|c|c|c|c|c|}
\hline
&\multicolumn{2}{c|}{$m_1=0$}& \multicolumn{2}{c|}{$m_3=0$} \\
\hline
Texture &${m_2\over m_3}$ & status & ${m_2\over m_1}$ & status   \\
\hline
&&&&\\
$M_{\n\;11}=0$ & ${t_z^2\over s_x^2}$ & viable & ${1\over t_x^2}$ & unviable
\\
&&&&\\
\hline
&&&&\\
$M_{\n\;22}=0$ & ${s_y^2\,c_z^2\over s_x^2\,s_y^2\,s_z^2 + c_x^2\,c_y^2 -2\,s_x\,c_x\,s_y\,c_y\,s_z\,c_\d}$
& unviable & ${c_x^2\,s_y^2\,s_z^2 +s_x^2\,c_y^2 + 2\, s_x\,c_x\,s_y\,c_y\,s_z\,c_\d\over s_x^2\,s_y^2\,s_z^2
+ c_x^2\,c_y^2 -2\,s_x\,c_x\,s_y\,c_y\,s_z\,c_\d}$ & viable\\
&&&&\\
 & $\approx {t_y^2\over c_x^2}\,\left(1+ 2\, t_x\,t_y\, c_\d\,s_z\right) + O(s_z^2)$& &
 $\approx t_x^2\,\left( 1 + 2{c_\d t_y\over s_x\, c_x}\,s_z\right) + O(s_z^2)$& \\
 \hline
 &&&&\\
 $M_{\n\;33}=0$ & ${c_y^2\,c_z^2\over s_x^2\,c_y^2\,s_z^2 + s_x^2\,s_y^2 + 2\,s_x\,c_x\,s_y\,c_y\,s_z\,c_\d}$ &
  unviable &
  ${c_x^2\,c_y^2\,s_z^2 +s_x^2\,s_y^2 - 2\, s_x\,c_x\,s_y\,c_y\,s_z\,c_\d\over s_x^2\,c_y^2\,s_z^2
  + c_x^2\,s_y^2 +2\,s_x\,c_x\,s_y\,c_y\,s_z\,c_\d}$ & viable
\\
&&&&\\
& $\approx {1\over t_y^2 c_x^2}\,\left(1- 2\, {t_x\, c_\d\,s_z\over t_y}\right) + O(s_z^2)$& &
 $\approx t_x^2\,\left( 1 - 2{c_\d \over t_y s_x\, c_x}\,s_z\right) + O(s_z^2)$& \\
 \hline
 &&&&\\
 $M_{\n\;23}=0$ & ${s_y\,c_y\,c_z^2\over \sqrt{s_x^2\,s_y^2\,s_z^2 + c_x^2\,c_y^2 - 2\,s_x\,c_x\,s_y\,c_y\,s_z\,c_\d}\sqrt{s_x^2\,c_y^2\,s_z^2 + c_x^2\,s_y^2 + 2\,s_x\,c_x\,s_y\,c_y\,s_z\,c_\d}}$ & unviable &$ { \sqrt{c_x^2\,s_y^2\,s_z^2 + s_x^2\,c_y^2 + 2\,s_x\,c_x\,s_y\,c_y\,s_z\,c_\d}\sqrt{c_x^2\,c_y^2\,s_z^2 + s_x^2\,s_y^2 - 2\,s_x\,c_x\,s_y\,c_y\,s_z\,c_\d}\over \sqrt{s_x^2\,s_y^2\,s_z^2 + c_x^2\,c_y^2 - 2\,s_x\,c_x\,s_y\,c_y\,s_z\,c_\d}\sqrt{s_x^2\,c_y^2\,s_z^2 + c_x^2\,s_y^2 + 2\,s_x\,c_x\,s_y\,c_y\,s_z\,c_\d}}$ & unviable
\\
 &&&&\\
& $\approx {1\over  c_x^2}\,\left(1- 2\, {t_x\, c_\d\,s_z\over t_{2\,y}}\right) + O(s_z^2)$& &
 $\approx t_x^2\,\left( 1 - 2{c_\d s_z\over t_{2\,y}\,s_x\, c_x}\right) + O(s_z^2)$& \\
 &&&&\\
 \hline
&&&&\\
$M_{\n\;12}=0$ & ${s_z\,s_y\over s_x\,\sqrt{s_x^2\,s_y^2\,s_z^2 + c_x^2\,c_y^2 -2\,s_x\,c_x\,s_y\,c_y\,s_z\,c_\d}}$ & viable & ${\sqrt{c_x^2\,s_y^2\,s_z^2 +s_x^2\,c_y^2 + 2\, s_x\,c_x\,s_y\,c_y\,s_z\,c_\d}\over t_x\,\sqrt{s_x^2\,s_y^2\,s_z^2 + c_x^2\,c_y^2 -2\,s_x\,c_x\,s_y\,c_y\,s_z\,c_\d}}$ & viable
\\
&&&&\\
& $\approx {t_y\,s_z\over s_x\, c_x}\,\left(1 + c_\d\,t_y\, t_x\, s_z\right) + O(s_z^3)$& &
 $\approx  1 + {t_y\,c_\d s_z\over s_x\, c_x} -{s_z^2\,t_y^2\,\left(c_\d^2\,\left(c_x^2 - 3 \,s_x^2\right) - c_{2x}\right) \over 2 c_x^2\,s_x^2}+ O(s_z^3)$& \\
&&&&\\
\hline
$M_{\n\;13}=0$ & ${s_z\,c_y\over s_x\,\sqrt{s_x^2\,c_y^2\,s_z^2 + c_x^2\,s_y^2 +2\,s_x\,c_x\,s_y\,c_y\,s_z\,c_\d}}$ & viable & ${\sqrt{c_x^2\,c_y^2\,s_z^2 +s_x^2\,s_y^2 - 2\, s_x\,c_x\,s_y\,c_y\,s_z\,c_\d}\over t_x\,\sqrt{s_x^2\,c_y^2\,s_z^2 + c_x^2\,s_y^2 +2\,s_x\,c_x\,s_y\,c_y\,s_z\,c_\d}}$ & viable\\
&&&&\\
& $\approx {s_z\over t_y\,s_x\, c_x}\,\left(1 - {c_\d\, t_x\, s_z\over t_y} \right) + O(s_z^3)$& &
 $\approx  1 - {c_\d s_z\over t_y\,s_x\, c_x}  -{s_z^2\,\left(c_\d^2\,\left(c_x^2 - 3 \,s_x^2\right) - c_{2x}\right) \over 2 c_x^2\,s_x^2\,t_y^2} + O(s_z^3)$& \\
\hline
 \end{tabular}
\end{center}
 \caption{\small  The mass ratio formulae for the singular texture. Exact and approximate expressions for ratios are included. The status of each singular texture is indicated.}
\label{singmass}
 \end{table}
 \end{landscape}
%%%%%%%%%%%%%%%%%%%%%%%%%%%%%%%%%%%%%%%%

\section{Symmetry realization}
All neutrino mass matrices with one zero-texture can be realized in a
simple way in models based on seesaw mechanism with a flavor Abelian
symmetry. The realization method relies on the necessity of at least two zeros in $M_R$ to be
reflected as two zeros in $M_\n$ through the seesaw relation in Eq.~\ref{see-saw} and on assuming that the Dirac
neutrino mass is diagonal. Relaxing this assumption, in that Dirac neutrino mass matrix can take a non-diagonal form, then one
sole zero
in $M_\n$ can be produced. The construction should be augmented with three Higgs doublet with appropriate
flavor symmetry transformation to generate non diagonal Dirac neutrino mass matrix while keeping charged
lepton mass matrix diagonal.

We need three right-handed neutrinos $\nu_{Rj}$, three right-handed charged leptons $l_{Rj}$ and three
left-handed lepton doublets $D_{Lj}=(\nu_{Lj} , l_{Lj})^{T}$, where $j$ is the family index. Also we need the
three standard model (SM) Higgs doublets, plus other scalar singlets. We follow a similar procedure to that
of ~\cite{Lavoura} and assume a $Z_8\times Z_2$ underlying symmetry. For the sake of illustration, let us take
the case of $M_{\n\;11}=0$. Under the action of $Z_8$ factor, the leptons (the right singlets and the
components of  left doublets) of the first, second, and third families are multiplied by
$(1,-1,\om =e^{{i\,\pi/ 4}})$ respectively. As to the three SM Higgs doublets $\Phi_1$, $\Phi_2$ and
$\Phi_3$ , the first two do not change under $Z_8$, while the third is multiplied by $\omega^5$.
Under the factor $Z_2$, all fields stay invariant except $\n_{Rj}$, $\Phi_2$ and $\Phi_3$ which change sign.

The bilinears of $\overline{D}_{Li} l_{Rj}$ and $\overline{D}_{Li} \n_{Rj}$, relevant for Dirac mass matrix of
neutrino and charged leptons transform under the factor $Z_8$  as
\bea
\overline{D}_{Li}\, l_{Rj} \cong  \overline{D}_{Li}\, \n_{Rj}  & \cong &
\left(
\begin {array}{ccc}
1&\omega^4& \omega\\
\omega^4& 1& \omega^5\\
\omega^7&\omega^3&1
\end {array}
\right).
\label{M11bi}
\eea
The assigned symmetry transformation for three Higgs doublets guarantee diagonal mass matrix for charged
leptons and Dirac neutrino mass matrix of the form,
\be
M_{\n\,D}=\left(
\begin {array}{ccc}
\times &0& 0\\
0 & \times& \times\\
0 & 0& \times
\end {array}
\right),
\label{MD11}
\ee
where the cross sign denotes a non vanishing entry.

The bilinears $\nu_{Ri} \nu_{Rj}$, relevant for the Majorana neutrino mass matrix
$M_R$, transform under the factor $Z_8$  as
\bea
\nu_{Ri}\, \nu_{Rj} &\cong &
\left(
\begin {array}{ccc}
1&\omega^4& \omega\\
\omega^4& 1& \omega^5\\
\omega&\omega^5&\omega^2
\end {array}
\right),
\label{biMR11}
\eea
The $(1,1)$ and $(2,2)$ matrix elements of $M_R$ are $Z_8\times Z_2$
invariant, hence their corresponding mass terms are directly present in the
Lagrangian. We require a Yukawa coupling to a
real scalar singlet ($\chi_{12}$) which changes sign under the factor $Z_8$ to generate the
$(1,2)$ matrix element in $M_R$, when acquiring a vacuum expectation value (VEV) at the seesaw scale. The $(1,3)$
matrix element is equally generated by the Yukawa coupling to a complex
scalar singlet ($\chi_{13}$) with a multiplicative number $\omega^7$ under the factor $Z_8$.
The resulting right-handed Majorna mass matrix can be casted in the form,
\be
M_R=\left(
\begin {array}{ccc}
\times &\times& \times\\
\times& \times& 0\\
\times&0&0
\end {array}
\right),
\label{MR11}
\ee
The resulting effective Majorana mass matrix of light neutrino takes the form,
\bea
M_\n = M_D M_R^{-1} M_D^T & =&
\left(
\begin {array}{ccc}
0 &\times& \times\\
\times& \times& \times\\
\times&\times&\times
\end {array}
\right).
\label{Mn11}
\eea
which is of the required structure.

For the other patterns, they can be generated in a similar way summarized in the
Table~(\ref{oner}).
\clearpage
\begin{table}[hbtp]
\begin{center}
\begin{tabular}{|c||c||c|c|c|c|c|c|c|c|c|c|c|}
\hline
\multicolumn{13}{|c|}{$ \mbox{Symmetry under}\; Z_8\; \mbox{factor}$}\\
\hline
   \mbox{Pattern}& $\Phi_1$  & $\Phi_2$ & $\Phi_3$ & $1_F $ &$2_F $   & $3_F $ & $\chi_{11}$  &
   $\chi_{12}$& $\chi_{13}$ & $\chi_{22}$ & $\chi_{23}$ & $\chi_{33}$  \\
\hline
 ${\bf M_{\n\;11}=0}$ & $1$ & $1$ &$\omega^5$ & $1$ & $-1$ &  $\omega$ & $\times$   &
$-1$ & $\omega^7$ & $\times$ & $\times$ & $\times$
\\
\hline
 ${\bf M_{\n\;22}=0 }$ & $1$ & $1$ &$\omega$ & $1$ & $-1$ & $\omega$ & $\times$   &
 $-1$ & $\times$ & $\times$ & $\omega^3$ & $\times$\\
 \hline
 ${\bf M_{\n\;33}=0}$ & $1$ & $1$ & $\omega^5$ & $\omega$ & $-1$ &  $1$ & $\times$   &
 $\times$ & $\omega^7$ & $\times$ & $-1$ & $\times$\\
\hline
 ${\bf M_{\n\;12}=0}$ & $1$ & $1$ & $\omega$ & $1$ & $-1$ &  $\omega$ & $\times$   &
 $\times$ & $\times$ & $\times$ & $\omega^3$ & $\omega^6$\\
\hline
 ${\bf M_{\n\;13}=0}$ & $1$ & $1$ & $\omega^5$ & $\omega$ & $-1$ &  $1$ & $\omega^6$   &
 $\times$ & $\times$ & $\times$ & $-1$ & $\times$\\
\hline
${\bf M_{\n\;23}=0}$ &$1$ & $1$& $\omega$ & $1$ & $-1$ & $\omega$ & $\times$   &
 $-1$ & $\times$ & $\times$ & $\times$ & $\omega^6$\\
\hline
\multicolumn{13}{|c|}{$ \mbox{Symmetry under}\; Z_2\; \mbox{factor}$}\\
\hline
\multicolumn{13}{|c|}{$\Phi_2, \Phi_3, \n_{R1}, \n_{R2}, \n_{R3}\;\mbox{change sign, while the rest doesn't change}$}\\
\hline
\end{tabular}
\end{center}
 \caption{\small  The $Z_8\times Z_2$ symmetry realization for six patterns of
 one zero texture. The index $1_F$ indicates the lepton first
 family and so on. The $\chi_{kj}$ denotes a scalar singlet which produces
 the entry $(k,j)$ of the right-handed Majorana mass matrix when acquiring a VEV
 at the see-saw scale. The transformation properties, under the
 specified group, is listed below each family and needed scalar singlet for each
 Pattern. The cross sign indicates the absence of the corresponding scalar field.
 $\omega$ denotes $e^{{i\,\pi/4}}$. }
\label{oner}
 \end{table}

The one-zero texture singular models, with vanishing $m_1$ or $m_3$, can be accommodated
within the seesaw schemes by incorporating a singular neutrino Dirac mass matrix. As evident from
the seesaw formula in Eq.~(\ref{see-saw}), $M_R$ can not be singular; otherwise, the seesaw mechanism
would not work and thus we are obliged to introduce a singular neutrino Dirac mass matrix. The symmetry
realization can be  done similarly to the non-singular case, but we need more zeros in $M_D$
in order to make it singular. These further zeros can emerge as a result of enhancing the symmetry form
$Z_8\times Z_2$ to $Z_{12} \times Z_2$. This generic choice of the group, $Z_{12} \times Z_2$ was discussed
in \cite{Grimus}, and in the present work we use it to realize the case of singular one-zero texture
models, although it might not be the most economic choice. The assigned transformations, under the action of
the factor $Z_{12}$, for leptons are
\bea
l_{R1} \rightarrow \theta\, l_{R1} &  \nu_{R1} \rightarrow \theta\, \nu_{R1} &
\overline{D}_{L1} \rightarrow  \theta\, \overline{D}_{L1},\nn\\
l_{R2} \rightarrow \theta^2\, l_{R2} &  \nu_{R2} \rightarrow \theta^2\, \nu_{R2} &
\overline{D}_{L2} \rightarrow  \theta^3\, \overline{D}_{L2},\nn\\
l_{R3} \rightarrow \theta^5\, l_{R3} &  \nu_{R3} \rightarrow \theta^5\, \nu_{R3} &
\overline{D}_{L3} \rightarrow  \theta^8\, \overline{D}_{L3},\nn\\
\eea
 where $\t= e^{i \pi/6}$ identifying the $Z_{12}$ symmetry. The bilinears  $\overline{D}_{Lj}\,l_{Rk}$,
 $\overline{D}_{Lj}\,\nu_{Rk}$ and $\nu_{Rj}\,\nu_{Rk}$ transform as
 \bea
\overline{D}_{Lj}\, l_{Rk} \cong  \overline{D}_{Lj}\,\n_{Rk}  & \cong &
\left(
\begin {array}{ccc}
\t^2&\t^3& \t^6\\
\t^4& \t^5& \t^8\\
\t^9&\t^{10}&\t
\end {array}
\right),\;
\nu_{Rj}\, \nu_{Rk} \cong
 \left(
\begin {array}{ccc}
\t^2&\t^3& \t^6\\
\t^3& \t^4& \t^7\\
\t^6&\t^{7}&\t^{10}
\end {array}
\right).
\label{Mllsing}
\eea
The diagonal charged lepton mass matrix can be achieved by introducing only three Higss doublets which
are denoted by $\Phi_{11}$, $\Phi_{22}$ and $\Phi_{33}$ and get, respectively,  multiplied by
$\t^{10}$, $\t^7$ and $\t^{11}$ under the action of $Z_{12}$. The other demanded scalars
(singlets and doublets) and their transformation properties are dependent on the required singular one-zero
texture. As an example, we consider the singular case of $M_{\n\;11}=0$, and the remaining other cases can be
done in a similar fashion. The required forms for $M_{\nu\;D}$ -singular one- and $M_{R}$ are
\bea
M_{\nu\;D}  & = &
\left(
\begin {array}{ccc}
\times &0& 0\\
0& 0& \times\\
0 & 0 &\times
\end {array}
\right),\;
M_R =
 \left(
\begin {array}{ccc}
\times & 0 & \times\\
0 & \times & 0\\
\times & 0 & 0
\end {array}
\right),
\label{Mtex11s}
\eea
where the cross sign denotes a non vanishing element. These forms are by no means exclusive but they are
just mere possibilities. The effective singular Majorana mass matrix for the light neutrino assumes the needed form,
\bea
M_\n = M_D M_R^{-1} M_D^T & =&
\left(
\begin {array}{ccc}
0 &\times& \times\\
\times& \times& \times\\
\times&\times&\times
\end {array}
\right),
\label{Mn11s}
\eea
The non-vanishing elements of $M_\nu$ are not merely constrained due to $M_\nu$ being a symmetric matrix
but are also further constrained due to its non-invertibility. The required form of $M_{\nu\;D}$ can be
achieved by introducing scalar doublets which are denoted by $\tilde{\Phi}_{11}$, $\tilde{\Phi}_{23}$ and
$\tilde{\Phi}_{33}$. These scalars are getting respectively multiplied by $\t^{10}$, $\t^4$ and $\t^{11}$
under the action of $Z_{12}$. As to the Majorana mass matrix, $M_R$, it can be constructed by introducing
the set of scalar singlets $\chi_{11}$, $\chi_{13}$ and $\chi_{22}$, which under the action of $Z_{12}$ are
 respectively multiplied by $\t^{10}$, $\t^{6}$ and $\t^{8}$. The symmetry realization for all singular
 patterns are summarized in the
Table~(\ref{onesr}).
%\clearpage
\begin{table}[hbtp]
\begin{center}
\begin{tabular}{|c||c|c|c|c|c|c|c|c|c|c|c|c|c|c|c|}
\hline
\multicolumn{16}{|c|}{$ \mbox{Symmetry under}\; Z_{12}\; \mbox{factor}$}\\
\hline
&&&&& &&&&& &&&&&\\
   \mbox{Pattern}& $\tilde{\Phi}_{11}$  & $\tilde{\Phi}_{12}$ & $\tilde{\Phi}_{13}$ &$\tilde{\Phi}_{21}$&
   $\tilde{\Phi}_{22}$& $\tilde{\Phi}_{23}$ & $\tilde{\Phi}_{31}$& $\tilde{\Phi}_{32}$ & $\tilde{\Phi}_{33}$&
   $\chi_{11}$  & $\chi_{12}$& $\chi_{13}$ & $\chi_{22}$ & $\chi_{23}$ & $\chi_{33}$  \\
\hline
 ${\bf M_{\n\;11}=0}$ & $\t^{10}$ & $\times$ &$\times$ & $\times$ & $\times$ & $\t^4$ & $\times$&
$\times$ & $\t^{11}$ & $\t^{10}$ & $\times$ & $\t^6$ & $\t^8$ & $\times$ & $\times$
\\
\hline
 ${\bf M_{\n\;22}=0 }$ & $\times$ & $\times$ &$\t^6$ & $\times$ & $\t^7$ & $\times$ & $\times$&
$\times$ & $\t^{11}$ & $\t^{10}$ & $\times$ & $\times$ & $\t^8$ & $\t^5$ & $\times$
\\
 \hline
 ${\bf M_{\n\;33}=0}$ & $\t^{10}$ & $\times$ &$\times$ & $\t^8$ & $\times$ & $\times$ & $\times$&
$\times$ & $\t^{11}$ & $\times$ & $\times$ & $\t^6$ & $\t^8$ & $\times$ & $\t^2$\\
\hline
 ${\bf M_{\n\;12}=0}$ & $\t^{10}$ & $\times$ &$\times$ & $\times$ & $\times$ & $\t^4$ & $\t^3$&
$\times$ & $\t^{11}$ & $\t^{10}$ & $\times$ & $\times$ & $\t^8$ & $\times$ & $\t^2$\\
\hline
 ${\bf M_{\n\;13}=0}$  & $\t^{10}$ & $\times$ &$\times$ & $\t^8$ & $\times$ & $\t^4$ & $\times$&
$\times$ & $\t^{11}$ & $\t^{10}$ & $\times$ & $\times$ & $\t^8$ & $\times$ & $\t^2$\\
\hline
${\bf M_{\n\;23}=0}$ & $\t^{10}$ & $\times$ &$\times$ & $\times$ & $\times$ & $\t^4$ & $\t^3$&
$\times$ & $\t^{11}$ & $\t^{10}$ & $\times$ & $\times$ & $\t^8$ & $\times$ & $\t^2$\\
\hline
\multicolumn{16}{|c|}{$ \mbox{Symmetry under}\; Z_2\; \mbox{factor}$}\\
\hline
\multicolumn{16}{|c|}{}\\
\multicolumn{16}{|c|}{$\tilde{\Phi}_{jk}\; \mbox{and}\;  \n_{Rj} \;\mbox{change sign, while the
remaining fields do not change sign}$}\\
\hline
\end{tabular}
\end{center}
 \caption{\small  The $Z_{12}\times Z_2$ symmetry realization for $6$ patterns of singular
 one zero texture. The $\tilde{\Phi}_{kj}$ denotes a scalar dublet which produces
 the entry $(k,j)$ of the Dirac neutrino mass matrix when acquiring a VEV at the electroweak scale.
 The $\chi_{kj}$ denotes a scalar singlet which produces
 the entry $(k,j)$ of the right-handed Majorana mass matrix when acquiring a VEV at the see-saw scale.
 The transformation properties, under the
 specified group, are listed below each needed scalar singlet or doublet for each
 pattern. The cross sign indicates the absence of the corresponding scalar field.
 $\t$ denotes $e^{{i\,\pi/
6}}$. }
\label{onesr}
 \end{table}

Two points are in order here. The first one is that the models constructed above would display flavor-changing neutral Yukawa interactions,
since various Higgs doublets provide different entries of each mass matrix. However, the effects are calculable in the
models and it should be possible to suppress processes, like the decay $\mu \rightarrow e \g$, by adjusting Yukawa couplings
in such a way that the relevant elements of $Y^{\dagger}_\n Y_\n$ are quite small \cite{Grimus,Claudia}.
The second point concerns the stability of the zero-texture under radiative corrections, where the contribution of
off-diagonal elements of the neutrino Yukawa couplings to the renormalization group (RG) running of the neutrino mass matrix
can then replace the texture zero by a non-vanishing entry. Here, again, one can avoid situations, where the RG running spoils
at low energy the acceptability of a zero-texture at high energy, by choosing smaller Yukawa couplings, or a smaller mass for the LNM,
or different values for Majorana phases, all of which are not constrained strongly by the data \cite{Claudia}.

Another interesting possibility to realize the one-zero textures is achieved by using the type-II seesaw mechanism \cite{seesaw2}, where the standard model is extended by
introducing several $SU(2)_L$  scalar triplets $H_a$, $(a=1,2,\cdots N)$,
\be
H_a\equiv \left[H_a^{++}, H_a^+, H_a^0\right].
\ee
The gauge invariant Yukawa interaction relevant for neutrino mass takes the form,
\be
\mathcal{L}_{H,L} = \sum_{i,j=1}^3\sum_{a=1}^N\,Y_{ij}^a\,\left[ H_a^0 \n_{Li}^T\, \mathcal{C}\, \n_{Lj} + H_a^+ \left(\n_{Li}^T\, \mathcal{C}\, l_{Lj}
+ l_{Lj}^T \,\mathcal{C}\, \n_{Li}\right) + H_a^{++} l_{Li}^T \,\mathcal{C}\, l_{Lj}\right],
\ee
where $Y_{ij}^a$ are the corresponding Yaukawa coupling constants, the indices $i,j$ are flavor ones, and $\mathcal{C}$ is the charge conjugation matrix.

The field $H_a^0$ could acquire a small expectation value, $\langle H_a^0\rangle_0 $ that gives rise to a Majorana neutrino mass
matrix of the following form,
\be
M_{\n\;ij} =  \sum_{a=1}^N\,Y_{ij}^a\,\langle H_a^0\rangle_0.
\ee
The smallness of the vacuum $\langle H_a^0\rangle_0$ is attributed to the largeness of the triplet scalar mass scale\cite{seesaw2}.

Using type-II seesaw mechanism as described above, all possible one-zero textures can be generated by using four Higgs triplets
augmented by $Z_5$ flavor symmetry. For the purpose of illustration, the case of $M_{\n\;11}=0$ is detailed as follows. We assume that the
leptons (the right charged singlets and the components of the left doublets) of the first, second and third
families are multiplied  by $(1, \Omega=e^{i {2 \pi\over 5}}, \Omega^2)$. As to the four Higgs triplets $(H_1, H_2, H_3, H_4)$ they are
multiplied by $(\Omega^4, \Omega^3, \Omega^2, \Omega)$ respectively.  The bilinear  of $\n_{Li}\,  \n_{Lj}$ relevant for Majorana mass matrix
transforms under $Z_5$ as,
 \bea
\nu_{Li}\, \nu_{Lj} &\cong &
\left(
\begin {array}{ccc}
1&\Omega & \Omega^2\\
\Omega & \Omega^2 & \Omega^3\\
\Omega^2 &\Omega^3 &\Omega^4
\end {array}
\right).
\label{biseesaw2}
\eea
The assigned symmetry transformation for $\n_{Li}$ and $H_a$'s enforces the absence of the term $\sum_{a=1}^4\, H_a^0 \n_{L1}^T\, \mathcal{C}\, \n_{L1}$, and thus, after $H_a^0$ acquiring a vev, the resulting mass matrix assumes the form:
\bea
M_\n & =&
\left(
\begin {array}{ccc}
0 &\times& \times\\
\times& \times& \times\\
\times&\times&\times
\end {array}
\right).
\label{Mnseesaw2}
\eea
The charged lepton mass matrix is guaranteed to be diagonal provided the standard model Higgs doublet is singlet under flavor
symmetry. In our case we need only one Higgs doublet.  The other one-zero textures can be generated in a similar way, which is summarized
in Table~(\ref{seesaw2tab}).
\begin{table}[hbtp]
\begin{center}
\begin{tabular}{|c||c||c|c|c|c|c|c|}
\hline
\multicolumn{8}{|c|}{$ \mbox{Symmetry under}\; Z_5\; \mbox{factor}$}\\
\hline
   \mbox{Pattern}& $H_1$  & $H_2$ & $H_3$ & $H_4 $& $1_F $ &$2_F $   & $3_F $    \\
\hline
 ${\bf M_{\n\;11}=0}$ & $\Omega^4$ & $\Omega^3$ &$\Omega^2$ & $\Omega$ & $1$ &  $\Omega$ & $\Omega^2$
\\
\hline
 ${\bf M_{\n\;12}=0 }$ & $1$ & $\Omega^3$ &$\Omega^2$ & $\Omega$ & $1$ & $\Omega$ & $\Omega^2$   \\
 \hline
 ${\bf M_{\n\;23}=0}$ & $1$ & $\Omega^4$ &$\Omega^3$ & $\Omega$ & $1$ & $\Omega$ & $\Omega^2$   \\
 \hline
 ${\bf M_{\n\;33}=0}$ &  $1$ & $\Omega^4$ &$\Omega^3$ & $\Omega^2$ & $1$ & $\Omega$ & $\Omega^2$   \\
 \hline
 ${\bf M_{\n\;13}=0}$ & $\Omega^3$ & $\Omega^4$ &$1$ & $\Omega$ & $\Omega$ & $1$ & $\Omega^2$   \\
 \hline
${\bf M_{\n\;22}=0}$ & $\Omega^3$ & $\Omega^4$ &$1$ & $\Omega^2$ & $\Omega$ & $1$ & $\Omega^2$   \\
 \hline
\end{tabular}
\end{center}
 \caption{\small  The $Z_5$ symmetry realization for six patterns of
 one zero texture. The index $1_F$ indicates the lepton first
 family and so on. The $H_a$ denotes the a-th scalar triplet. The transformation properties, under the
 specified group, is listed below each family and needed scalar triplet for each
 pattern. $\Omega$ denotes $e^{i {2 \pi\over 5}}$. }
\label{seesaw2tab}
 \end{table}
 \\
One can see from Tables~(\ref{oner}, \ref{onesr} and \ref{seesaw2tab}), that the type-II seesaw construction, compared to the corresponding one in
type-I scheme, is simpler
in the sense of a minimal added particle content and a smaller $Z_n$ flavor symmetry group. However there is no clear way
for symmetry realization of singular Majorana mass matrix in case of type-II seesaw in contrast to the type-I case presented earlier. Nonetheless,
there are two advantages of this Type-II seesaw construction. First, the Majorana
neutrino mass matrix is renormalized multiplicativelly\cite{renseesaw} and thus the zero textures are stable under
renormalization effects. Second, the flavor changing neutral current due to the triplet is highly suppressed due to the
heaviness of the triplet mass scale, or equivalently the smallness of the neutrino masses.

\section{ Discussion and conclusions}

Taking into account the recent oscillation data that confirms relatively large value of $\t_z$, we have presented an updated
comprehensive phenomenological analysis for all the  possible patterns of Majorana neutrino mass matrices with one-zero element.
All the six possible cases allow to accommodate the current data without need to adjust the input parameters. For the chosen
acceptable parameter points, all the matrices are complex displaying CP violation effects.

Non-invertible mass matrices with one-zero texture occur only in the cases:
$M_{11}$(N), $M_{12}$ and $M_{13}$ (N and I), $M_{22}$ and $M_{33}$ (I), where the normal hierarchy (N) corresponds
to $m_1=0$ and the inverted hierarchy (I) implies $m_3=0$.

All the nonsingular textures can produce all the three types of hierarchy, except the texture $M_{32}$ which can not produce
the normal type in contrast to the texture $M_{11}$ which produces just this type. The mixing and phase angles cover their
whole experimentally allowed regions. However, in the normal type of $M_{33}$ ($M_{22}$), the angle $\t_y$ is constrained to be
larger than $49^o$ (smaller than $41^o$). Also, in the inverted type of $M_{13}$ ($M_{12}$), the phase $\d$ is bound to lie inside
the interval union $[40.72^o,140^o]\cup [221.9^o,320.81^o]$ ( $[43.77^o,136^o]\cup [223.7^o,316.82^o]$), and the other phase
angle $\r$ is forbidden to be around  $\frac{\pi}{2}$ in the texture $M_{11}$.

There exists  a  linear correlation between $\d$ and $\r$ or $\s$, for textures $M_{22}$ and $M_{33}$
in case of normal and degenerate hierarchies and for  texture $M_{23}$ in case of degenerate and inverted hierarchies.
In case of $M_{13}$ and $M_{12}$ textures there is a clear correlation but is not a linear one.  This induces a clear correlation of $J$ against
these two phases. As to the correlation ($\r,\s$), it is linear in the textures $M_{23}$ ,$M_{13}$ and $M_{12}$ in case of normal and
degenerate hierarchies but it has a kite-like shape in case of $M_{11}$. We see also that imposing a zero-texture would not in general force
the parameter $\mee$ to take  very tiny values, except for the texture $M_{11}$ where it is identically zero. Another less obvious exception is the case of normal hierarchy of the patterns
$M_{12}$ and $M_{13}$ where $\mee$ can attain very tiny values of $O(10^{-5})\,\mbox{eV}$  substantially smaller than the allowed
lower bounds in the other cases.

These features might help in distinguishing between the four independent cases ($M_{33},M_{13},M_{32},M_{11}$, say). If by
measuring the masses we note a normal hierarchy, then $M_{32}$ is excluded. We look then for the hierarchy intensity ($m_3/m_2$).
If it is larger than $2$ (in the 3-$\s$ precision level) then the pattern $M_{33}$, where the range of the ratio is $[1.4,1.9]$,
is excluded. If, then, the measurement of $m_{ee}$ gives a non-zero value then the pattern $M_{13}$ (or its T-equivalent $M_{12}$) is
the only possible pattern. On the other hand, if $m_3/m_2$ is less than $2$ then the pattern $M_{11}$, where this mass ratio
lies in $[3.3,6.2]$, is excluded and we need to choose between the pattern $M_{33}$ and the pattern $M_{13}$ (where the corresponding
interval is $[1.43,6]$). If we have $m_3/m_2 > 2$ or if $\t_y <49^o$, then
$M_{33}$ is excluded. Contrariwise, if $m_3/m_2 < 2$ and $\t_y>49^o$, then measuring ($\s,\d$) can decide which of the two patterns fit by comparing to the `narrow'
dotted regions in the plots (e-N,L) of Figs.\ref{m33fig1},~\ref{m13fig1}.

However, if the mass spectrum corresponds to inverted hierarchy type, then the acceptable patterns are ($M_{32}, M_{33}, M_{13}$).
Here, if the mass ratio ($m_2/m_3$) (in the 3-$\s$ precision level) is strong ($> 100$) then the pattern
 $M_{32}$ is excluded since this ratio in it does not exceed the value 4.5. Moreover, the measurement of $\d$ can exclude $M_{13}$ if it lies outside
 $[40.72^o,140^o]\cup [221.9^o,320.81^o]$, whereas, in the other case, the narrow linear band of the correlation ($\r,\s$) in
 Figure~\ref{m13fig1} (plot e-I,R) can help in this exclusion. This last correlation, nonetheless, does not help, when needed in the case $m_2/m_3 <100$,
 in differentiating between $M_{32}$ and $M_{33}$ (see the `blurred' plots (e-I,R) in Figs.~\ref{m33fig1},~\ref{m32fig1}).

In the degenerate case, the possible patterns are also ($M_{32}, M_{33}, M_{13}$). Although the mere measurements of $\d$
can not in its own exclude one of the textures, however, the knowledge of all the phase angles jointly, and referring to the
`narrow' bands of the correlations ($\r,\s$) for the $M_{32}, M_{13}$ cases and the linear correlations of ($\d,\s$) for $M_{33},M_{32}$
cases can help in deciding which texture does fit the data.

We note in what preceded a strong similarity with the vanishing minor analysis given in \cite{LashinChamoun2}, in line with the mapping
(Eq.~\ref{mapping-minor-zero}) swapping inverted and normal hierarchies. However, as we indicated before, this correspondence is qualitative
and can not be used to deduce the actual allowed values of the parameters defining the one-zero textures starting from those of the vanishing
minor case, due largely to the fact that the experimental bounds we have are not invariant under this mapping. Moreover, there are
singular models for the one-zero textures which are rich in phenomenology, and which are not related to any model of the vanishing
minor case, be it singular or not.

All the singular viable one-zero textures do not allow the angle $\t_z$ to be vanishing, in contrast to the  singular vanishing one
minor texture\cite{LashinChamoun2} which implies vanishing $\t_z$ and thus becomes unviable according to the recent oscillation data.
There are three viable singular textures with
normal-type hierarchy ($m_1=0$): $M_{11}, M_{13}, M_{12}$. The singular texture $M_{11}$ is viable only at the 3-$\s$ level and
is characterized by $\s=\frac{\pi}{2}$ and that $\t_z \in  [11.71^o , 12.92^o]$, whereas $\t_x <34.69^o$. In the $T_1$-related normal-type singular one-zero
textures $M_{13}$ and $M_{12}$, the difference between $2\,\s$ and $\d$ is equal to zero and  $\pi$ respectively, up to $O(s_z)$, whereas $\t_z<7.8^o$, which helps in distinguishing
them from the singular $M_{11}$ pattern.

There are four viable singular textures with inverted-type hierarchy ($m_3$=0): $M_{12}, M_{13}, M_{22}, M_{33}$. In the $T_1$-related inverted-type
singular textures ($M_{13},M_{12}$), we have $\r\approx\s$ and $\d \in [82.22^0,97.39^o]\cup [262.2^o,276.47^o]$.

The $T_1$-related inverted-type
singular textures ($M_{33},M_{22}$) are viable for all precision levels except at the 1-$\s$ level where $M_{33}$ can be accommodated  while $M_{22}$ fails. The Majorana phases are constrained to satisfy   $|\r-\s| \simeq \frac{\pi}{2}$. As
to $\d$, it is bound to be in the $1^{st}$ and $4^{th}$ quadrants outside $[62^o,300^o]$ for  $M_{22}$, whereas in $M_{33}$ it lies in the $2^{nd}$ and $3^{rd}$ quadrants inside the interval $[117.13^o,241.19^o]$. Thus the difference between $\r$ and $\s$ and the quadrants in which $\d$
lies can help in distinguishing between the four inverted-type singular one-zero textures.

In singular textures where $m_1=0$, it might be possible to attribute the  smallness of $\t_z$  to the smallness of $R_\n$ through the relations in Eq.~(\ref{m23con}). This is clear in the case of $(m_1=0, M_{\n\, 11}=0)$ which implies $t_z^2 \approx \sqrt{R_\n}\,s_x^2$. However, this will not work in the other viable patterns of $m_1=0$, because there will be a dependency on $\d$ besides the mixing angles.

All the one-zero patterns can be realized in the framework of flavor Abelian discrete
symmetry implemented in seesaw schemes. In the case of type-I scheme, the group is ($Z_8 \times Z_2$), and we need three scalar doublets, and two additional scalar singlets in the invertible patterns transforming appropriately.
As to the singular patterns for the neutrino mass matrix, and although they enjoy a larger symmetry ($Z_{12} \times Z_2$), their realizations required more scalar fields. Actually, we needed three scalar singlets and at most seven scalar doublets. As to the type-II seesaw scheme, we presented alternative simple realizations based on introducing several Higgs triplets without introducing right handed neutrinos. The flavor symmetry turns out to be $Z_5$ which is simpler than those required for the case of type-I seesaw.

Finally, and compared to preceding analyses of the one-zero textures (\cite{rod1} and references therein), we can summarize the new points carried out in our work as follows. First, we have updated the viability testing for the textures considering the new oscillation data. Second, we have pointed out a relation between the one-zero textures and the one-vanishing-minor textures analyzed in \cite{LashinChamoun2} and used it to cross-check the calculations. Third, we signaled the importance of the $T_1$-symmetry relating different patterns with each other, and used it to check the consistency of the results in the related patterns. Fourth, and due to the random sampling method we used, we could cover the whole parameter space rather than limiting the study to representative points. Last, we have presented symmetry realizations theoretically motivated by the seesaw mechanism.

\section*{{\large \bf Acknowledgements}}
Part of the work was done
within the associate scheme and short visits program of ICTP.\\  N.C. acknowledges funding provided by the Alexander von Humboldt Foundation.

\end{document}